\documentclass[aps,prd,superscriptaddress,eqsecnum,nofootinbib,preprint,tightenlines,floatfix,titlepage]{revtex4-2}

\usepackage{bm}
\usepackage{graphicx}
\usepackage{float}
\usepackage[pdftex]{color}
\usepackage{amsmath, amssymb}
\usepackage{array}
\usepackage{rotating}
\usepackage{mathtools}
\usepackage{multirow}
\usepackage{changepage}
\usepackage[normalem]{ulem}
\usepackage{verbatim}

\usepackage[sort&compress]{natbib}

\usepackage[hyperfootnotes=false]{hyperref}
\hypersetup{
    colorlinks=true,     
    linkcolor=blue,      
    citecolor=blue,      
    filecolor=blue,      
    urlcolor=blue        
}

\usepackage[capitalise]{cleveref}

\crefname{section}{Sec.}{Secs.}
\Crefname{section}{Section}{Sections}
\crefrangeformat{equation}{Eqs.~(#3#1#4)--(#5#2#6)}
\Crefrangeformat{equation}{Equations~(#3#1#4)--(#5#2#6)}
\crefrangeformat{figure}{Figs.~#3#1#4--#5#2#6}
\Crefrangeformat{figure}{Figures~#3#1#4--#5#2#6}

\newcommand{\bi}{\begin{itemize}}
\newcommand{\ei}{\end{itemize}}

\newcommand{\ben}{\begin{enumerate}}
\newcommand{\een}{\end{enumerate}} 

\newcommand{\be}{\begin{equation}}
\newcommand{\ee}{\end{equation}}

\newcommand{\bea}{\begin{eqnarray}}
\newcommand{\eea}{\end{eqnarray}}

\newcommand{\nn}{\nonumber}

\newcommand{\chpt}{$\chi$PT}

\newcommand{\amu}{a_\mu}
\newcommand{\amuL}{a_\mu^{ll}(\mathrm{conn.})}
\newcommand{\amuHVP}{a_\mu^{\mathrm{HVP,LO}}}

\newcommand{\amuW}{a^{\mathrm W}_{\mu}}

\newcommand{\amuWTwo}{a^{\mathrm{W2}}_{\mu}}

\newcommand{\amuLW}{a^{ll,{\mathrm W}}_{\mu}(\mathrm{conn.})}
\newcommand{\amuLWTwo}{a^{ll,\mathrm {W2}}_{\mu}(\mathrm{conn.})}

\newcommand{\amuLWRes}{a^{ll,{\mathrm W}}_{\mu}(\mathrm{conn.}) = 206.6(1.0) \times 10^{-10}}
\newcommand{\amuLWTwoRes}{a^{ll,\mathrm {W2}}_{\mu}(\mathrm{conn.}) = 100.7(3.2) \times 10^{-10}}

\newcommand{\gmtwo}{$g-2$} 

\newcommand{\pr}{\operatorname{pr}}

\newcommand{\coloaf}{Department of Physics, University of Colorado, Boulder, Colorado 80309, USA}
\newcommand{\fnalaf}{Theory Division, Fermi National Accelerator Laboratory, Batavia, Illinois, 60510, USA}
\newcommand{\iuaf}{Department of Physics, Indiana University, Bloomington, Indiana 47405, USA}
\newcommand{\mitaf}{Center for Theoretical Physics, Massachusetts Institute of Technology, Cambridge, \\MA 02139, USA}
\newcommand{\msuaf}{Department of Computational Mathematics, Science and Engineering, and Department of Physics and Astronomy, Michigan State University, East Lansing, Michigan 48824, USA}

\newcommand{\ugraf}{CAFPE and Departamento de Física Teórica y del Cosmos, Universidad de Granada, \\E-18071 Granada, Spain}
\newcommand{\uiucaf}{Department of Physics, University of Illinois, Urbana, Illinois, 61801, USA}
\newcommand{\icasuuiaf}{Illinois Center for Advanced Studies of the Universe, University of Illinois, Urbana, Illinois, \\61801, USA}
\newcommand{\unizar}{Departmento de Física Teórica, Universidad de Zaragoza, 50009 Zaragoza, Spain}
\newcommand{\utahaf}{Department of Physics and Astronomy, University of Utah, Salt Lake City, UT 84112, USA}
\newcommand{\glasaf}{SUPA, School of Physics and Astronomy, University of Glasgow, Glasgow, G12 8QQ, \\United Kingdom}
\newcommand{\cornaf}{Laboratory for Elementary-Particle Physics, Cornell University, Ithaca, NY 14853, USA}
\newcommand{\plyaf}{Centre for Mathematical Sciences, University of Plymouth, Plymouth PL4 8AA, \\United Kingdom}

\begin{document}

\preprint{FERMILAB-PUB-23-006-T}

\title{Light-quark connected intermediate-window contributions to the muon \texorpdfstring{\boldmath\gmtwo}{g-2}\ hadronic vacuum polarization from lattice QCD}

\author{Alexei~Bazavov}\affiliation{\msuaf}
\author{Christine Davies}\affiliation{\glasaf}
\author{Carleton~DeTar}\affiliation{\utahaf}
\author{Aida~X.~El-Khadra}\affiliation{\uiucaf}\affiliation{\icasuuiaf}
\author{Elvira~Gámiz}\affiliation{\ugraf}
\author{Steven~Gottlieb}\affiliation{\iuaf}
\author{William~I.~Jay}\affiliation{\mitaf}
\author{Hwancheol~Jeong}\affiliation{\iuaf}
\author{Andreas~S.~Kronfeld}\affiliation{\fnalaf}
\author{Shaun~Lahert}\email{slahert@illinois.edu}\affiliation{\uiucaf}\affiliation{\icasuuiaf}
\author{G.~Peter~Lepage}\affiliation{\cornaf}
\author{Michael~Lynch}\affiliation{\uiucaf}\affiliation{\icasuuiaf}
\author{Andrew~T.~Lytle}\affiliation{\uiucaf}\affiliation{\icasuuiaf}
\author{Paul~B.~Mackenzie}\affiliation{\fnalaf}
\author{Craig~McNeile}\affiliation{\plyaf}
\author{Ethan~T.~Neil}\affiliation{\coloaf}
\author{Curtis~T.~Peterson}\email{curtis.peterson@colorado.edu}\affiliation{\coloaf}
\author{Gaurav~Ray}\affiliation{\plyaf}
\author{James~N.~Simone}\affiliation{\fnalaf}
\author{Ruth~S.~\surname{Van~de~Water}}\affiliation{\fnalaf}
\author{Alejandro~Vaquero}\affiliation{\utahaf}\affiliation{\unizar}

\collaboration{Fermilab Lattice, HPQCD, and MILC Collaborations}
\noaffiliation

\date{\today}

\begin{abstract}
We present a lattice-QCD calculation of the light-quark connected contribution to window observables associated with the leading-order hadronic vacuum polarization contribution to the anomalous magnetic moment of the muon, $\amuHVP$. We employ the MILC Collaboration's isospin-symmetric QCD gauge-field ensembles, which contain four flavors of dynamical highly-improved-staggered quarks with four lattice spacings between $a\approx 0.06$--$0.15$~fm and close-to-physical quark masses. We consider several effective-field-theory-based schemes for finite-volume and other lattice corrections and combine the results via Bayesian model averaging to obtain robust estimates of the associated systematic uncertainties. After unblinding, our final results for the intermediate and ``W2'' windows are $\amuLWRes$ and $\amuLWTwoRes$, respectively.
\end{abstract}

\maketitle

\raggedbottom
\allowdisplaybreaks

\section{Introduction}\label{sec:intro}

In April 2021, the Fermilab muon \gmtwo\ experiment, E989, released their first result for the muon's anomalous magnetic moment $a_\mu\equiv(g_{\mu}-2)/2$ based on Run-1 data collected in 2018~\cite{PhysRevLett.126.141801}.
When combined with the previous measurement from Brookhaven National Lab experiment E821~\cite{Bennett:2006fi},
the new result for the muon's anomalous magnetic moment increases the disagreement with the Standard Model (SM) theory prediction~\cite{Aoyama:2020ynm}\footnote{The SM prediction is based on a large body of theoretical work~\cite{Aoyama:2012wk, Aoyama:2019ryr, Czarnecki:2002nt, Gnendiger:2013pva, Davier:2017zfy, Keshavarzi:2018mgv, Colangelo:2018mtw, Hoferichter:2019mqg, Davier:2019can, Keshavarzi:2019abf, Kurz:2014wya, Melnikov:2003xd, Masjuan:2017tvw, Colangelo:2017fiz, Hoferichter:2018kwz, Gerardin:2019vio, Bijnens:2019ghy, Colangelo:2019uex, Blum:2019ugy, Colangelo:2014qya}, and reflects the consensus of the muon $g-2$ theory community.} from 3.7$\sigma$ to 4.2$\sigma$.
Because the anomaly arises from loop effects, it is sensitive to the contributions of yet-undiscovered particles that could give rise to small deviations from the theoretical prediction.
Increased precision is now essential to say conclusively if this substantial difference is from physics beyond the SM.

The error on the experimental average of the muon's anomalous magnetic moment is now 0.35 parts per million (ppm), and is limited by statistics. 
Fermilab E989 continues to collect data and improve the experimental apparatus, and ultimately aims to measure $a_\mu$ to a precision at or below 0.14~ppm by the end of its lifetime.
Additionally, a new complementary experiment to measure the muon's anomalous magnetic moment and electric dipole moment is planned for later this decade at J-PARC in Japan~\cite{Abe:2019thb,E34webpage}.  
The J-PARC E34 experiment will employ a different method to determine $a_\mu$ than the ``magic momentum'' approach of both Fermilab E989 and BNL E821 and aims for a precision of 0.45~ppm from its initial run in 2027~\cite{Abe:2019thb,Keshavarzi:2021eqa}.

Corresponding theoretical efforts are underway to reduce the uncertainty on the prediction for $a_\mu$ in the SM, which currently stands at 0.37~ppm~\cite{Aoyama:2020ynm}. At present, over 90\% of the SM theory error comes from the leading-order hadronic vacuum polarization (HVP) contribution to the anomaly, $\amuHVP$. 
The HVP contribution is difficult to determine precisely because the bulk of it comes from the low-energy, nonperturbative regime of quantum chromodynamics (QCD). 
To date, the most precise theoretical results for $\amuHVP$ are obtained from a data-driven, dispersive approach~\cite{Bouchiat:1961lbg,PhysRev.168.1620} using experimental measurements of the total cross section for $e^+e^-\to\text{hadrons}$ (the so-called $R$~ratio) as input.
These data-driven determinations have achieved around 0.5\% precision on $\amuHVP$ corresponding to 0.34~ppm uncertainty on $a_\mu$~\cite{Davier:2017zfy,Keshavarzi:2018mgv,Colangelo:2018mtw,Hoferichter:2019mqg,Davier:2019can,Keshavarzi:2019abf} and are the basis of the Muon $g-2$ Theory Initiative's SM prediction for $\amuHVP$~\cite{Aoyama:2020ynm}.

Lattice QCD provides an alternative, {\it ab initio}, approach for calculating the leading-order HVP contribution that is independent of experimentally measured cross sections.\footnote{A small number of experimentally-measured quantities are employed in lattice-QCD calculations to fix the quark masses and lattice spacing in the QCD Lagrangian.} 
The most precise lattice QCD calculation of $\amuHVP$ to~date (and the first with sub-percent precision) comes from the BMW collaboration~\cite{Borsanyi:2020mff}.  
Although BMW's result implies a SM value for $\amu$ that is within $1.5\sigma$ of experiment, it differs from the $R$-ratio based prediction of Ref.~\cite{Aoyama:2020ynm} by 2.1$\sigma$. 
Independent lattice-QCD calculations with commensurate precision are therefore urgently needed to address this theoretical discrepancy.

The leading-order HVP contribution to the muon's anomalous magnetic moment is computed in lattice QCD as a weighted integral over Euclidean time of the two-point correlation function of the quarks' electromagnetic vector current~\cite{Blum:2013xva, Bernecker:2011gh}.
By judiciously restricting the integration range (or ``window''), one can construct sub-quantities of $\amuHVP$ that avoid problematic statistical and/or systematic effects~\cite{RBC:2018dos,Aubin:2022hgm,Davies:2022epg}. 
The same Euclidean-time window observables can also be obtained from $R$-ratio data by including a suitable weight function in the dispersive integral for $\amuHVP$~\cite{Colangelo:2022vok}.
Because Euclidean-time windows allow for more detailed and sensitive comparisons between independent $\amuHVP$ calculations, they are a valuable tool both for diagnosing sources of disagreement between lattice-QCD results and for quantifying differences (if any) between data-driven and lattice determinations. 

Various Euclidean-time windows with differing positive features and drawbacks have been proposed~\cite{RBC:2018dos,Aubin:2022hgm,Davies:2022epg}.
In 2018, the RBC and UKQCD Collaborations separated the Euclidean-time integral into contributions from ``short''  ($t \lesssim 0.4$~fm), ``intermediate'' ($0.4 \lesssim t \lesssim 1.0$~fm), and ``long'' ($t \gtrsim 1.0$~fm) times~\cite{RBC:2018dos}.
The intermediate window observable $\amuW$ can be computed in lattice QCD with high statistical precision.  Hence, it has been adopted by the muon $g-2$ theory community as a benchmark quantity.
Several independent three- and four-flavor lattice-QCD calculations of $\amuW$ are now available~\cite{RBC:2018dos,Borsanyi:2020mff,Ce:2022kxy,Alexandrou:2022amy}, but the results are not fully consistent.
RBC/UKQCD's initial intermediate-window result~\cite{RBC:2018dos} is within about 1$\sigma$ of the determination from $R$-ratio data~\cite{Colangelo:2022vok}.  
More recent lattice-QCD calculations of $\amuW$ by the BMW~\cite{Borsanyi:2020mff}, Mainz/CLS~\cite{Ce:2022kxy}, and ETM~\cite{Alexandrou:2022amy} collaborations, however, are all more than 3$\sigma$ higher than the data-driven value.\footnote{The RBC/UKQCD collaboration's update \cite{Blum:2023qou}, which appeared on arXiv on the same day as our paper, is in good agreement with these recent results.} 
Further scrutiny of the intermediate window is therefore needed to clarify the picture. 

In this work, we calculate the intermediate-window contribution to $\amuHVP$ in four-flavor lattice QCD.
Using the same methods, we also calculate the ``W2'' window observable introduced by Aubin {\it et al.}~\cite{Aubin:2022hgm}, which corresponds to the Euclidean-time range $t \in [1.5, 1.9]$~fm. As pointed out in that work,
although $\amuWTwo$ is statistically noisier than $\amuW$, effective-field theory (EFT) estimates of finite-volume, lattice-discretization, and pion-mass corrections are more reliable at larger times. 
We focus exclusively on the connected contribution from light (up and down) quarks in the isospin-symmetric limit, which accounts for about 90\% of $\amuHVP$. 
Calculations of the heavier quark flavors, isospin-breaking corrections and quark-disconnected contributions are in progress~\cite{FermilabLattice:2017wgj, Yamamoto:2018cqm, FermilabLattice:2019dbx, FermilabLattice:2021hzx}.   

Our calculation of the intermediate and W2 window observables in this work builds upon our 2019 calculation of $\amuHVP$~\cite{Davies:2019efs}.
As before, we employ the MILC collaboration's dynamical-QCD gauge-field configurations~\cite{MILC:2012znn} with four flavors of highly-improved-staggered quarks (HISQ)~\cite{Follana:2006rc}.  Our numerical simulations are again performed at the physical pion mass and with four lattice spacings ranging from about $0.15$ to $0.06$~fm. 
Since our earlier work, however, we have increased statistics significantly at our three finest lattice spacings. 
The new data give better control of the lattice-dependence of the Euclidean-time window observables and enable stringent tests of the EFT-based corrections, which inform our analysis of the associated systematic errors. 
We estimate the uncertainties on $\amuW$ and $\amuWTwo$ from making different, reasonable analysis choices for finite-volume corrections and treating discretization effects, among others, via Bayesian model averaging~\cite{Jay:2020jkz,Neil:2022joj}.
Finally, to avoid confirmation bias, the Euclidean-time window observables were blinded until the analysis and error budgets were finalized. (See Sec.~\ref{sec:blinding} for details.) 

This paper is organized as follows. First, in \cref{sec:windows} we provide analytic expressions for $\amuW$ and $\amuWTwo$ in terms of the Euclidean-time vector-current correlation function.  Next, in \cref{sec:isosymQCD} we define the isospin-symmetric QCD limit employed here and describe our numerical correlator computations in \cref{sec:lat}. In \cref{sec:analysis} we present a detailed description of our analysis procedures, starting with how blinding was applied and removed in \cref{sec:blinding}. Briefly, in \cref{sec:integrals}, we use the lattice correlators to calculate the Euclidean-time windows corresponding to our numerical-simulation parameters. We then correct these ``raw'' window values on each ensemble for the finite lattice volume, slight mistunings of the simulated pion mass, and (optionally) remove taste-breaking discretization effects in \cref{sec:lat_corrections}.
Next, we extrapolate the corrected window values to zero lattice spacing in \cref{sec:cont_extrap}.
Sections~\ref{sec:BMA} and~\ref{sec:errors} describe our procedure for Bayesian model averaging and the resulting systematic error budget.
We conclude in \cref{sec:conclusion} by presenting our final results for $\amuW$ and $\amuWTwo$ and comparing them with previous determinations. Appendices~\ref{sec:appendix:fit} and~\ref{sec:chiralModel} provide additional details on obtaining the Euclidean-time windows from staggered correlators and on computing corrections to the windows using the chiral model of pions, photons, and $\rho$ mesons introduced in Ref.~\cite{Chakraborty:2016mwy}, respectively.
Progress reports on related ongoing work can be found in Refs.~\cite{Yamamoto:2018cqm, FermilabLattice:2019dbx, FermilabLattice:2021hzx, Lahert:2021xxu, Ray:2022ycg}.

\section{Preliminaries}

\subsection{Definitions of Euclidean-time window observables}\label{sec:windows}


The hadronic vacuum polarization function $\Pi(Q^2)$ can be obtained from Euclidean vector-current correlation functions through the equations
\begin{align}
\Pi^{\mu \nu}(Q^{2})&= \left(\delta^{\mu \nu} Q^{2}-Q^{\mu} Q^{\nu}\right) \Pi(Q^{2})=\int \mathrm{d}^{4} x e^{i Q x}\left\langle J^{\mu}(x) J^{\nu}(0)\right\rangle, \\
J^{\mu}(x)&= \sum_{f} q_{f} \bar{\psi}_{f}(x) \gamma^{\mu} \psi_{f}(x), \label{eq:vecCurrent}
\end{align}
where $J^{\mu}(x)$ is the electromagnetic current summed over quark flavors $f=\{u,d,s,c,b,t\}$, $q_f$ are the corresponding electric charges in units of $e$, and $\langle J^{\mu}(x) J^{v}(0)\rangle$ includes both quark-line connected and disconnected Wick contractions. The HVP contribution to the muon's anomalous magnetic moment can then be obtained from a weighted integral of $\hat{\Pi}(Q^2) \equiv \Pi(Q^2)-\Pi(0)$ via \cref{em-rep}.

It is now standard for lattice-QCD HVP calculations, however, to employ the alternative time-momentum representation introduced by Bernecker and Meyer~\cite{Bernecker:2011gh}. This formulation is more convenient for an inherently space-time approach such as lattice QCD and allows the construction of Euclidean-time windows. Starting with the spatial-vector-current correlation function $C(t)$, defined as 
\begin{align} 
    C(t)&= \frac{1}{3} \sum_{\mathbf{x}, k}\left\langle J^{k}(\mathbf{x}, t) J^{k}(0)\right\rangle ,
    \quad\quad\quad
    k=1,2,3,
    \label{eq:corrFunc2pt}
\end{align}
$\amuHVP$ is obtained via 
\begin{align}
    a_{\mu}^{\mathrm{HVP}, \mathrm{LO}} &= 4\alpha^{2} \int_{0}^{\infty} \mathrm{d} t\, C(t)  \tilde K(t),
    \label{eq:amuTint}\\
    \tilde{K}(t) &= 2 \int_{0}^{\infty} \frac{\mathrm{d} Q}{Q}\, K_{E}(Q^{2})
        \left[Q^{2} t^{2}-4 \sin ^{2}\left(\frac{Q t}{2}\right)\right],
    \label{eq:Ktilde}
\end{align}
where $K_E(Q^2)$ is given in \cref{eq:KE}. 

\begin{figure}
\centering
\includegraphics[scale=0.73]{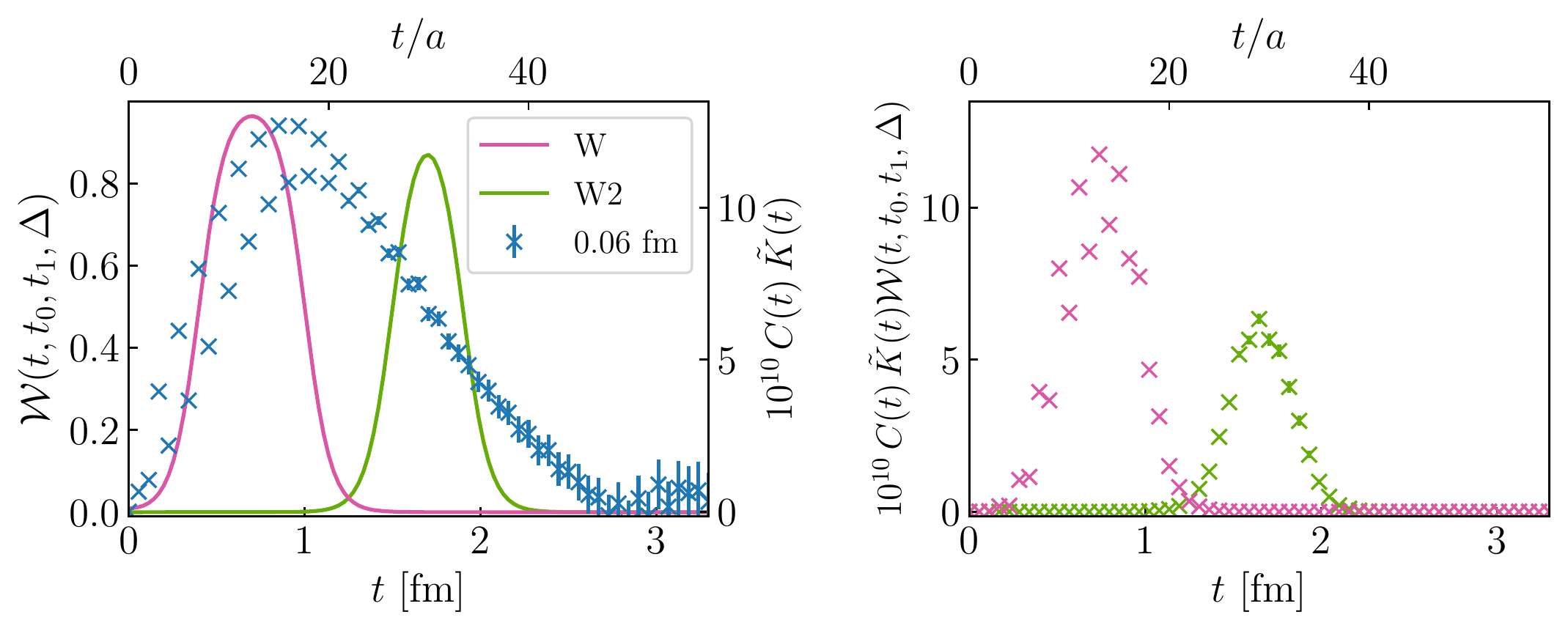}
\vspace{-5mm}
\caption{(Left) The W (magenta) and W2 (green) window functions (corresponding to the parameters in \cref{eqn:WDef,eqn:WTwoDef}) overlaid with raw lattice data for the integrand of \cref{eq:amuTint} (blue crosses) from our finest ensemble. (Right)  The windowed integrand of \cref{eq:amuTintWin} for the corresponding window functions using the lattice data in the left panel.}
\label{fig:window06Data}
\end{figure}

The window observables are then easily obtained by introducing the window function $\mathcal{W}$, limiting the Euclidean-time region over which $C(t)$ is integrated~\cite{RBC:2018dos}:
\begin{align}
a_\mu^{\mathrm{win}(t_0, t_1, \Delta)} &= 4 \alpha^{2} \int_{0}^{\infty} \mathrm{d} t\, C(t)\tilde{K}(t)   \mathcal{W}\left(t, t_0, t_1, \Delta\right), \label{eq:amuTintWin}\\
\mathcal{W}\left(t, t_0, t_1, \Delta\right)&=\frac{1}{2}\left[\tanh \left(\frac{t-t_{0}}{\Delta}\right)-\tanh \left(\frac{t-t_{1}}{\Delta}\right)\right] + \left(t \to -t\right). \label{eq:windofunc}
\end{align}
The parameters $t_0$ and $t_1$ of $\mathcal{W}$ control the location of the window's boundaries, while $\Delta$ controls the sharpness of its edges. In this work, we consider two such windows, the intermediate window W,
\begin{align}
\amuW &\equiv  a_\mu^{\mathrm{win}(\textrm{0.4, 1, 0.15})} \label{eqn:WDef} ,
\end{align}
and W2,
\begin{align}
\amuWTwo &\equiv  a_\mu^{\mathrm{win}(\textrm{1.5, 1.9, 0.15})} \label{eqn:WTwoDef} , 
\end{align} 
with the parameters in fm. We plot $\mathcal{W}$ in \cref{eq:windofunc} for these window regions in the left panel of \cref{fig:window06Data}. 

It is convenient in lattice-QCD calculations of $\amuHVP$ to separately compute and then sum up the contributions from each quark flavor and from connected and disconnected Wick contractions. Here we focus on the light-quark connected contribution to the Euclidean-time windows, $\amuLW$ and $\amuLWTwo$, in the isospin-symmetric limit. Therefore, our electromagnetic vector current $J^{\mu}(x)$ includes only the terms for light quarks with $m_l = (m_u + m_d)/2$ and our correlation function $C(t)$ includes only the connected contractions. 
\Cref{fig:window06Data}, left, shows the light-quark connected contribution to the integrand [${\tilde K}(t)C(t)$ in \cref{eq:amuTint}] using the lattice correlation function obtained on our finest ensemble (see \cref{sec:lat}). Figure~\ref{fig:window06Data}, right, shows the corresponding window integrands [$C(t)\tilde{K}(t)   \mathcal{W}\left(t, t_0, t_1, \Delta\right)$ in \cref{eq:amuTintWin}] for the W and W2 windows. 

\subsection{Prescription for isospin-symmetric QCD}\label{sec:isosymQCD}

Both the MILC HISQ gauge-field configurations and light-quark connected correlators employed in this work correspond to the isospin symmetric limit of QCD, {\it i.e.}, a pure-QCD world with equal-mass up- and down-quark masses and without electromagnetism. 
Following the prescription introduced (for three flavors) in Refs.~\cite{HPQCD:2004hdp,MILC:2004qnl} and later extended to include the charm quark in Ref.~\cite{FermilabLattice:2014tsy}, we set the light-quark masses and lattice scale in physical units using the pion mass and decay constant.
We then set the strange- and charm-quark masses using the kaon and $D_s$-meson masses, respectively.

Prior to the tuning procedure, however, electromagnetic effects must be removed from the experimental inputs.
Neither $M_{\pi^0}$ nor $f_{\pi^+}$ are affected significantly by the quarks' electric charges, so their pure-QCD values are defined to be $M_{\pi} \equiv M_{\pi^0}$ and $f_{\pi} \equiv f_{\pi^+}$.
In the Fermilab Lattice and MILC Collaboration's most recent analysis of pseudoscalar-meson masses and decay constants~\cite{Bazavov:2017lyh}, the numerical values for these inputs were taken from the 2016 Particle Data Group:
$M_{\pi^0} = 134.977$~MeV and $f_{\pi^+}=130.50(1)_{\rm exp.}(3)_{V_{ud}}(13)_\text{EM}$~MeV~\cite{ParticleDataGroup:2016lqr}.
The remaining pure-QCD meson masses employed in Ref.~\cite{Bazavov:2017lyh} are $M_{K^0}=497.567$~MeV, $M_{K^+}=491.405$~MeV, and $M_{D_s}=1967.02$~MeV.
Details on how these values were obtained can be found in Sec.~IV of that work and references therein.
For the inputs in isospin-symmetric QCD we use the same values for $M_{\pi^0}$ and $f_{\pi^+}$ as above, while the kaon mass is defined as the average of $M_{K^0}$ and $M_{K^+}$, giving $M_K = 494.486$~MeV.

\subsection{Lattice-QCD ensembles and correlation functions}\label{sec:lat}

\begin{figure}[t]
\centering
\includegraphics[scale=0.75]{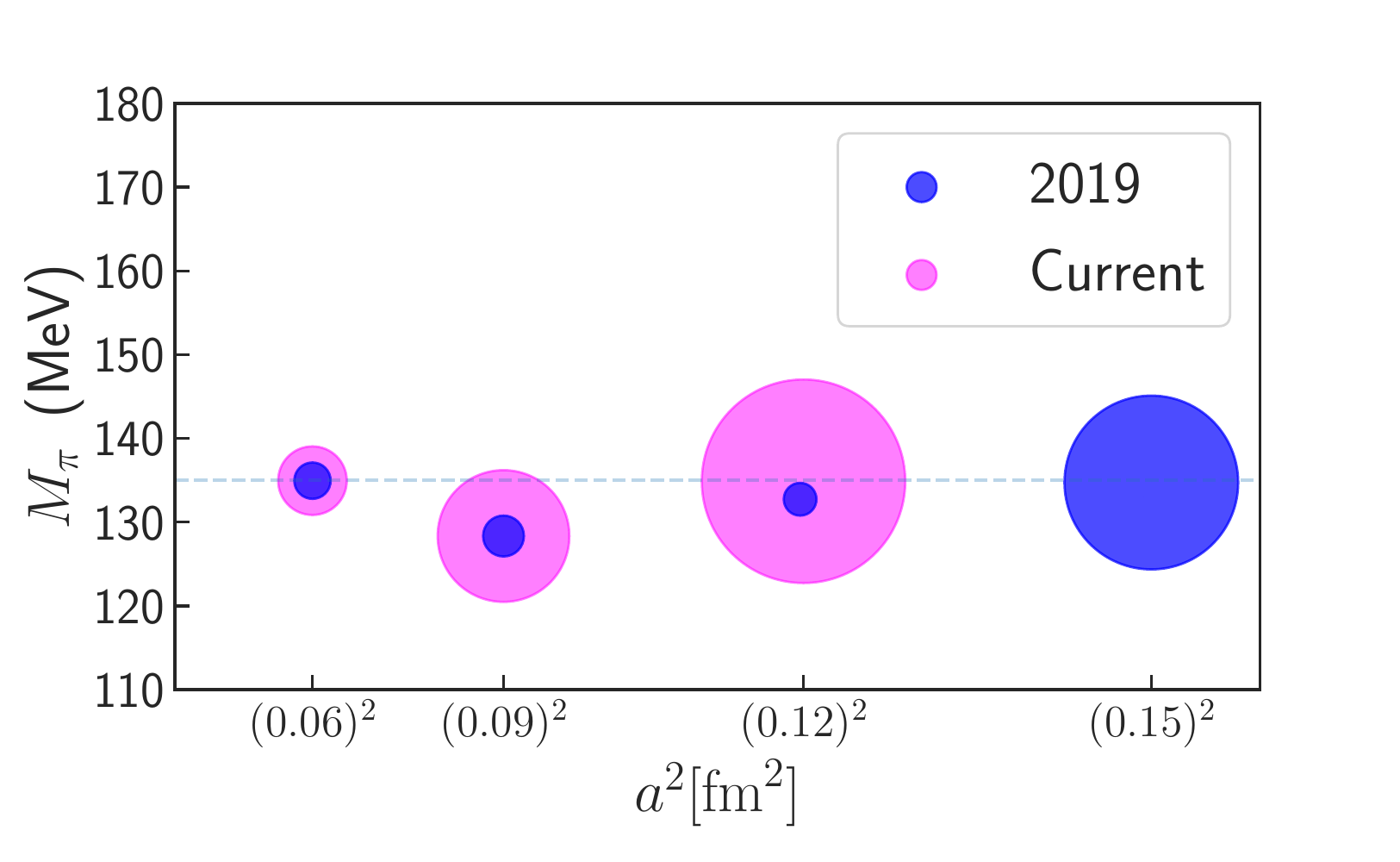}
\vspace{-6mm}
\caption{Visualization of the ensemble parameters and statistics employed in this work (labeled ``Current'') and in our previous $\amuHVP$ calculation~\cite{Davies:2019efs} (labeled ``2019''). Each disk is centered at the corresponding ensemble's squared lattice spacing and pion mass ($a^2$ and $M_{\pi5}$ in \cref{table:enssolves}), while the disk areas are proportional to the size of each data set ($N_{\text{conf}} \times N_{\text {loose sources}}$ in \cref{table:enssolves}).}
\label{fig:ensembles} 
\end{figure}

Our calculation employs the MILC Collaboration's four-flavor lattice-QCD configurations with dynamical up, down, strange, and charm quarks. The ensembles use the HISQ action~\cite{Follana:2006rc} for the sea quarks, a Symanzik-improved gauge action~\cite{Symanzik:1983gh,Luscher:1984xn,Luscher:1985zq,Alford:1995hw,Hart:2008sq} that includes the plaquette, the $1\times2$ rectangle, and the so-called bent-chair 6-link term for the gluon fields as well as tadpole improvement~\cite{Lepage:1992xa} based on the plaquette. Details of the configuration generation can be found in Ref.~\cite{MILC:2010pul}.

In this work, we employ a subset of the available MILC ensembles, for which the quark masses are well tuned to their physical values.  We include ensembles at four lattice spacings $a\approx 0.15$, 0.12, 0.09, and 0.06~fm.
The high-statistics ensemble at $a \approx 0.15$~fm is unchanged from Ref.~\cite{Davies:2019efs}, where additional details can be found.
The ensemble at $a\approx 0.12$~fm was generated specifically for our muon $g-2$ project, and has better tuned sea-quark masses compared with the ensemble with the same bare coupling used previously~\cite{Davies:2019efs}. 
It contains about 10,000 configurations. 
We also extended the ensemble with $a\approx 0.09$~fm~\cite{MILC:2012znn} to include over 5,000 configurations; this is a factor of roughly $3.5$ beyond what was used in Ref.~\cite{Davies:2019efs}. 
The pion mass for this ensemble is less accurately tuned than for the other three ensembles used in our study, which were generated more recently using quark masses obtained from a detailed analysis of pseudoscalar mesons and their decay constants~\cite{Bazavov:2017lyh}.
Finally, we increased the number of configurations in our finest ensemble with lattice spacing $a \approx 0.06$~fm~\cite{Bazavov:2017lyh} by about a factor of two compared with Ref.~\cite{Davies:2019efs}. We are continuing to extend this ensemble in anticipation of future needs. 
Our ensemble set is visualized in Fig.~\ref{fig:ensembles} and detailed in~\cref{table:enssolves}.

\begin{table*}[t]
\centering
\caption{Ensemble parameters used in this work. The first column lists the approximate lattice spacings in~fm. The second column gives the spatial length $L$ of the lattices in~fm. The third column lists the volumes of the lattices in number of space-time points. The fourth column gives the sea-quark masses in lattice-spacing units. The fifth column lists the ratios of the gradient-flow scale $w_0$~\cite{Borsanyi:2012zs} to the lattice spacing, where we take these values from Ref.~\cite{Davies:2019efs} except for the newer ensemble with $a\approx 0.12$~fm. To convert simulation results to physical units, we take $w_0=0.1715(9)$~fm from Ref.~\cite{Dowdall:2013rya}. The sixth column gives the taste-Goldstone pion masses \cite{MILC:2012znn}. The seventh column lists the renormalization factors for the local vector current, taken from Ref.~\cite{Chakraborty:2017hry}. The second-last column lists the number of configurations analyzed. The last column gives the number of loose-residual solves per configuration used in the truncated solver method \cite{TSM1, TSM2}.\vspace{1mm}}
\label{table:enssolves}
\begin{tabular}{lllclllcc}
\hline \hline
$\approx a/\mathrm{fm}$ & $L/\mathrm{fm}$ & $N_s^3 \times N_t$ & $a m_{l}^\text{sea} / a m_{s}^\text{sea} / a m_{c}^\text{sea}$ & \multicolumn{1}{c}{$w_{0} / a$} & $M_{\pi_{5}}/\mathrm{MeV}$ & \multicolumn{1}{c}{$Z_V$} & $N_{\text{conf}}$ & $N_{\text {loose}}$ \\ 
\hline
$0.15$ & $4.85$ & $ 32^3 \times 48$ & 0.002426/0.0673/0.8447 & $1.13215(35)$ & 134.73(71) & 0.9881(10) & 9362 & 48 \\ 
$0.12$ & $5.83$ & $48^3 \times 64$ & 0.001907/0.05252/0.6382 & 1.41110(59) & 134.86(71) & 0.9922(4) & 9637 & 64 \\ 
$0.09$ & $5.62$ & $64^3 \times 96$ & 0.00120/0.0363/0.432 & 1.95180(70) & 128.34(68) & 0.9940(5) & 5384 & 48 \\ 
$0.06$ & $5.46$ & $96^3\times128$ & 0.0008/0.022/0.260 & 3.0170(23) & 134.95(72) & 0.9950(6) & 2621 & 24 \\ 
\hline \hline
\end{tabular}
\end{table*}
The tuned quark masses listed in \cref{table:enssolves} are determined from the analysis in Ref.~\cite{Bazavov:2017lyh} in which pseudoscalar-meson masses and decay constants were computed using 24 gauge ensembles with six lattice spacings ranging from $\approx 0.15$ to 0.03~fm.  The pion decay constant is used to set the scale and the meson masses used to determine the up, down, strange, and charm masses are given in \cref{sec:isosymQCD}. Further details may also be found in Ref.~\cite{FermilabLattice:2014tsy} which used fewer configurations for a similar study.

The light-quark propagators from which the correlation functions $C(t)$ are constructed are computed using the HISQ action and truncated solver method (TSM) \cite{TSM1, TSM2}. Using random-wall sources, we compute one fine-residual conjugate gradient solve and the number of loose-residual solves ($N_{\text {loose}}$) shown in the last column of \cref{table:enssolves}. Compared with Ref.~\cite{Davies:2019efs}, we have increased the number of loose sources per configuration by factors of 4, 3, and 1.5 for the ensembles at $a \approx 0.12, 0.09, 0.06$~fm, respectively. Exploiting time-reversal invariance, we further increase statistics by averaging the correlator values at times $t$ and $N_t -t$ on each configuration. For the electromagnetic-current operator, we use the same local taste-vector vector current as in Ref.~\cite{Davies:2019efs}. To match the local vector current to continuum QCD, we use the nonperturbatively computed renormalization factors obtained by the HPQCD Collaboration in Ref.~\cite{Chakraborty:2017hry}. Specifically, for the $a \approx 0.15, 0.12$, and 0.09~fm ensembles, we take the ``H-H'' $Z_{V^4}$ values from Table IV, while for the $a\approx 0.06$~fm ensemble, we use the extrapolated value at this lattice spacing given in Appendix~B of that work. To test our corrections for pion-mass mistuning (see \cref{sec:DeltaMpi}), we generated additional vector- and pseudoscalar-current correlation functions with unphysical valence-quark masses on our two coarsest ensembles. \Cref{table:PQsims} lists the valence-quark masses used in these partially quenched simulations.

\begin{table}
\centering
\caption{Additional valence-quark masses used to study the pion-mass dependence of $\amuW$ and $\amuWTwo$. Simulation parameters not listed are the same as in Table~\ref{table:enssolves}.\vspace{1mm}}
\label{table:PQsims}
\begin{tabular}{lcc}
\hline \hline
$\approx a/\mathrm{fm}$ & $am_{q}^\text{val}$ & $M_{\pi_{5}}^\text{val}/\mathrm{MeV}$ \\
\hline
$0.15$ & 0.001524 & 107.28(56) \\
 & 0.003328 & 157.16(83) \\
\hline
$0.12$ & 0.001190 & 107.05(56) \\
 & 0.002625 & 157.65(83) \\
\hline \hline
\end{tabular}
\end{table}

In the course of our current analysis, we discovered two mistakes in Ref.~\cite{Davies:2019efs} pertaining to the $a \approx 0.06$~fm ensemble.  First, a small subset ($\approx$ 5\%) of this ensemble's correlation functions were affected by a software bug in the data processing script. Second, the renormalization factor employed for this ensemble was taken from the arXiv version of Ref.~\cite{Chakraborty:2017hry}, and differs from the published result by $\approx 0.1\%$. 
The latter error was realized after we unblinded our analysis (see \cref{sec:blinding}). Hence, while keeping the analysis procedure frozen, we now use the published value of $Z_{V^4}$ at $a\approx 0.06$~fm from Ref.~\cite{Chakraborty:2017hry} in our determinations of the window observables.
We are preparing errata to Refs.~\cite{Davies:2019efs, Davies:2022epg}, but do not expect the results for $\amuL$, $\amuHVP$, or the one-sided Euclidean-time windows to change significantly.

\section{Data analysis}\label{sec:analysis}

Here we present the analysis procedure to obtain the window observables $\amuLW$ and $\amuLWTwo$ at the physical, isospin-symmetric, pion mass and in the continuum and infinite-volume limits. First, as discussed in \cref{sec:blinding}, we blinded the analyses of all components of $\amuHVP$ to avoid unintentional introduction of bias. Second, because the correlation functions $C(t)$ are obtained at discrete Euclidean times, the integral in \cref{eq:amuTintWin} must be approximated by a discrete integration rule. As described in \cref{sec:integrals}, we use both the trapezoidal and Simpson's rules to quantify the associated discretization effects. 

The resulting lattice data for $\amuLW$ and $\amuLWTwo$ must then be corrected for finite-volume effects, pion-mass mistunings and (optionally) taste-breaking effects. Our estimates of these lattice corrections to the intermediate and W2 window observables in \cref{sec:lat_corrections} are based on the use of EFTs and EFT-inspired models that capture the dominant low-energy, two-pion physics contribution to $\amuHVP$.
In particular, we consider variations obtained from four different approaches: next-to-leading-order (NLO) and next-to-next-to-leading order (NNLO) chiral perturbation theory (\chpt)  \cite{Aubin:2015rzx,Bijnens:2017esv,Aubin:2020scy,Aubin2019,Borsanyi:2020mff,Aubin:2022hgm}; the Chiral Model (CM) \cite{Chakraborty:2016mwy} employed in Ref.~\cite{Davies:2019efs}; the Meyer-Lellouch-L\"uscher-Gounaris-Sakurai (MLLGS) approach \cite{Gounaris:1968mw, Luscher:1991cf, Luscher:1990ux,Lellouch:2000pv, Lin:2001ek, Meyer:2011um,Francis:2013fzp,DellaMorte:2017dyu}; and the relativistic pion EFT approach of Hansen and Patella (HP) \cite{Hansen:2020whp}. In the absence of data-driven guidance, the spread of EFT-based corrections provides an especially important indicator of the underlying uncertainties.

As is well known, staggered actions include additional, unphysical degrees of freedom (so-called ``tastes''), yielding a 16-fold enlarged meson spectrum at finite lattice spacing~\cite{Aubin:2004wf,MILC:2009mpl,MILC:2010pul,MILC:2012znn}.
The splittings between the tastes are a lattice artifact that vanishes in the continuum limit.  At finite lattice spacing, taste splittings of the pion masses are a significant discretization effect in $\amuHVP$ observables (so-called taste-breaking effects). As discussed in \cref{sec:lat_corrections}, the splittings also affect the pion-mass and finite-volume dependencies, resulting in an interplay between them. In the case of pseudoscalar meson masses and weak matrix elements, discretization effects due to the taste-splittings are well-described by staggered chiral perturbation theory~\cite{Lee:1999zxa,Bernard:2001yj}, providing an additional handle on the continuum extrapolations. 

Our continuum extrapolation analysis in \cref{sec:cont_extrap} includes a comprehensive study of taste breaking and other discretization effects. For each of the two window observables, we perform continuum extrapolations with and without first correcting for taste-breaking effects. In addition, we vary the fit function used for the continuum extrapolations, and we also include continuum-limit fits dropping the data at the coarsest lattice spacing ($a\approx 0.15$~fm). The fit function used in our continuum extrapolation contains the strong coupling constant $\alpha_s$.  In this work, following Ref.~\cite{Bazavov:2017lyh} we use $\alpha_s = \alpha_V(2/a)$ and take $\alpha_V (n_f= 4, \mu=5.0~{\rm GeV}) = 0.2530(38)$ from Ref.~\cite{Chakraborty:2014aca}, where we evolve the coupling using the four-loop beta function.

Our lattice-QCD calculations of $\amuLW$ and $\amuLWTwo$ entail numerous analysis choices. As described in \cref{sec:BMA}, we incorporate the systematic uncertainties due to these variations using Bayesian model averaging (BMA)~\cite{Jay:2020jkz,Neil:2022joj}. The remaining uncertainties in the corrected data sets include the statistical errors from the Monte-Carlo integration and parametric errors from $w_0$, $w_0/a$, and $Z_V$, which are propagated through the analysis as Gaussian random variables.
Our final results for $\amuLW$ and $\amuLWTwo$ and error budgets from the respective BMA analyses are presented and discussed in \cref{sec:errors}.

\subsection{Blinding}\label{sec:blinding}

To avoid confirmation bias, we blinded this analysis until the systematic error budgets were finalized. The analysis was then frozen and used to generate the unblinded results and figures presented here. We employ a software blinding procedure, in which each observable is multiplied by an unknown random factor, chosen from a uniform distribution between $[0.7,1.3]$. As the correlation function on the $a \approx 0.15$~fm ensemble is unchanged from Ref.~\cite{Davies:2019efs}, we additionally blinded the results from this ensemble by adding to it an offset equal to  its standard deviation times an undisclosed random number between $[-1,1]$. Each observable receives its own unique blinding factor, which is kept the same for all lattice spacings, except 0.15~fm. This procedure allows us to unblind specific sub-quantities, such as the intermediate window observables discussed here, without unblinding other quantities for which our analyses are ongoing.

\subsection{Extraction of window observables}\label{sec:integrals}

After the light-quark-connected vector-current correlation functions $C(t)$ are obtained on each ensemble as described in \cref{sec:lat}, lattice values for $\amuLW$ and $\amuLWTwo$ are computed from these correlators via \cref{eq:amuTintWin}, using a chosen numerical integration scheme.

Because we employ a single-time-slice vector-current operator in our computations, the spectral representation of our staggered correlation functions consists of a sum of positive contributions from states with the desired parity and, additionally, contributions that oscillate in time as $(-1)^{t/a}$ from opposite-parity states (see \cref{eq:corrfitfunc} of \cref{sec:appendix:fit}). These oscillations are discretization effects, and should in principle be removed via the continuum extrapolation in \cref{sec:cont_extrap}. To quantify any residual uncertainty or bias on $\amuLW$ and $\amuLWTwo$ from oscillations in our correlation functions, we also perform our full analysis using the fit-reconstructed correlator without oscillations, $C_{\textrm{no osc.}}(t)$, which is defined in \cref{eq:fitReconOsc}. \cref{sec:appendix:fit} also presents a number of alternate schemes for removing the unwanted oscillating terms from $C(t)$.  

Numerical integration of the lattice correlators introduces additional discretization errors that depend upon the method used. Here we consider two integration schemes: the trapezoidal rule and Simpson's rule, which is formally higher order in the lattice spacing. Given a Euclidean-time correlator $C(t)$, windows of $\amu$ are obtained with the trapezoidal rule via
\begin{align}
a^{\mathrm{win}(t_0, t_1, \Delta)}_{\mu, \text{ Trap.}} &= 4 \alpha^{2} a\sum_{t=1}^{N_t/2-1} C(t)\tilde{K}(t)   \mathcal{W}\left(t, t_0, t_1, \Delta\right), \label{eq:amuTsumWin}
\end{align}
where the integration kernel $\tilde{K}(t)$ and window function $\mathcal{W}(t, t_0, t_1, \Delta)$ are given in \cref{eq:Ktilde,eq:windofunc}, respectively, and the boundary terms are omitted as $\tilde{K}(t)\mathcal{W}(t, t_0, t_1, \Delta)=0$ for these cases. Similarly, Euclidean-time windows are obtained with Simpson's rule via
\begin{align}
a^{\mathrm{win}(t_0, t_1, \Delta)}_{\mu, \text{ Simp.}} &= 4 \alpha^{2} \frac{a}{3}\left[\left(4 \sum_{t \in\left\{ t_{\text{odd}} \right\} }^{N_t/2-1} +2 \sum_{t \in \left\{ t_{\text{even}} \right\} }^{N_t/2-1}\right) C(t)\tilde{K}(t)\mathcal{W}\left(t, t_0, t_1, \Delta\right) \right]. \label{eq:amuTsumWinSimp}
\end{align}

\begin{figure}[t]
\centering
\includegraphics[width=0.496\textwidth]{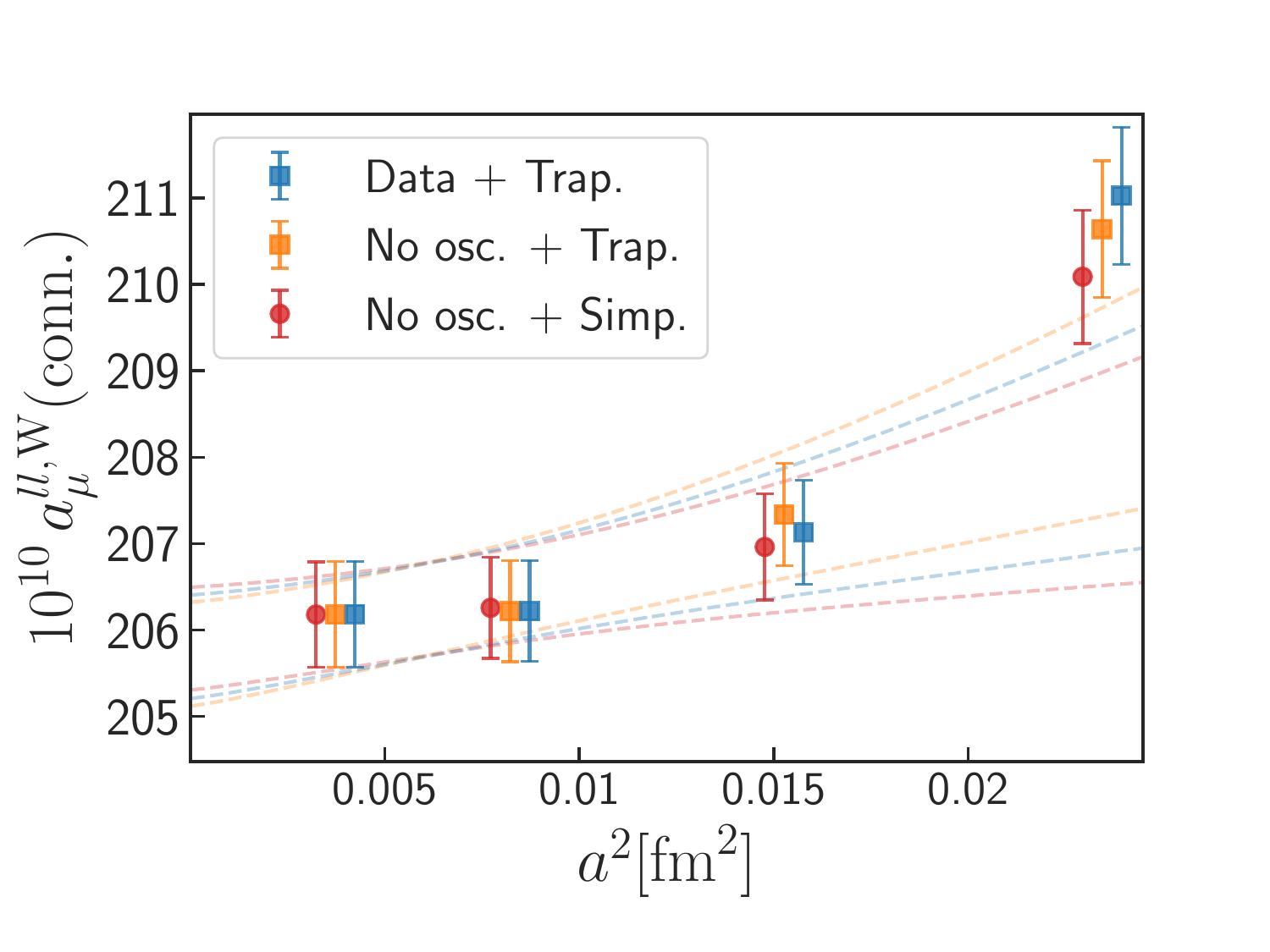} \includegraphics[width=0.496\textwidth]{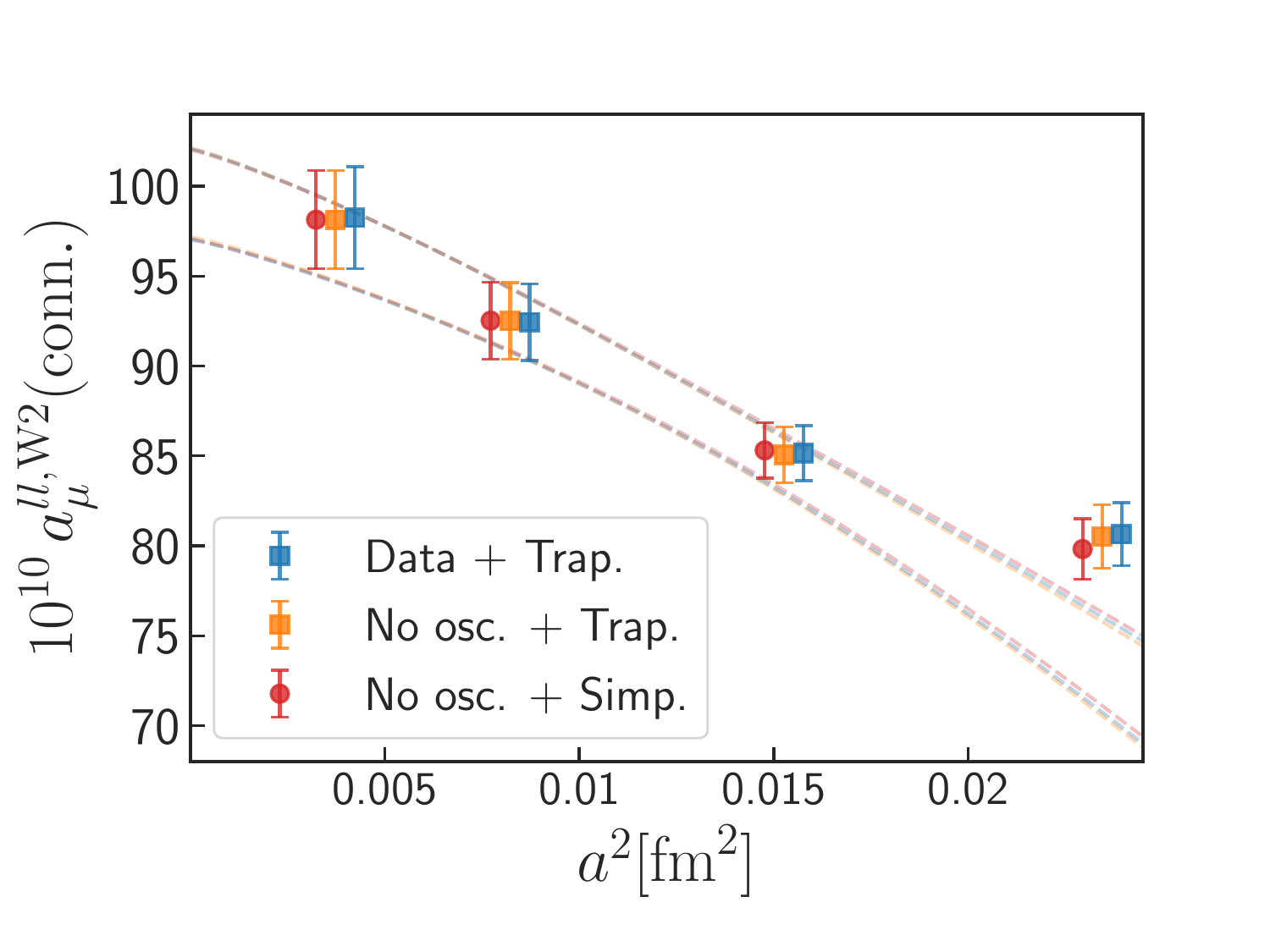}
\vspace{-10mm}
\caption{Comparison of results for $\amuLW$ (left) and $\amuLWTwo$ (right) from integrating the raw lattice correlator $C(t)$ with the trapezoidal rule (blue squares), fit reconstruction with oscillating-state contributions removed $C_{\textrm{no osc.}}(t)$ with the trapezoidal rule (orange squares), and $C_{\textrm{no osc.}}(t)$ with Simpson's rule (red circles). Data at the same lattice spacing are offset horizontally for visibility. As described in \cref{sec:DeltaFV,sec:DeltaMpi}, each point is corrected for finite-volume effects using the CM and pion-mass mistuning effects using the data-driven approach. For each integration scheme, we fit the data for the three finest ensembles to a function linear in $\alpha_s a^2$. The dashed curves show the fits' error bands, with colors matching the corresponding plot symbols.}
\label{fig:oscRemovalAnalysis}
\end{figure}

Figure~\ref{fig:oscRemovalAnalysis} compares lattice data for $\amuLW$ (left) and $\amuLWTwo$ (right) obtained by integrating $C(t)$  using the trapezoidal rule (blue squares), integrating $C_{\textrm{no osc.}}(t)$ using the trapezoidal rule (orange squares), and integrating $C_{\textrm{no osc.}}(t)$ using Simpson's rule (red circles). To enable meaningful comparisons between $\amuLW$ (or $\amuLWTwo$) at different lattice spacings, all data in these plots include corrections for the finite spatial volumes and pion-mass mistuning using the CM. (See \cref{sec:lat_corrections} for details.) 

The impact of temporal oscillations in our staggered lattice correlators on $\amuLW$ and $\amuLWTwo$ can be assessed by comparing the data sets obtained from integrating both $C(t)$ and $C_{\textrm{no osc.}}(t)$ using the trapezoidal rule (blue and orange squares in \cref{fig:oscRemovalAnalysis}, respectively). For $\amuLW$, the trapezoidal-rule data sets are statistically indistinguishable at our two finest lattice spacings (see \cref{table:OscVSNoOsc} in \cref{sec:appendix:fit}, which provides the correlated pairwise differences). Further, the differences between them decrease rapidly with the lattice spacing.  For $\amuLWTwo$, which corresponds to a later Euclidean time range, oscillations in the correlator from heavier opposite-parity states have largely died out (see \cref{fig:window06Data}). Consequently, the trapezoidal-rule data sets are statistically consistent on all ensembles and there is no clear lattice-spacing dependence in their correlated differences. As, for both $\amuLW$ and $\amuLWTwo$, the continuum extrapolations of the trapezoidal-rule data sets are in excellent agreement, we therefore conclude that temporal oscillations are an insignificant source of discretization error in our calculation.

Similarly, discretization errors stemming from the numerical integration can be estimated comparing the data sets obtained by integrating $C_{\textrm{no osc.}}(t)$ with either the trapezoidal rule or Simpson's rule (orange squares and red circles in \cref{fig:oscRemovalAnalysis}, respectively). As is displayed in the figure and quantified in \cref{table:trapVSsimp}, on our coarse ensembles the differences between integration schemes are statistically significant for both $\amuLW$ and $\amuLWTwo$. These differences, however, decrease with lattice spacing much faster than $a^2$ and are already at the per-mille level at our finest lattice spacing. We therefore conclude that lattice artifacts from the choice of numerical integration scheme are negligible compared to the leading discretization terms in the Symanzik effective Lagrangian, which are of ${\mathcal O}(\alpha_s a^2)$ (see \cref{sec:cont_extrap} for details). 

Based on these observations, we generate two data sets for each of the two observables ($\amuLW$ and $\amuLWTwo$). The first is obtained from integrating the original correlation function data $C(t)$ with the trapezoidal rule, and the second is obtained from integrating the reconstructed correlation function data $C_{\textrm{no osc.}}(t)$ with Simpson's rule. The inclusion of both data sets in the subsequent analysis accounts for any residual systematic effects due to both $\mathcal{O}(a^2)$ artifacts induced by the trapezoidal rule as well as the oscillating contributions that remain after the continuum extrapolation.  For each observable, the two data sets are taken as inputs in the next step of the analysis, where the corrections are applied to the $\amu$ data.

\subsection{Lattice corrections}\label{sec:lat_corrections}

Before taking the continuum limit, we correct our lattice $\amuLW$ and $\amuLWTwo$ data in two (or three) separate steps: first for finite volume (FV), second for pion-mass mistuning ($M_\pi$), and (sometimes) third for the effects of taste splittings (TB).  The last step is optional, since changing discretization effects will only alter the window observables' lattice-spacing dependence, not their continuum-limit values. Further, the pion-mass corrections to the intermediate and W2 windows are only numerically significant on the $a \approx 0.09$~fm ensemble for which the simulated pion mass is $\sim 5\%$ below the physical value (see \cref{table:PQsims}).

Mathematically, our correction scheme is defined via the following equations:
\begin{equation}
a_{\mu}\left(L_{\infty}, M_{\pi_{\text {phys }}}\right)=a_{\mu}\left(L_{\text {latt}}, M_{\pi_{\text {lat }, \xi_{1}}}, \cdots, M_{\pi_{\text {lat }, \xi_{16}}}\right)+\Delta_{\mathrm{FV}}+\Delta_{M_{\pi}}+\Delta_{\mathrm{TB}},  \label{eq:corrScheme}
\end{equation}
where
\begin{flalign}
\Delta_{\mathrm{FV}}&=a_{\mu}\left(L_{\infty}, M_{\pi_{\mathrm{lat}, \xi_{1}}}, \cdots, M_{\pi_{\mathrm{lat}, \xi_{16}}}\right)-a_{\mu}\left(L_{\text {latt }}, M_{\pi_{\mathrm{lat}, \xi_{1}}}, \cdots, M_{\pi_{\mathrm{lat}, \xi_{16}}}\right), \label{eq:corrFV} \\
\Delta_{M_{\pi}}&=a_{\mu}\left(L_{\infty}, M_{\pi_{\mathrm{phys}, \xi_{1}}}, \cdots, M_{\pi_{\mathrm{phys}, \xi_{16}}}\right)-a_{\mu}\left(L_{\infty}, M_{\pi_{\mathrm{lat} , \xi_{1}}}, \cdots, M_{\pi_{\mathrm{lat}, \xi_{16}}}\right), \label{eq:corrMpi} \\
\Delta_{\mathrm{TB}}&=a_{\mu}\left(L_{\infty}, M_{\pi_{\mathrm{phys}}}\right)-a_{\mu}\left(L_{\infty}, M_{\pi_{\mathrm{phys}, \xi_{1}}}, \cdots, M_{\pi_{\mathrm{phys}, \xi_{16}}}\right).  \label{eq:corrTB}
\end{flalign}
``$a_{\mu}$'' is shorthand for either $\amuLW$ or $\amuLWTwo$. The first term on the right-hand-side of \cref{eq:corrScheme} is the window observable on each ensemble obtained in \cref{sec:integrals}.
The three corrections in \cref{eq:corrFV,eq:corrMpi,eq:corrTB} are evaluated for each ensemble and added to the lattice values for $\amu$. The first correction, $\Delta_{\mathrm{FV}}$ in \cref{eq:corrFV}, takes $\amuLW$ or $\amuLWTwo$ from the simulated spatial volume, indicated by $L_{\rm latt}$, to the infinite-volume limit, denoted by $L_{\infty}$. 
The second correction, $\Delta_{M_{\pi}}$, takes the simulated taste-Goldstone pion mass to the physical value, while the final correction, $\Delta_{\mathrm{TB}}$, removes the effects of the pion taste-splittings, $a^2\Delta_{\xi_{i}}$. In practice, the lattice and ``physical'' staggered-pion masses $M_{\pi_{\mathrm{lat}, \xi_{i}}}$ and $M_{\pi_{\mathrm{phys}, \xi_{i}}}$ in \cref{eq:corrFV,eq:corrMpi,eq:corrTB} are calculated via the leading-order staggered \chpt\ relationship $M^2_{\pi_{\xi_{i}}} = M^2_{\pi_{\xi_{1}}} + a^2\Delta_{\xi_{i}}$ with  $M_{\pi_{\mathrm{lat}, \xi_{1}}}$ and $M_{\pi_{\mathrm{phys}, \xi_{1}}}$ fixed to the taste-Goldstone pion mass (column six of \cref{table:enssolves}) and the experimentally-measured $\pi^0$ mass, respectively.

The order in which the finite-volume, pion-mass, and taste-breaking corrections are applied impacts the form of the corrections. In our case, we apply the corrections in \cref{eq:corrScheme} from left to right.  Therefore, the finite-volume and pion-mass mistuning corrections in \cref{eq:corrFV,eq:corrMpi} must preserve the taste splittings. The left-hand-side of \cref{eq:corrScheme} is the infinite-volume, physical pion-mass and finite-lattice-spacing $\amu$ with correct physical parameters, which are inputs to the continuum extrapolations described in \cref{sec:cont_extrap}.

We employ four different effective-field-theory-based schemes for the finite-volume and taste-breaking corrections:
\begin{itemize}
\item Chiral Perturbation Theory (\chpt) at next-to-leading order (NLO) and next-to-next-to-leading order (NNLO) \cite{Bijnens:2017esv,Borsanyi:2020mff, Aubin2019}. 
The staggered NNLO \chpt\ expressions of Refs.~\cite{Aubin2019,Borsanyi:2020mff,Aubin:2022hgm} are derived for a taste-singlet vector current, which couples to taste-diagonal pion pairs. We adapt the expressions of Ref.~\cite{Aubin:2022hgm} to the taste-vector vector current employed here, which couples to taste-nondiagonal two-pion states. We replace the pion energies in Eq.~(3.3) of Ref.~\cite{Aubin:2022hgm} with averages of the energies of the two pions in the two-pion states which contribute to the taste-vector vector current.\footnote{For the term labeled NNLO,4 in the finite-volume correction of \cite{Aubin2019,Borsanyi:2020mff}, we substitute the average masses of the two-pion state into the energies instead of substituting the average energies. This avoids a numerical instability in this term for the case of unequal energies.} We test this approximation for the case of NLO \chpt\ (and for the CM and MLLGS approaches discussed below) where we have exact formulas for the taste-vector current. These tests reveal at most sub-percent differences in the corrections computed using the exact approach versus the approximation.

\item The Chiral Model (CM) is an extension of \chpt, where the $\rho$ meson is included explicitly through a massive spin-1 vector field.
This model was introduced by Jegerlehner and Szafron to study $\rho-\gamma$ mixing in $e^+e^- \to \pi\pi$ scattering~\cite{JegerlehnerModel}. It was first applied to Euclidean-space lattice-QCD calculations of the muon $g-2$ HVP (with modifications to incorporate the staggered-pion mass spectrum) by the HPQCD Collaboration~\cite{Chakraborty:2016mwy}.\footnote{In Refs.~\cite{Borsanyi:2020mff} and~\cite{Aubin:2022hgm}, the staggered Chiral Model is denoted ``SRHO."}
The equations for the CM corrections computed in this analysis differ from Refs.~\cite{Chakraborty:2016mwy,Davies:2019efs} slightly in that the effects of taste-breaking are included as specified in \crefrange{eq:discrete}{eq:sigma}.

\item The  Meyer-Lellouch-L\"uscher-Gounaris-Sakurai (MLLGS) approach combines the pion form-factor parameterization of Gounaris-Sakurai with the mapping (due to Meyer-Lellouch-L\"uscher~\cite{Gounaris:1968mw, Luscher:1991cf, Luscher:1990ux, Lellouch:2000pv, Lin:2001ek, Meyer:2011um, Francis:2013fzp, DellaMorte:2017dyu}) between the infinite-volume scattering amplitude and finite-volume energies and amplitudes of the two-pion states. We account for taste-breaking effects in the same fashion as Ref.~\cite{Borsanyi:2020mff}, by including contributions from two-pion states constructed with all 16 tastes of pions. Here, we modify the expressions of Ref.~\cite{Borsanyi:2020mff} to the case of the taste-vector vector current which couples to taste-nondiagonal two-pion states. As in \cite{Borsanyi:2020mff}, we fix the number of finite-volume states to $n=8$. 

\item The relativistic-pion effective-field-theory approach by Hansen and Patella (HP) for finite-volume effects \cite{Hansen:2020whp}. We obtain the correction defined in \cref{eq:corrFV} using the same replacement as described in the \chpt\ description above.
\end{itemize}

We describe the above schemes as ``effective-field-theory-based" because, in some parts of our analysis, they may be employed outside the schemes' ranges of validity. Except for the CM, these EFTs and phenomenological models include only the contributions to $\amuHVP$ observables from two-pion intermediate states.\footnote{Although \chpt, MLLGS, and HP do not treat the $\rho$ meson as a dynamical degree-of-freedom, they implicitly incorporate some resonance effects through parameters that are tuned to match experiment.}  Because contributions to the Euclidean-time correlation function fall off as $\exp(-Et)$ (see \cref{sec:appendix:fit}), those from low-lying $\pi\pi$ states are most important at large Euclidean times. Consequently, the correction schemes listed above should best describe the volume and pion-mass dependence of $C(t)$ and, hence, $a_\mu^{\mathrm{win}(t_0,t_1,\Delta)}$, for later time ranges. Indeed, we and other collaborations find that, for $t_0 \gtrsim 1.5$~fm (which includes the `W2' window), all of the higher-order correction schemes enumerated above ({\it i.e.}, excluding NLO \chpt) yield similar predictions for the finite-volume, pion-mass, and taste-breaking corrections to $a_\mu^{{\rm win}(t_0,t_1,\Delta)}$. Further, the estimates from these schemes for the sum of finite-volume, pion-mass, and taste-breaking corrections reasonably describe the observed differences between lattice data in this region.~\cite{DellaMorte:2017dyu,Gerardin:2019rua,Borsanyi:2020mff,Aubin:2022hgm}.
Therefore, they can be reliably used to calculate lattice corrections to $\amuLWTwo$. In the intermediate-window region, the predicted corrections from the EFT-based schemes display a wider variation. The sizes of the finite-volume and pion-mass corrections to $\amuLW$, however, are numerically small ($\Delta^{\rm W}_{{\rm FV},M_{\pi}} \lesssim 0.5\%$), and we incorporate the spread in $\amuLW$ results obtained with different  correction schemes in our systematic error estimate in \cref{sec:BMA}. 

\subsubsection{Finite-volume corrections}\label{sec:DeltaFV}

\begin{figure}[t]
\centering
\includegraphics[scale=0.85]{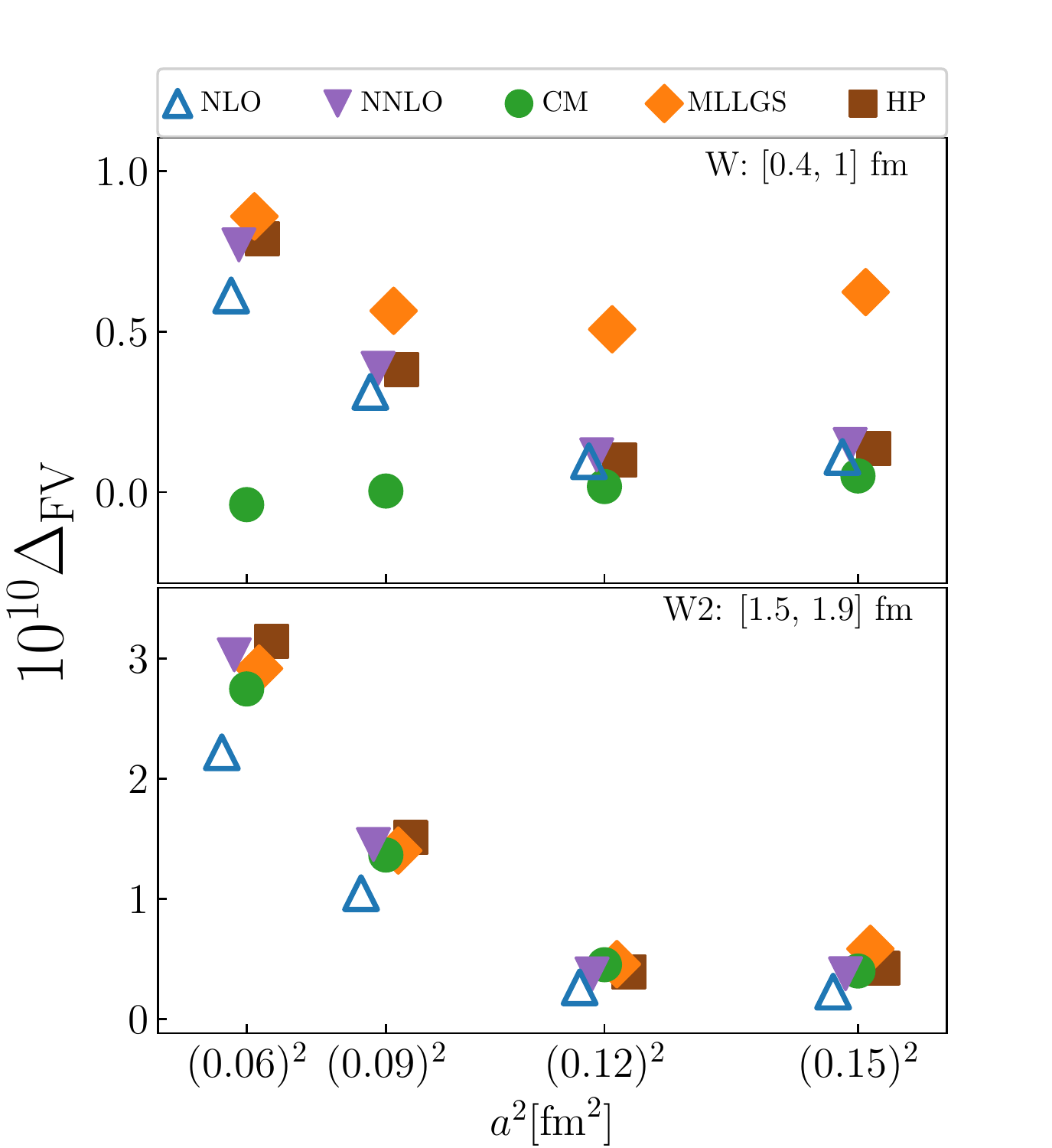}
\vspace{-3mm}\caption{Finite-volume corrections to $\amuLW$ (top) and $\amuLWTwo$ (bottom) obtained from NLO \chpt\ (open blue triangles), NNLO \chpt\ (purple downward triangles), CM (green circles), MLLGS (orange diamonds), and HP (brown squares). The data points at each lattice spacing are offset horizontally for visual clarity. Lattice spatial volumes are given in \cref{table:enssolves}.}
\label{fig:fvspreadStag}
\end{figure}

\Cref{fig:fvspreadStag} shows the finite-volume corrections to $\amuLW$ and $\amuLWTwo$ computed via \cref{eq:corrFV} for each ensemble listed in \cref{table:enssolves} and the four correction schemes discussed above. For $\amuLW$ (top panel), the finite-volume corrections $\Delta_{\rm FV}^{\rm W}$ are always less than 0.5\%. There is, however, a significant spread between the different schemes. In particular, the finite-volume corrections obtained from the CM (green circles) are close to zero on all ensembles. This is because in the CM, the renormalized vacuum polarization function, \cref{cmVacPol},  is comprised of two terms: the first is identical to NLO \chpt, while the second accounts for $\rho$-$\pi$-$\pi$ interactions. For $\amuLW$, the latter contribution produces a correction opposite in sign to the former. In contrast, in \chpt, the NLO (open blue triangles) and NNLO contributions to $\Delta_{\rm FV}$ have the same sign, making the total NNLO corrections (purple downward triangles) larger.  The spread between the finite-volume corrections in the top panel of \cref{fig:fvspreadStag} reflects the limitations of the correction schemes in the intermediate window region, as discussed earlier.

By design \cite{Aubin:2022hgm}, \chpt\ (and the other EFTs) should work better in the W2 region, for which contributions from low-lying $\pi\pi$ states are more important. Hence, we expect better consistency between the finite-volume corrections to $\amuLWTwo$ from the different approaches. Indeed, these expectations are borne out in the bottom panel of \cref{fig:fvspreadStag}, where the corrections from the higher-order schemes have a much-reduced (relative) spread compared to $\amuLW$. Additionally, the finite-volume corrections to $\amuLWTwo$ are larger than for $\amuLW$ ($\Delta_{\rm FV}^{\rm W2} \sim\,$3\% at the finest lattice spacing) due to the increased sensitivity to long-distance contributions at later Euclidean times.

Below $a \lesssim 0.12$~fm, the size of finite-volume corrections to $\amuLW$ and $\amuLWTwo$ decrease with increasing lattice spacing. This is because the pion taste splittings are larger on coarser lattices, and finite-volume corrections in systems with heavier masses are smaller. The finite-volume corrections at $a \approx 0.15$~fm are generally larger than at $a \approx 0.12$~fm, however, because the spatial volume of our coarsest ensemble is substantially smaller than the others (see \cref{table:trapVSsimp}).

In the absence of guidance from a direct finite-volume study, we take the range of finite-volume corrections for the schemes we consider here as an estimate of the associated systematic uncertainty. 
Some or all of the EFT-based models considered are of questionable reliability in the intermediate-window region. Motivated by this, we generate a second set of corrections to $\amuLW$ obtained from restricting the window to higher $t$, namely $[0.7,1.0]$~fm. The spread of these restricted corrections is $\sim 20\%$ smaller than the full W window case. Therefore, in total we include ten sets of finite-volume-corrected data for each input $\amuLW$ data set in our analysis: two each for NLO \chpt, NNLO \chpt, CM, MLLGS, and HP. For the W2 region,  the EFTs are on more solid theoretical footing and the higher-order schemes (NNLO \chpt, CM, MLLGS, and HP) yield consistent results. Therefore, in our analysis we include four sets of finite-volume-corrected data for $\amuLWTwo$, omitting NLO \chpt\ because NNLO \chpt\ should be more accurate in this region. These finite-volume-corrected data sets are inputs into the next step, and eventually feed into the BMA analysis of \cref{sec:BMA}.

\subsubsection{Pion-mass adjustment}\label{sec:DeltaMpi}

We next consider the effects of pion-mass mistuning on $\amuLW$ and $a_{\mu}^{ll,\mathrm{W}2}$(conn.) and estimate the pion-mass adjustments to these quantities, $\Delta_{M_\pi}$ in \cref{eq:corrMpi}, using a data-driven approach. As stated in \cref{sec:lat}, on our ensembles with $a\approx 0.15$ and 0.12~fm, in addition to the unitary correlation functions listed in \cref{table:enssolves}, we have partially quenched correlators (and hence lattice data for $\amuLW$ and $\amuLWTwo$) with valence-quark masses bracketing the physical light quark (see \cref{table:PQsims})\footnote{The $a \approx 0.15$~fm correlators were also employed in Ref.~\cite{FermilabLattice:2017wgj} to study strong-isospin-breaking effects in $\amuL$.}. Together with our unitary data at $a \approx 0.09$ and $0.06$~fm, this allows us to predict the size of pion-mass adjustments to $\amuLW$ and $\amuLWTwo$ on all of our ensembles as follows. 

\begin{figure}[t]
\centering
\includegraphics[width=0.95\textwidth]{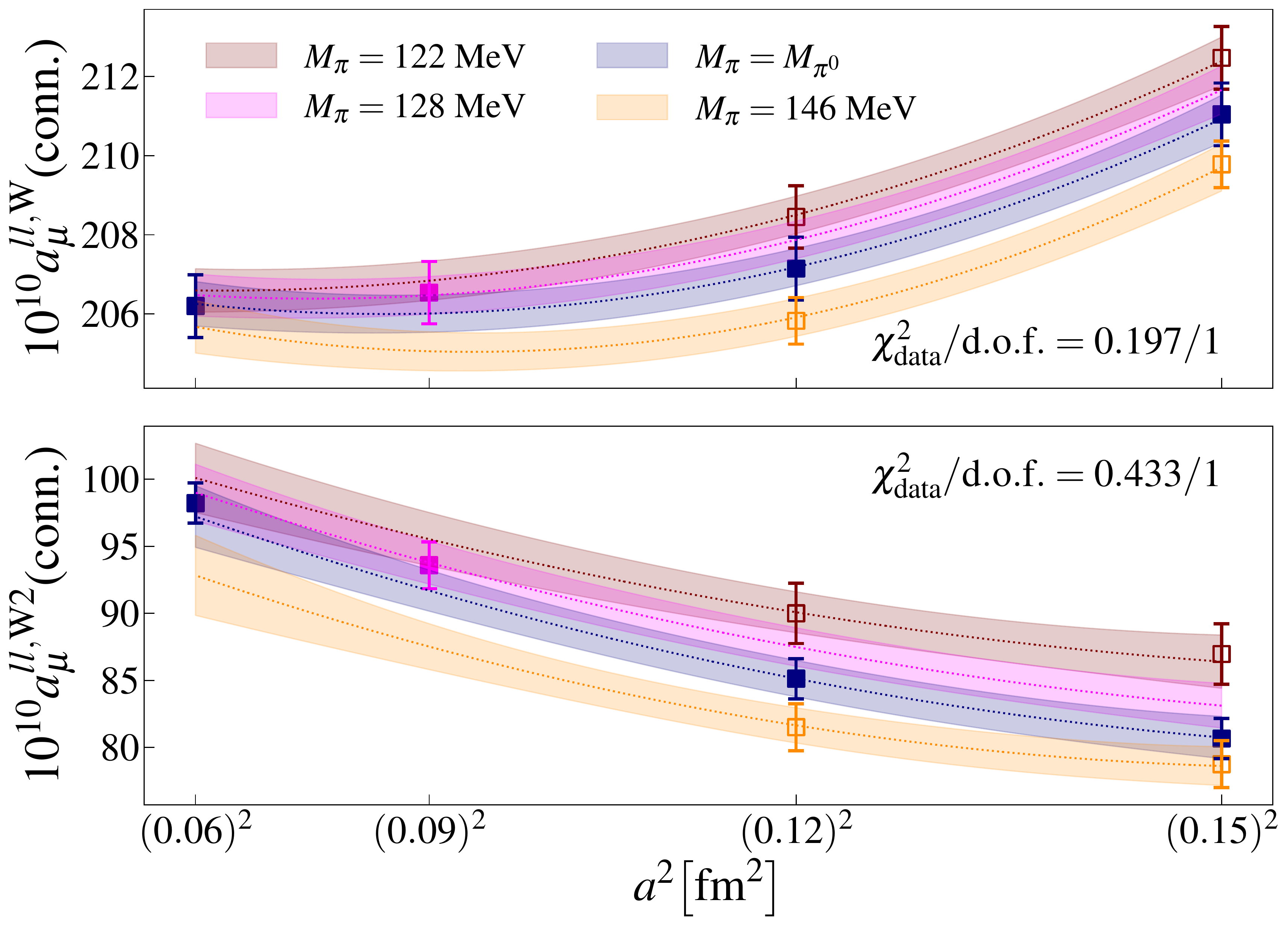}
\vspace{-3mm}
\caption{Data-driven estimate of pion-mass adjustments to $\amuLW$ (top) and $\amuLWTwo$ (bottom) on all ensembles. Lattice data (corrected for finite-volume effects using the CM scheme) are shown as open/filled squares with error bars, with each color denoting a different simulation pion mass: $M_{\pi^0}$ (blue), the taste-Goldstone pion mass at $a \approx 0.09$~fm (magenta), and the partially-quenched pion masses bracketing $M_{\pi^0}$ at $a\approx 0.12$ and $0.15$~fm (maroon and orange). The results of fitting these data to an interpolating function in $M_{\pi}^2$ and $a^2$ (specifically, \cref{eqn:data_driven_fit} with $n_{\{-1,0,1\}} = \{1,2,1\}$) are shown for fixed $M_\pi$ as dashed curves with error bands, and share the same color coding as the data points.}  
\label{fig:data_driven_interp}
\end{figure}

First, we correct our entire dataset for finite-volume effects as described in \cref{sec:DeltaFV}. We then fit the corrected $\amuLW$ and $\amuLWTwo$ data to an interpolating function of the form
\begin{align}
    a^{ll,\mathrm{win}}_{\mu}\big(a, M_{\pi}\big) &= \sum_{i=-1}^{1} c_i(a) (M_{\pi}/\Lambda)^{2i}; \label{eqn:data_driven_fit} \nn \\
    c_{i}(a) &= \sum_{j=0}^{n_{i}}c_{ij} (a\Lambda)^{2j},
\end{align}
where $\Lambda=500$ MeV (following Ref.~\cite{Davies:2019efs}) and $M_\pi$ is the taste-Goldstone valence-{\it sea} pion mass, which is what enters the leading-order pion loops in \chpt. The parametric dependence on $M_{\pi}$ in \cref{eqn:data_driven_fit} is motivated by \chpt, with an additional $1/M_{\pi}^2$ term accounting for the expected infrared-divergent behavior of $a_{\mu}^{ll}$ in the $M_{\pi}\rightarrow 0$ limit \cite{Chakraborty:2016mwy,Golterman:2017qtu}.
For each value of $i$ in \cref{eqn:data_driven_fit}, we consider several values for $n_i \geq 0$, requiring only that the $\{n_i\}$ are the same for both $\amuLW$ and $\amuLWTwo$, and that each fit has at least one degree-of-freedom (d.o.f.). (Note that if $n_{i}=0$, then $j=0$ and $c_{i}(a)$ is independent of $a\Lambda$.) Following Ref.~\cite{lsqfitDocs}, we account for correlations between the independent variables ($a$ and $M_\pi$), as well as between the independent and dependent variables ($\amuLW$ and $\amuLWTwo$), using Bayesian priors.
We monitor the $\chi^2_{\mathrm{data}}/\mathrm{d.o.f.}$ of each fit variation, preferring fits with $\chi^2_{\mathrm{data}}/\mathrm{d.o.f.}$ closest to 1. (A $\chi^2_{\mathrm{data}}/\mathrm{d.o.f.} \gg 1$ indicates that the fit function does not describe the data, while a  $\chi^2_{\mathrm{data}}/\mathrm{d.o.f.} \ll 1$ suggests that we are overfitting.)
After trying several combinations of $\{n_i\}$, we select $n_{-1} = 1$, $n_{0} = 2$, and $n_{1} = 1$ for our central analysis because this functional form gives the best interpolation of our data for both Euclidean-time windows simultaneously.

Once the coefficients $c_{ij}$ are determined for a given set of $\{n_i\}$, we can use \cref{eqn:data_driven_fit} to predict $\amuLW$ and $\amuLWTwo$ at the target physical pion mass, $M_{\pi,\mathrm{phys}} = M_{\pi^0}$ (see \cref{sec:integrals}) for each ensemble.  Figure~\ref{fig:data_driven_interp} shows our central fits for $\amuLW$ (upper panel) and $\amuLWTwo$ (lower panel). At each lattice spacing, we take the difference in $a^{ll,\mathrm{win}}_{\mu}$ between the fit prediction at the physical-pion mass (blue dashed curve) and the unitary lattice data (filled squares) as our data-driven estimate of the pion-mass adjustment $\Delta_{M_\pi, \mathrm{DD}}$.  As seen in \cref{fig:data_driven_interp}, our data-driven analysis finds that a correction to $\amuLW$ and $\amuLWTwo$ of about 1 sigma is needed  on the $a \approx 0.09$~fm ensemble, for which the simulation pion mass is about 5\% below the physical value.

\begin{figure}[t]
\centering
\includegraphics[width=\textwidth]{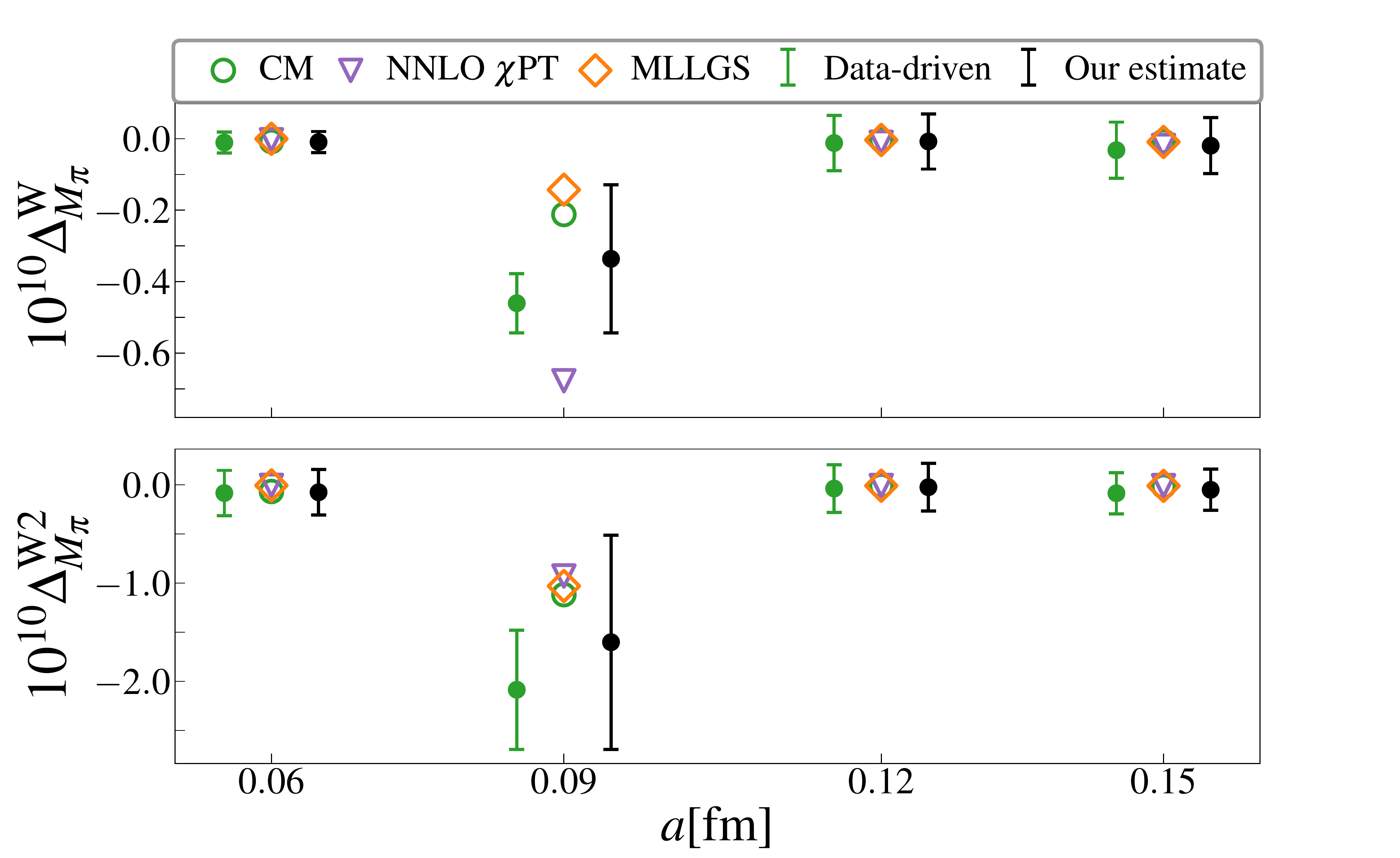}
\vspace{-8mm}
\caption{Comparison of predictions for the pion-mass adjustments with error bars show the predictions of our data-driven analysis, which employs finite-volume corrections from the CM. Open symbols show predictions from the CM (green circles), NNLO \chpt\ (empty purple upside down triangle), MLLGS (orange diamonds) correction schemes. Black points with error bars show our final estimates for the pion-mass adjustments on each ensemble, which account for the spread between predictions as described in the text.}\label{fig:data_driven_test}
\end{figure}

The $\amuLW$ and $\amuLWTwo$ data entering our central fits are corrected for finite-volume effects using the CM. Repeating this analysis using other finite-volume correction schemes yields almost identical predictions for the pion-mass adjustments. Replacing the $1/M_{\pi}^2$ term in \cref{eqn:data_driven_fit} with $\log\big(M_{\pi}^2\big)$ also leads to negligible changes in the predicted values for $\Delta_{M_\pi}$.

Figure~\ref{fig:data_driven_test} compares the pion-mass adjustments to $\amuLW$ (upper panel) and $\amuLWTwo$ (lower panel) obtained in our data-driven analysis (filled green circles with error bars) and those estimated within three of the EFT-based correction schemes introduced in \cref{sec:lat_corrections}: the CM (empty green circles),  NNLO $\chi$PT (empty purple upside down triangle), MLLGS (empty orange square triangles).  On the three ensembles for which the pion mass is well tuned ($a \approx 0.06$, 0.12, and 0.15~fm), the pion-mass adjustment $\Delta_{M_{\pi},\mathrm{DD}}$ is negligible in all correction schemes. At $a \approx 0.09$~fm, however, the picture is less clear. For $\amuLW$, the spread in model estimates is significantly larger than the error bar on the data-driven evaluation. For $\amuLWTwo$, the models agree with each other, but differ from the data-driven prediction by $\approx 1.5\sigma$.

In light of the differences between predicted corrections at $a \approx 0.09$~fm, we adopt the following conservative procedure to obtain our final estimates for $\Delta_{M_\pi}$ (black filled circles with error bars in \cref{fig:data_driven_test}). For the central value, we use the average of the data-driven and chiral-model predictions, {\it i.e.}, $\Delta_{M_{\pi}} \approx (\Delta_{M_{\pi},\mathrm{DD}}+\Delta_{M_{\pi},\mathrm{CM}})/2$. For the error on $a \approx 0.09$~fm, we add (linearly) to the uncertainty on the data-driven prediction $\sigma_{M_{\pi},\mathrm{DD}}$ an additional systematic uncertainty given by half the absolute difference between the data-driven and chiral-model predictions, {\it i.e.}, $\sigma_{M_{\pi}, \mathrm{DD}} + |\Delta_{M_{\pi}, \mathrm{DD}} - \Delta_{M_{\pi},\mathrm{CM}}|/2$. On $a \approx 0.06, 0.12$, and $0.15$~fm we take the uncertainty to be just the uncertainty on the data-driven prediction. As shown in \cref{fig:data_driven_test}, our final estimates for the pion-mass adjustment at $a \approx 0.06$, 0.12, and 0.15~fm are essentially those from our data-driven analysis.  At $a \approx 0.09$~fm, our final estimate for the pion-mass adjustment covers most (all) of the model spread for $\amuLW$ ($\amuLWTwo$).

\subsubsection{Taste-breaking corrections}\label{sec:DeltaTB}

The final lattice correction, $\Delta_{\rm TB}$ in \cref{eq:corrTB}, accounts for the mass differences at finite lattice spacing between staggered pions with different taste quantum numbers.  For the HISQ action, 
these taste splittings arise from discretization effects of $O(\alpha_s^2 a^2)$ and higher. It is well known, however, that the HISQ taste splittings do not scale linearly with $\alpha_s^2a^2$~\cite{MILC:2010pul,MILC:2012znn}.\footnote{See also Ref.~\cite{Aubin:2022hgm} for a discussion of the HISQ taste splittings as they pertain to calculations of $\amuHVP$.} As shown in Ref.~\cite{MILC:2012znn}, the HISQ pion-taste splittings decrease faster than naive expectations at lattice spacings below around 0.09~fm, while increasing more slowly at very coarse lattice spacings above roughly 0.12~fm. The former observation is likely due to the HISQ smearing~\cite{Follana:2006rc} suppressing the leading $\alpha_s^2a^2$ taste-breaking discretization contributions, making higher-order, {\it e.g.}, ${\mathcal O}(\alpha_s^3a^2, a^4)$, effects more prominent, while the latter indicates the presence of additional higher-order terms. 

\begin{figure}[t]
\centering
\includegraphics[width=0.95\textwidth]{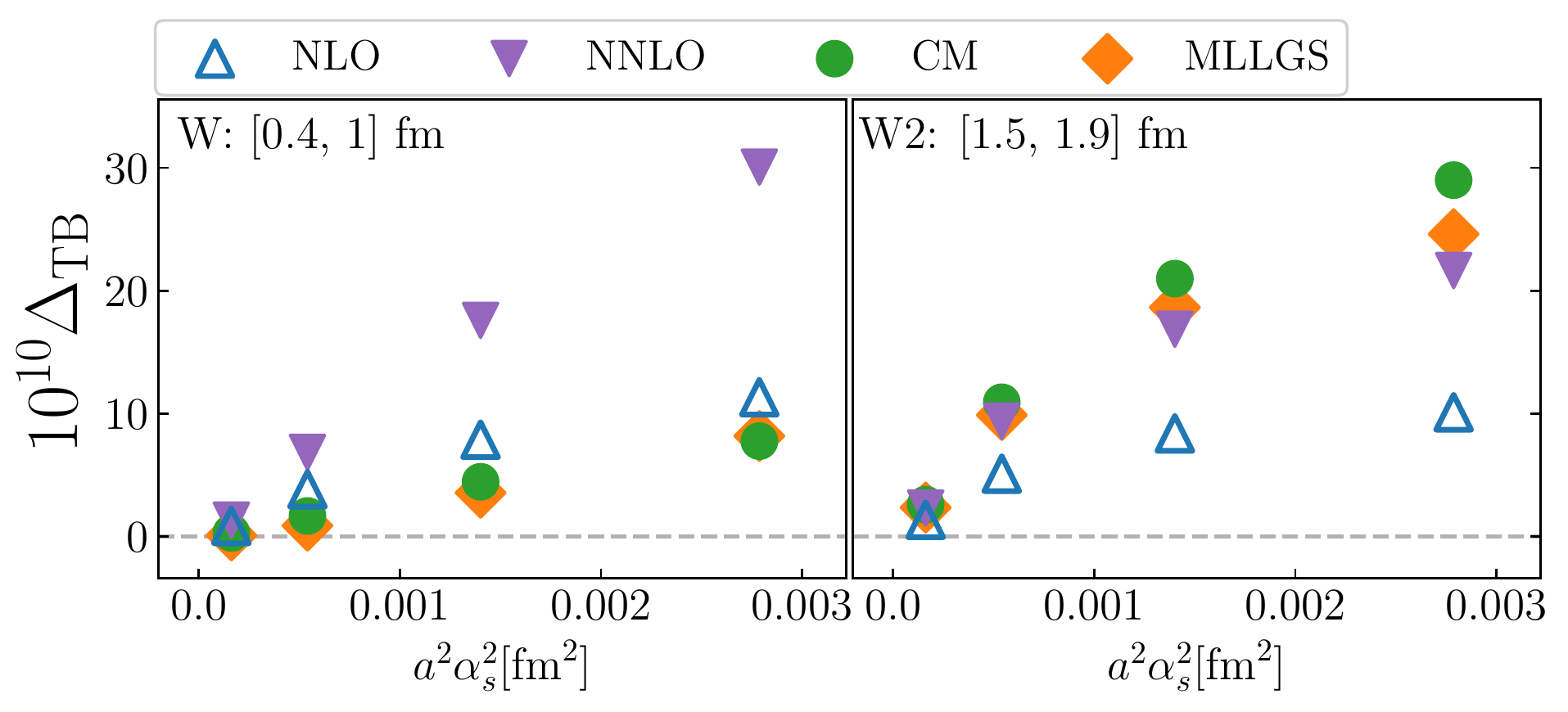}
\vspace{-3mm}
\caption{Taste-breaking corrections to $\amuLW$ (left) and $\amuLWTwo$ (right) obtained from NLO \chpt\ (open blue triangles), NNLO \chpt\ (purple downward triangles), CM (green circles), and MLLGS (orange diamonds).}
\label{fig:TSplitting}
\end{figure}

Figure~\ref{fig:TSplitting} shows the taste-breaking corrections to $\amuLW$ (left panel) and $\amuLWTwo$ (right panel) obtained within the \chpt, CM, and MLLGS correction schemes introduced at the beginning of \cref{sec:lat_corrections}. Qualitatively, they display the same behavior as the taste splittings, with $\Delta_{\rm TB}$ decreasing more rapidly at finer lattice spacings. Quantitatively, the estimated corrections span a wide range of values between $ 0 \lesssim \Delta_{\rm TB} \lesssim 30\times 10^{-10}$. This corresponds to corrections to $\amuLW$ and $\amuLWTwo$ on our coarsest ensemble of up to $\sim 15\%$ and $\sim 30\%$, respectively.  Differences between correction schemes and sizes of the corrections vanish at zero lattice spacing by construction. 

For $\amuLW$, the NLO \chpt, CM, and MLLGS results are in broad agreement, while the NNLO \chpt\ prediction is about 2--4 times larger for $a \gtrapprox 0.09$~fm. 
In contrast, for $\amuLWTwo$ the three higher-order schemes (NNLO \chpt, CM, and MLLGS) predict similar corrections, while NLO \chpt\ is the outlier. 
As with the finite-volume corrections (see \cref{sec:DeltaFV}), the spread of predicted taste-breaking corrections for $\amuLW$ is likely due to the correction schemes becoming less reliable at short distances.
Indeed, this expectation is borne out in Figs.~\ref{fig:continuumExtrap_W} and~\ref{fig:continuumExtrap_W2}, which show the continuum extrapolations of $\amuLW$ and $\amuLWTwo$, respectively, with and without taste-breaking corrections.  For $\amuLW$, applying taste-breaking corrections increases the lattice-spacing dependence. In contrast, for $\amuLWTwo$ the taste-breaking corrections computed in the three higher-order correction schemes all substantially reduce the lattice-spacing dependence, indicating that they capture the dominant discretization effects in this window.

Although the inclusion of taste-breaking corrections (and choice of scheme) will alter the lattice-spacing dependence of $\amuLW$ and $\amuLWTwo$, it should not change the continuum-limit values. Consequently, varying the treatment of taste-breaking in the continuum extrapolation provides an additional measure of the continuum-extrapolation error. For the analysis of the intermediate window, we generate taste-breaking-corrected data sets with NLO \chpt, NNLO \chpt, CM, and MLLGS corrections for each input $\amuLW$ data set from the previous sections. We follow the reasoning of \cref{sec:DeltaFV} and compute the corrections in two regions, the full intermediate window interval [0.4, 1]~fm  and the smaller interval of [0.7, 1]~fm, resulting in a total of eight taste-breaking-corrected data sets for each input set. For each input $\amuLWTwo$ data set we generate three sets of taste-breaking-corrected data, one each for NNLO \chpt, CM, and MLLGS, dropping NLO \chpt, as in \cref{sec:DeltaFV}. In both cases, we also keep the data sets uncorrected for taste-breaking effects. The corrected and uncorrected $\amuLW$ and $\amuLWTwo$ data sets are then taken as inputs into the continuum limit extrapolations, and feed ultimately into the Bayesian model averaging analysis of \cref{sec:BMA}.

\subsection{Continuum extrapolation}\label{sec:cont_extrap}

To perform our continuum extrapolation we consider fit functions of the form: 
\begin{align}
a_{\mu}^{l l}(a, \{m_f\})&= a_{\mu}^{l l}\left(1 + F^{\text{disc.}} (a) + F^m (\{\delta m_f\})\right), \label{eq:fitfunc}\end{align}
where
\begin{align}
 F^{\text{disc.}} (a) &= C_{a^{2},n}\left[(a\Lambda)^{2} \alpha_s^{n}\right]+C_{a^4} (a\Lambda)^{4} + C_{a^6} (a\Lambda)^{6} \label{eq:discfunc}\\\vspace{2mm}
 F^m (\{\delta m_f\}) &= C_{\text{sea}} \sum_{f=l, l, s} \delta m_{f} / \Lambda \label{eq:mfunc} .
\end{align}
The function $F^{\text{disc.}} (a)$ describes discretization effects and $F^m (\{\delta m_f\})$ accounts for quark mass differences in the sea, where $\delta m_{f}$ is the difference between the physical and the simulation quark masses (see \cref{sec:isosymQCD,sec:lat}). In $F^{\text{disc.}} (a)$ we include variations where the coefficient $C_{a^6}$ is set to zero and where the power of $\alpha_s$ in the $a^2$ term varies as $n=1,2$. For both variations of $n=1,2$, we label fit functions with $C_{a^6}=0$ as ``quadratic'' and fit functions where all terms listed in $F^{\text{disc.}} (a)$ are included as ``cubic''. Following Ref.~\cite{Davies:2019efs} we take $\Lambda = 500$~MeV and impose the Gaussian prior constraint  $C_{\text{sea}} = 0.0(3)$.  Here, the  $C_{\text{sea}}$ term accounts for residual light sea-quark mass miss-tuning effects, remaining after performing the correction in \cref{sec:DeltaMpi}, and also strange sea-quark miss-tuning effects. As in Ref.~\cite{Davies:2019efs}, the sea-quark masses in the ensembles employed here are so close to their physical values that our fits are insensitive to the $F^m (\{\delta m_f\})$ term and return posteriors for $C_{\text{sea}}$ with central values close to zero and uncertainties close to the initial prior width. This also means that higher-order terms involving $\delta m_f$ can be safely neglected. Additionally, to regulate the degrees of freedom in fits with an $a^6$ term, we constrain its coefficient with the Gaussian prior
\be
C_{a^6} = 0(2).
\ee 
This prior width conservatively accommodates instances among the over two thousand continuum fits when the posterior central values are close to or slightly larger than unity.
We also include fits to three ensembles, dropping the coarsest. In this case we include an additional prior constraint on the quadratic term $C_{a^4} = 0(2)$ with the same reasoning as above for $C_{a^6}$. We include continuum-extrapolation fit variations, both with and without including the $C_{\text{sea}}$ term in our Bayesian averaging process.

\begin{figure}[t]
\centering
\includegraphics[width=0.98\textwidth]{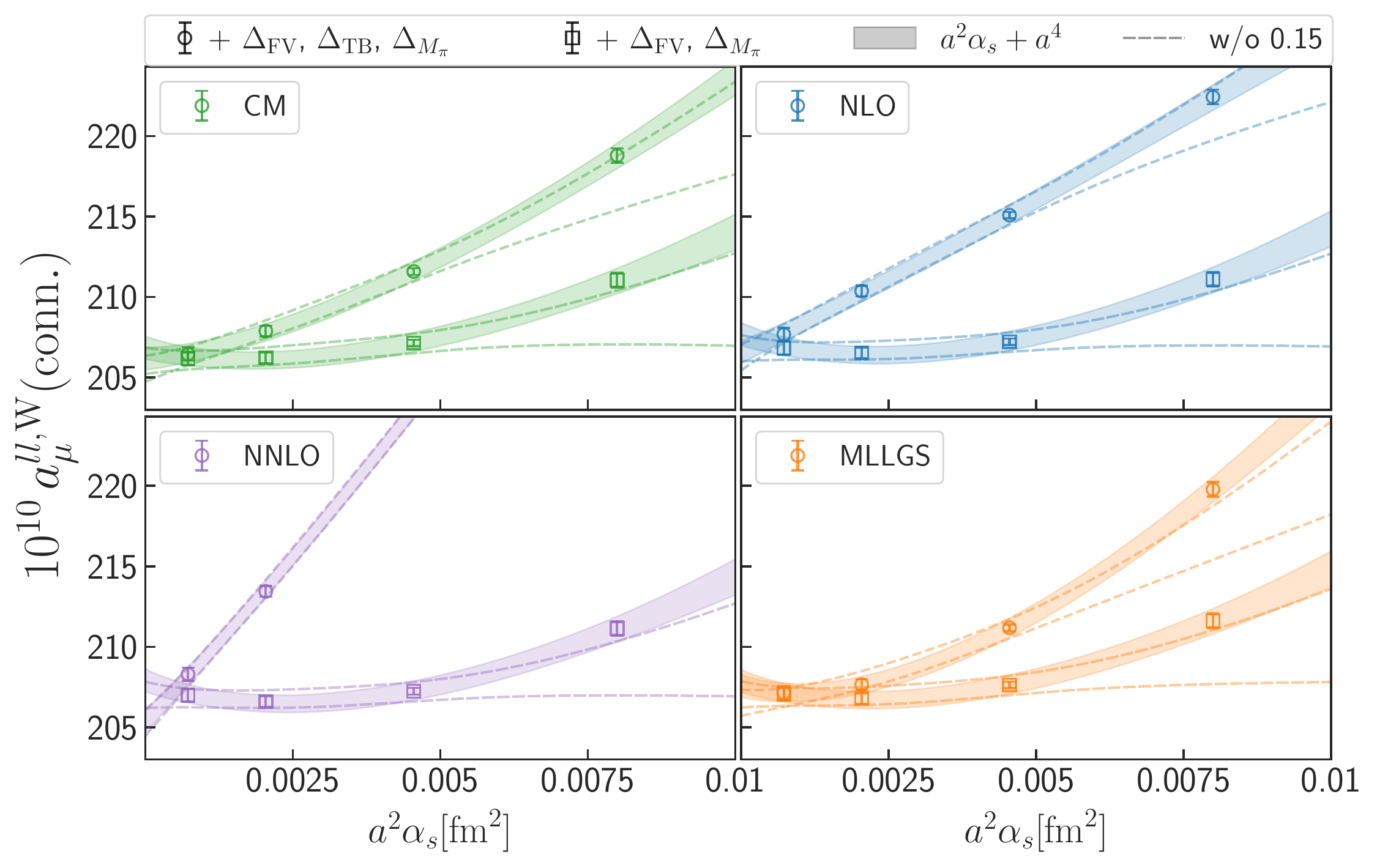}
\vspace{-4mm}
\caption{ Continuum extrapolations of $\amuLW$ using the CM (top left), NLO \chpt\ (top right), NNLO \chpt\ (bottom left) and MLLGS (bottom right) correction schemes.  All data are obtained from integrating the lattice correlator $C(t)$ using the trapezoidal rule, and corrected for finite-volume effects and adjusted for pion-mass mistuning. Data sets that also include taste-breaking corrections are shown as circles, while data without these optional corrections are shown as squares. All corrections come from the full window region. Solid bands (dashed lines) show the fit results of continuum extrapolations with (without) data at our coarsest lattice spacing (right-most point in each panel). All fits employ the same fit function, \cref{eq:fitfunc} with terms through $O(a^4)$.}
\label{fig:continuumExtrap_W}
\end{figure}

\begin{figure}[thb]
\centering
\includegraphics[width=0.98\textwidth]{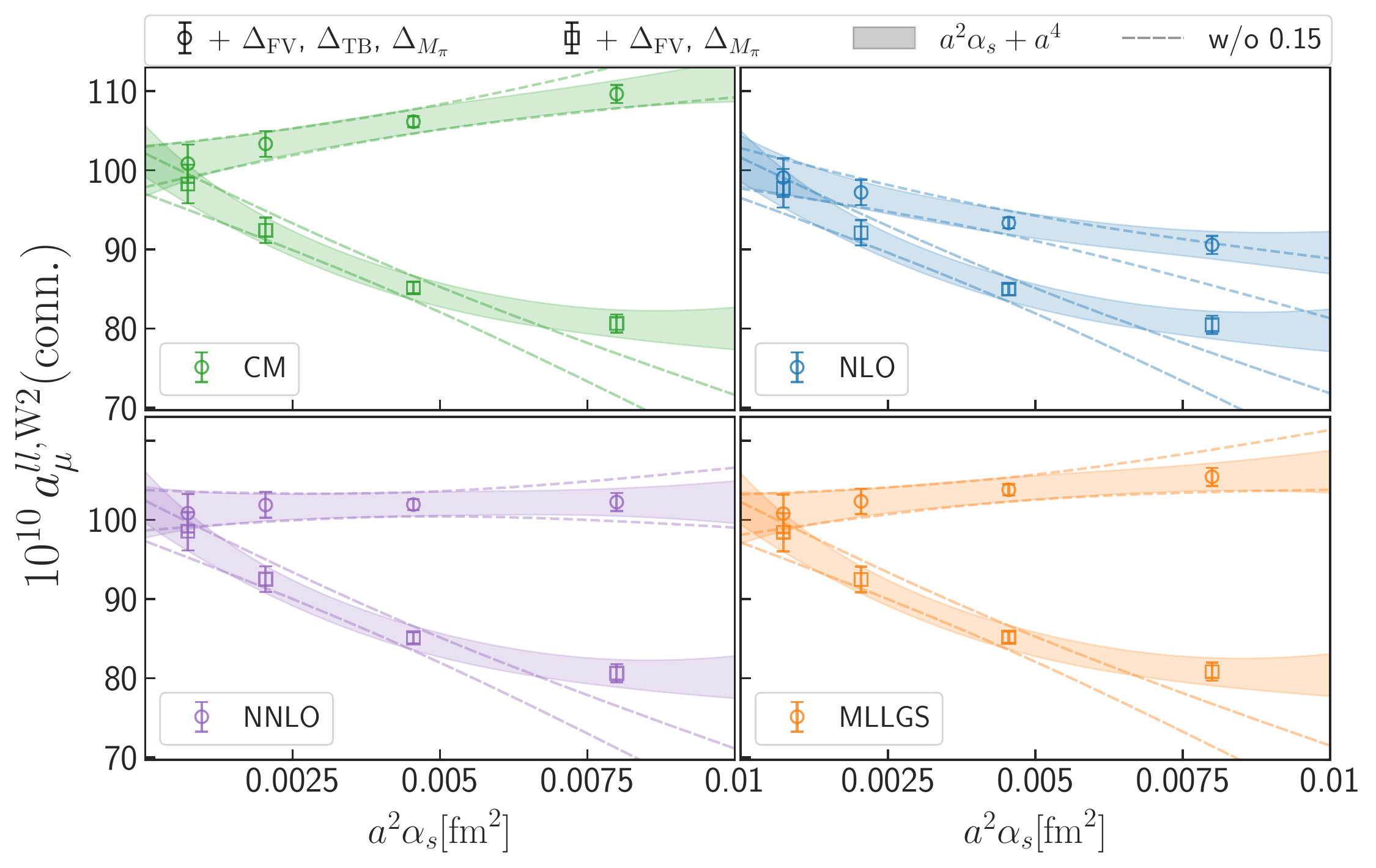}
\vspace{-4mm}
\caption{Continuum extrapolations of $\amuLWTwo$. Figure is described in the caption of \cref{fig:continuumExtrap_W}.}
\label{fig:continuumExtrap_W2}
\end{figure}

As illustration, in \cref{fig:continuumExtrap_W} we show results for quadratic continuum extrapolations of the $\amuLW$ data with $n=1$ in \cref{eq:fitfunc}. 
The four panels show finite-volume-corrected data computed from the CM (top left), NLO \chpt\ (top right), NNLO \chpt\ (bottom left) and MLLGS (bottom right). For each scheme, we compare continuum extrapolations of data with and without taste-breaking corrections, where we include fits to all four ensembles as well as fits to ensembles at only the three finest lattice spacings. 
For the data sets corrected with NLO \chpt\ or the CM, we find very good agreement between the four continuum extrapolated results, whereas the NNLO \chpt-\ and MLLGS-corrected data sets show larger spreads. Taking into account this variance, we find that the continuum results obtained with all four correction schemes are consistent with each other. The corresponding continuum extrapolations for $\amuLWTwo$ are shown in \cref{fig:continuumExtrap_W2}. In this case we find good agreement between the continuum extrapolated results, both within each scheme as well as across the different schemes, albeit with larger uncertainties. We observe, however, that the NLO \chpt\ taste-breaking corrections do a poor job at removing lattice spacing dependence compared to the higher-order schemes.

In summary, for each input data set, we perform twelve different continuum extrapolations, the results of which become inputs to the Bayesian model averaging analysis of \cref{sec:BMA}: in \cref{eq:fitfunc}, we take $n=1,2$ in the linear term, and include or don't include the $C_{\text{sea}}$ term\footnote{We find that the fit results are virtually unchanged when the $C_{\text{sea}}$ term is included and, in addition, that they are insensitive to the prior width of $C_{\text{sea}}$, after increasing it by up to a factor of ten.}. With these four variations, we perform fits to the data at four lattice spacings with and without the cubic ($a^6$) term in \cref{eq:fitfunc}, as well as with quadratic fits to data at the finest three lattice spacings.

Separately, as an independent analysis cross check of our continuum extrapolations and associated error estimate, we allow for higher-order terms in $\amuLW$ and $\amuLWTwo$ using the empirical Bayes (or maximum marginal likelihood) approach  described in Sec.~5.2 of Ref.~\cite{Lepage:2001ym}.
For this analysis, the discretization term $F^{\text{disc.}} (a)$ in \cref{eq:fitfunc} takes the form: 
\begin{align}
F^{\text{disc.}}_{\rm alt} (a) &= \sum_{j=1}^3 c_{1j} x \alpha_s^j  + \sum_{i=2}^5 \sum_{j=0}^3 c_{ij} x^i  \alpha_s^j + c_{60} x^6 , \quad \quad x \equiv \left(\frac{Q_{\textrm{eff}}}{\pi/a}\right)^2. 
\label{eq:GPLfunc}
\end{align}
The coefficients in \cref{eq:GPLfunc} are constrained with Gaussian priors $c_{ij} = 0(1)$, while the scale $Q_{\textrm{eff}}$ is chosen to maximize the Gaussian Bayes Factor, Eq.~(28) of Ref.~\cite{Lepage:2001ym}, 
 which is proportional to the marginal likelihood (model evidence). The results of this comparison are discussed in \cref{sec:BMA}.

\subsection{Bayesian model averaging}\label{sec:BMA}

In order to quantify the systematic uncertainty due to the analysis choices described in the previous sections, we employ Bayesian model averaging (BMA) \cite{Jay:2020jkz,Neil:2022joj}.  Summarizing these choices, we include variations of:

\begin{itemize}

\item \textbf{Observable extraction} - Two methods are used to extract the uncorrected values of $\amuLW$ and $\amuLWTwo$ from the correlation function data, as described in \cref{sec:integrals}:
\begin{itemize}
\item Raw correlation function data, $C(t)$, integrated with the trapezoidal rule.
\item Fit-reconstructed correlation function data $C_{\rm no osc.} (t)$  integrated with Simpson's rule.
\end{itemize}

\item \textbf{Finite-volume correction} - All correction schemes discussed in \cref{sec:DeltaFV} above: \chpt, CM, MLLGS, and HP. We include the NLO $\chi$PT variation for $\amuLW$ but not $\amuLWTwo$.

\item \textbf{Taste-breaking correction} - We include \chpt, CM, and MLLGS as well as $a_\mu$ data sets which are not corrected for taste-breaking effects prior to continuum extrapolation.

\item \textbf{Correction region} - For $\amuLW$, we include a variation on the corrections where they are computed from the range $[0.7, 1]$~fm instead of over the full W window interval. 

\item \textbf{Continuum fit} - We perform continuum extrapolations using all 12 fit function variations described in \cref{sec:cont_extrap} including fits to the three finest ensembles.

\end{itemize}

In the context of BMA, a ``model'' $M$ is defined as the set of analysis choices that yield a given result for the desired continuum, infinite-volume, physical observable from a single data set $D$. In our case, $M$ is given by a set of choices from the options listed above, while $D$ consists of the unmodified correlation function data.\footnote{Note that for BMA, the single data set $D$ is held fixed even in variations where ensembles are dropped since this is treated as a model change (see the discussion of data subset selection in \cite{Jay:2020jkz,Neil:2022joj}).  Also note that throughout this work, we also use the more colloquial definition of ``data set'' outside the context of BMA to refer to any set of $\amu$ data points before continuum extrapolation.} In order to carry out the averaging, each $M$ is assigned a probability weight given by
\be
\operatorname{pr}(M \mid D) \equiv \pr(M) \exp \left[-\frac{1}{2}\left(\chi_{\rm data}^{2}\left(\mathbf{a}^{\star}\right)+2 k+2 N_{\mathrm{cut}}\right)\right]. \label{modelProb}
\ee
This is the ``Bayesian Akaike information criterion'' (BAIC) as defined in \cite{Neil:2022joj}.  Here, $\chi_{\rm data}^{2}$ is the standard chi-squared function, {\it not} including the contribution of the priors, and $\mathbf{a}^{\star}$ is the posterior mode ({\it i.e.}, the best-fit point for the vector of fit parameters $\mathbf{a}$ when optimized against the augmented chi-squared function \cite{Lepage:2001ym}.)  $N_{\mathrm{cut}}$ is the number of data points cut from a data set---in this case, the number of ensembles omitted from a given extrapolation.  The parameter $k$ is the number of independent parameters in a given fit function. The factor $\pr(M)$ is the prior probability of a given $M$; we adopt a flat prior, so that this factor is a constant over all analysis variations and drops out of the model averaging results. 
The BMA mean and variance are then obtained from the following formulas: 

\begin{align}
\left\langle a_{\mu}\right\rangle &= \sum_{i=1}^{N_{M}}\left\langle a_{\mu}\right\rangle_{i} \operatorname{pr}\left(M_{i} \mid D\right), \label{eq:BMAMean} \\
\sigma_{a_{\mu}}^{2}&= \sum_{i=1}^{N_{M}} \sigma_{a_{\mu}, i}^{2} \mathrm{pr}\left(M_{i} \mid D\right)+\sum_{i=1}^{N_{M}}\left\langle a_{\mu}\right\rangle_{i}^{2} \mathrm{pr}\left(M_{i} \mid D\right)-\left(\sum_{i=1}^{N_{M}}\left\langle a_{\mu}\right\rangle_{i} \mathrm{pr}\left(M_{i} \mid D\right)\right)^{2}. \label{eq:BMAVar} 
\end{align}
The first term on the right-hand side of \cref{eq:BMAVar} is a weighted average over the variances of the individual results. The second and third terms reflect the spread in results obtained with different analysis choices (in our case, correction schemes and fit functions).  Because they encapsulate the systematic uncertainty due to analysis choices, we refer to their sum as the ``model variance.''

\begin{figure}[t]
\centering
\includegraphics[width=.95\textwidth]{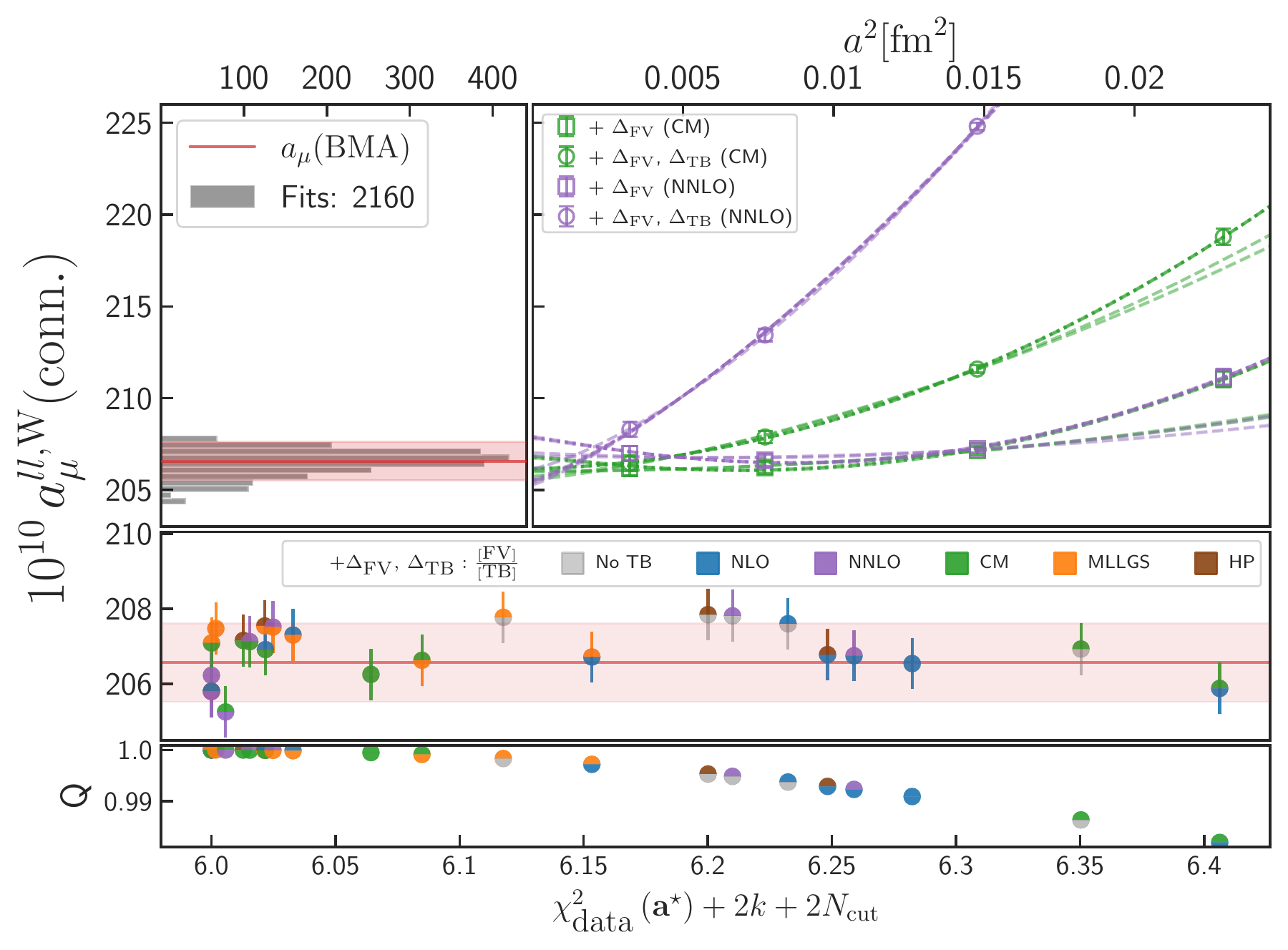}
\vspace{-3mm}
\caption{Results of the Bayesian model averaging (BMA) procedure applied to $\amuLW$. Top left: Histogram of all continuum extrapolations used in the BMA, the light-red band is the BMA result. Top right: The subset of data sets and extrapolations corresponding to correcting the data with the CM and NNLO \chpt. Data (without) with taste-breaking corrections are shown as (squares) circles. Different extrapolations correspond to variations of the fit function and ensembles included. Lower panels: The best fits according to model probability, \cref{modelProb}. The middle panel shows the fit results, while the bottom one shows the corresponding Q-values \cite{FermilabLattice:2016ipl}. In both panels, the correction schemes employed for $\Delta_{\rm FV}$ and $\Delta_{\rm TB}$ are indicated by the symbols' top and bottom colors, respectively.
}
\label{fig:BMA_W}
\end{figure}

\begin{figure}[thb]
\centering
\includegraphics[width=.95\textwidth]{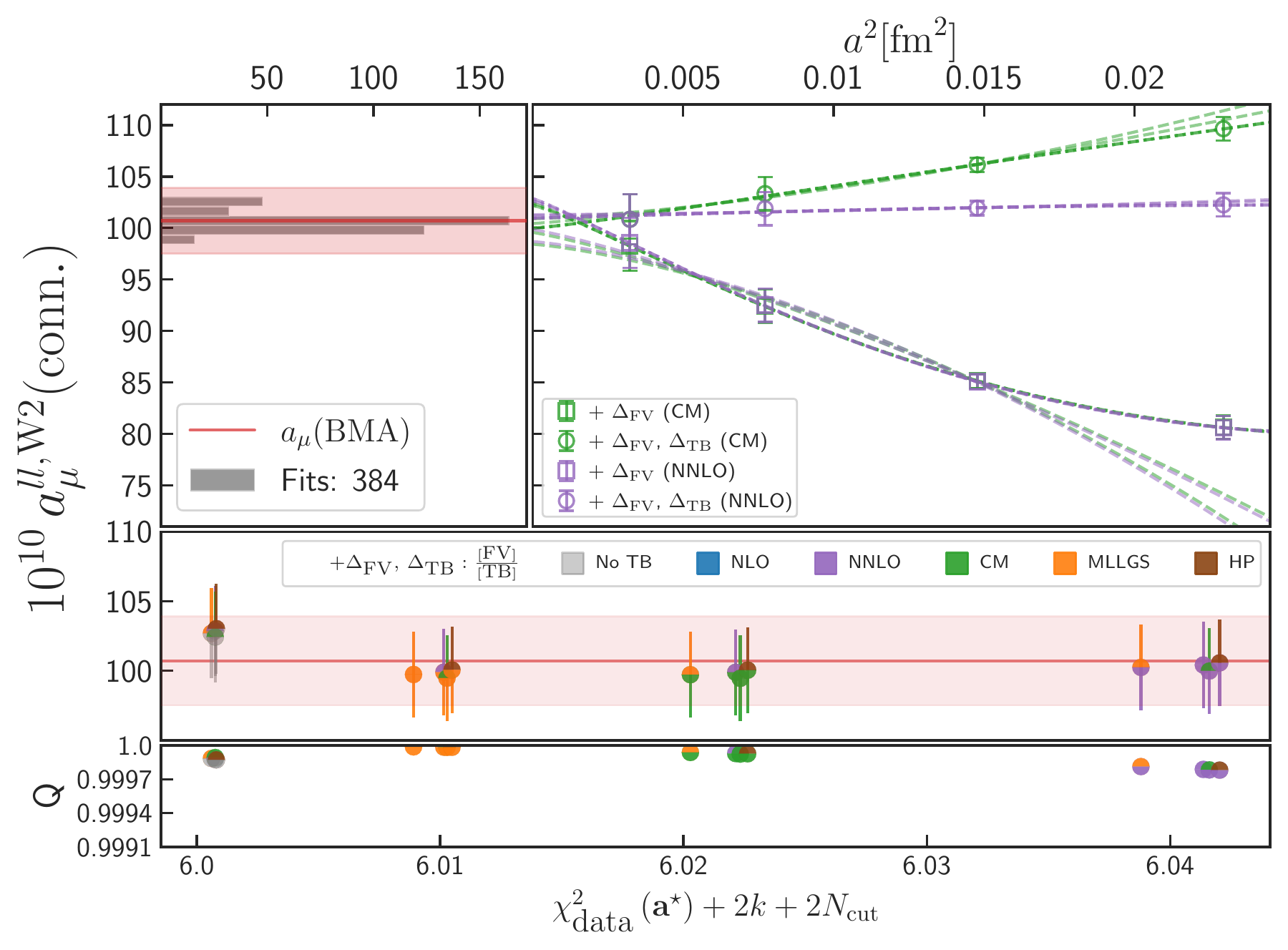}
\vspace{-3mm}
\caption{Results of the Bayesian model averaging (BMA) procedure applied to $\amuLWTwo$. Figure is described in the caption of \cref{fig:BMA_W}.}
\label{fig:BMA_W2}
\end{figure}

In Fig.~\ref{fig:BMA_W}, we show the results of the Bayesian model average for $\amuLW$. The top-right panel illustrates the continuum extrapolations on two data sets, the first corrected with NNLO \chpt\ and the second corrected with the CM, in both cases computed from the full W window interval. The dashed lines indicate the continuum extrapolations for each data set. In total, we include over two thousand separate fit results in the model average. The resulting distribution is shown in the top left panel of \cref{fig:BMA_W}, where it is overlaid on the BMA result (red line and error band) obtained using \cref{eq:BMAMean,eq:BMAVar}. The middle panel shows the results from the 24 best individual fits for each correction choice, ordered by the BAIC, in comparison to the BMA result, while the bottom panel gives the associated $Q$ values \cite{FermilabLattice:2016ipl} computed from $\chi_{\rm aug}^{2}$. We find that our best fits, as determined by the $Q$ value, tend to have the smallest BAIC and hence largest model probability. We also note that the continuum results for data sets corrected for taste breaking using NNLO \chpt\ tend to be smaller than those from the other variations and also return some of the largest model probabilities (points in middle panel with lower half purple).

Figure~\ref{fig:BMA_W2} shows the analogous BMA result for $\amuLWTwo$.  Here we include 384 fit results, which is fewer than for $\amuLW$. This stems from the absence of NLO \chpt\ corrections and from employing only a single correction region. The general features of this figure are the same as for \cref{fig:BMA_W}.  In the top left panel, we note that the BMA uncertainty for $\amuLWTwo$ is larger than the spread of the histogram.  This is because the bulk of the uncertainty in this case comes from the first term in \cref{eq:BMAVar} with relatively large statistical and scale-setting uncertainty contributions.

\begin{figure}[tbh]
\centering
\includegraphics[width=0.95\textwidth]{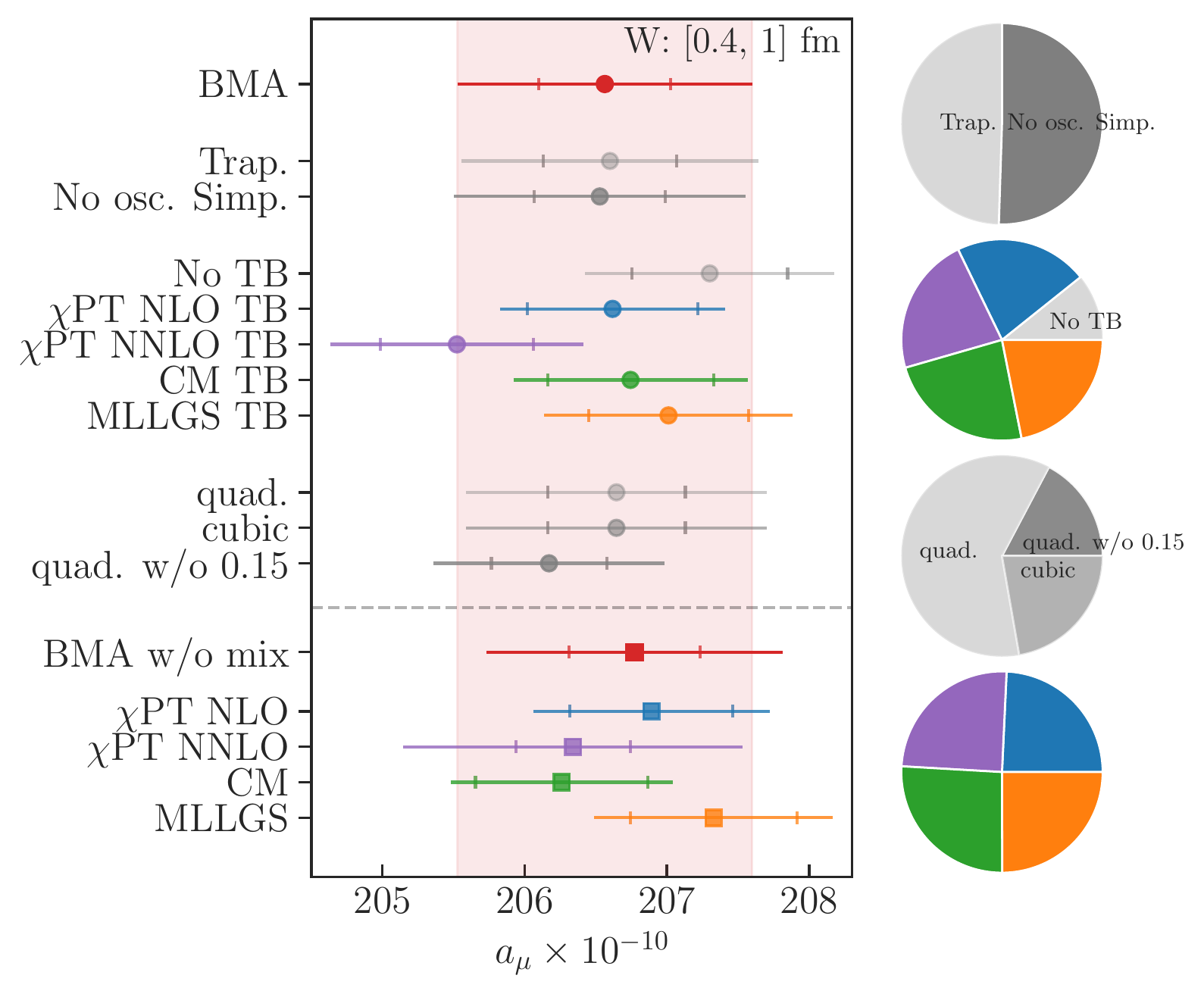}
\vspace{-3mm}
\caption{Breakdown of the results from the Bayesian model averaging applied to $\amuLW$. Left: From top to bottom, the first, main result (BMA) includes all data sets, schemes, and other variations. The next two use data integrated with either the trapezoidal rule (Trap.) or Simpson's rule (No osc. Simp.). The following five results are obtained from subsets with specific taste-breaking corrections. The next three are subsets with specific continuum fit functions: quadratic, cubic, or quadratic without the $0.15$~fm ensemble. The last block of results (below the dashed line) uses the same scheme for finite-volume and taste-breaking corrections. The top (``BMA w/o mix'') includes all four schemes; the final four are breakdowns using only a single correction scheme in the BMA. The inner error bar on the data points corresponds to the first term in \cref{eq:BMAVar}, while the outer is the total error. Right: Pie-charts showing the contributions to the BMA corresponding to the breakdowns in the left panel. The percentages are computed by summing over \cref{modelProb} for the particular subsets.}
\label{fig:BMACompareW}
\end{figure}

\begin{figure}[t]
\centering
\includegraphics[width=0.99\textwidth]{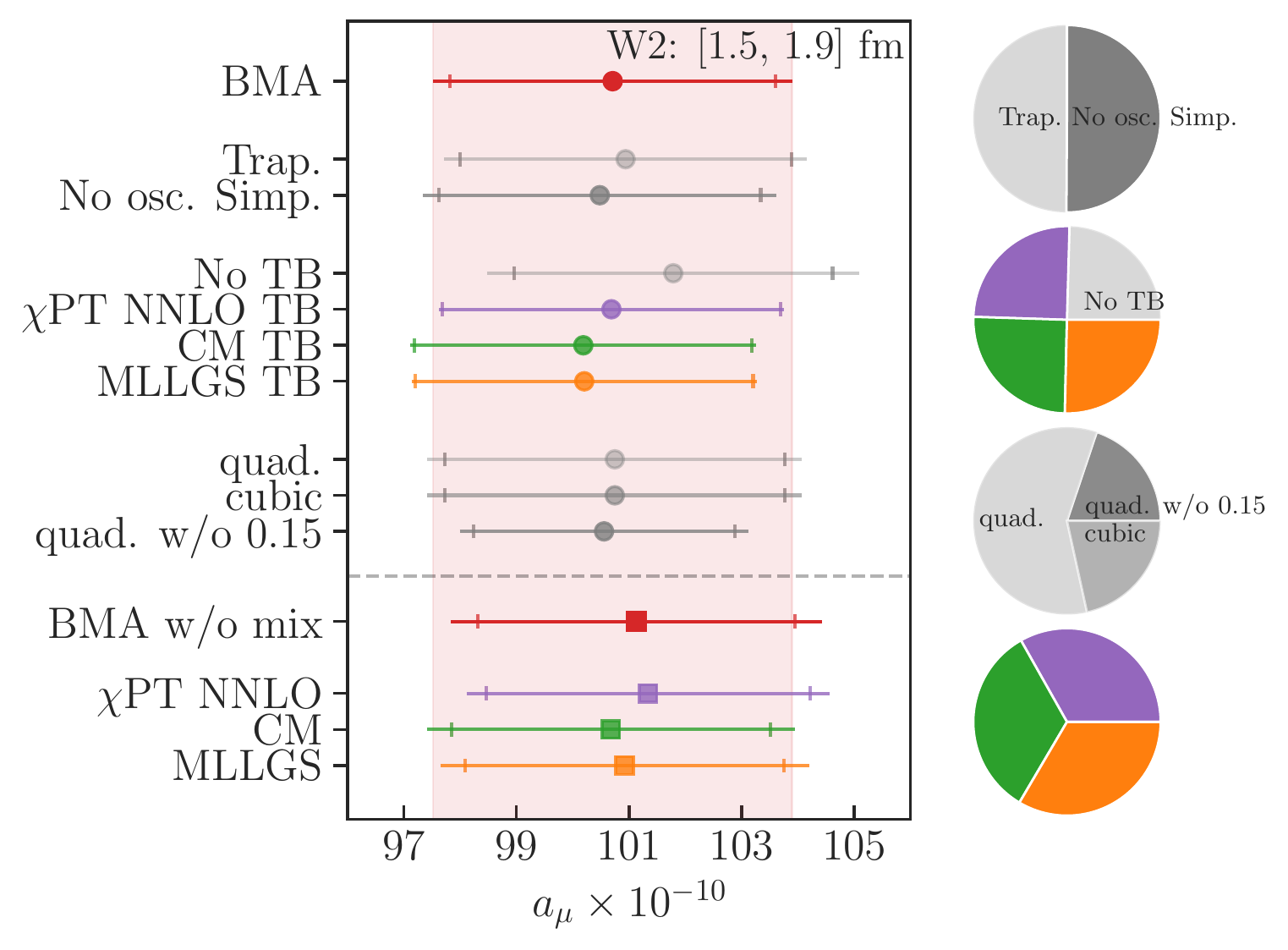}
\vspace{-3mm}
\caption{Breakdown of the results from the Bayesian model averaging applied to $\amuLWTwo$. Figure is described in the caption of \cref{fig:BMACompareW} (with ``four results" replacing ``five results" because NLO \chpt\ is not employed here).}
\label{fig:BMACompareW2}
\end{figure}

In order to better understand and  test the model-averaging results, we also perform Bayesian model averages on specific subsets of the variations. That is, we fix one of the analysis choices but vary the rest as usual. The results of these subset averages for $\amuLW$ are shown in \cref{fig:BMACompareW} (left). The top data point is our BMA result from \cref{fig:BMA_W}. The two data points below it, show the BMA results for the two observable extraction choices described in \cref{sec:integrals}. They are in excellent agreement with each other and with the full BMA result,  signifying, as expected, that residual effects of oscillating contributions and of $O(a^2)$ errors of the trapezoidal rule are negligibly small. The next five data points are the BMA results obtained from subsets with specific taste-breaking correction schemes. While these results are statistically consistent with the overall average, the differences in the central values contribute significantly to the systematic uncertainty through the latter two terms of \cref{eq:BMAVar} (outer uncertainty of the BMA result). In particular, as shown in \cref{fig:BMA_W}, the fit results obtained from NNLO~\chpt\ corrected data tend to lie below the average. The following three data points are BMA results obtained from subsets of specific continuum-extrapolation fit functions, which agree well with each other and with the full BMA result.
The last block of data points (below the dashed line) are BMA results from subsets that use the same schemes for finite-volume and taste-breaking corrections, where the top data point (BMA w/o mix) averages all four schemes (NLO \chpt, NNLO \chpt, CM, MLLGS), followed by results from the subsets corresponding to each single scheme, all of which are consistent with the full BMA result with small variations in central values.  

The probability weights defined in \cref{modelProb} can be used to assess the relative weight of specific analysis choices in the BMA. Comparison of these weights can identify if one particular choice of observable extraction method, correction scheme or fit-function variation is preferred by the averaging procedure.  More specifically, letting $S$ denote a subset of the full space of models $\{M\}$.  We can define the ``subset probability'' of $S$ by the relative posterior probability of the variations contained in $S$: 
\begin{equation}
\pr(S | D) = \sum_{M_i \in S} \pr(M_i | D).
\end{equation}
The subset probability encapsulates the relative weight of the models in a given subset compared to the whole model space, informed by the data.  For example, we can estimate the subset probability of using NNLO \chpt\ for taste-breaking and finite-volume corrections as
\begin{equation}
\pr(\mathrm{NNLO} | D) = \sum_{M_i \in \mathrm{NNLO}} \pr(M_i | D).   
\end{equation}
where ``$M_i \in \mathrm{NNLO}$'' denotes the subset of models ({\it i.e.}, analysis choices) in which NNLO \chpt\ is used for both corrections. Using this definition, we show the relative probabilities of the subsets  considered above as pie charts in \cref{fig:BMACompareW} (right). From the top pie chart, for the two methods of observable extraction, we find roughly equal contributions to the overall BMA result, indicating no preference by the BMA procedure. The second pie-chart from the top shows the subset probabilities for specific taste-breaking corrections. The probability of the subset in which the data are not corrected prior to continuum extrapolation is smaller by slightly more than a factor of two compared with the other subsets. This is because the taste-breaking corrections are computed in two window regions, [0.4, 1] and [0.7, 1]~fm in addition to the continuum fits to data without taste-breaking corrections having larger $\chi^2$ values, indicating a preference for data corrected for taste-breaking.
The third pie chart shows that quadratic continuum fits to the full set of four ensembles are preferred over cubic fits or fits to just three ensembles. In the case of the fits to three ensembles the smaller subset probability can be traced back to the penalty incurred, $N_{\mathrm{cut}}$, in \cref{modelProb} due to dropping a data point. 
For subsets in which the same correction scheme is used for finite volume and taste splittings (bottom pie chart) we find a slight preference for NNLO~\chpt\ and slight disinclination for NLO~\chpt.

\begin{table}
\centering
\caption{Comparison of results for $\amuLW$ and $\amuLWTwo$ obtained from the BMA analysis with an empirical Bayes approach. Both analyses use data sets corrected in the CM scheme. The third and fifth columns show $\amu$ results obtained in the empirical Bayes approach from fits to data sets without and with first correcting for taste splittings, respectively. The fourth and sixth columns list the effective scales $Q_{\rm eff}$ obtained by maximizing the Bayes Factor log(GBF).\vspace{1mm}}
\label{table:empBayesCrossCheck}
\setlength{\extrarowheight}{4pt}
\begin{tabular}{l|r|r|r|r|r}
\hline \hline
 & \multirow{2}*{\quad BMA: CM \quad}  & \multicolumn{4}{c}{\quad Empirical Bayes\quad} \\ 
 & & \multicolumn{1}{c|}{$\Delta_{\rm TB} = 0$} & \multicolumn{1}{c|}{$Q_{\rm eff}/$GeV} & \multicolumn{1}{c}{$\Delta_{\rm TB} \neq 0$} & \multicolumn{1}{|c}{$Q_{\rm eff}/$GeV} \\
\hline
$\amuLW$ & 206.28(81) \quad & \quad 206.52(69) \quad & \quad 1.8 \quad & \quad 205.83(67) \quad  & \quad 1.7 \quad \\
$\amuLWTwo$ & 100.9(3.3) \quad & \quad 98.7(2.8) \quad &  \quad 2.9 \quad & \quad 102.0(2.0) \quad & \quad 1.9 \quad \\
\hline \hline
\end{tabular}
\end{table}

\Cref{fig:BMACompareW2} shows the BMA subsets for $\amuLWTwo$, with results similar to those for $\amuLW$. As expected, there is greater consistency among the subset averages from specific correction schemes compared to the $\amuLW$ case, with the largest variation in central value coming from continuum extrapolations to data not corrected for taste-breaking effects. The pie charts in the right panel of \cref{fig:BMACompareW2} reveal roughly equal subset probabilities in each case, except for the third (from the top) pie chart, which illustrates that here too quadratic continuum fits to all four ensembles are preferred for the same reasons as above.

The AIC criterion used in Ref.~\cite{Borsanyi:2020mff} differs from \cref{modelProb} in that the weight assigned to cutting data points is given as $N_{\rm cut}$ instead of $2N_{\rm cut}$. In order to test the robustness of the model-averaging procedure, we repeat the analysis by replacing $2N_{\rm cut}$ with $N_{\rm cut}$ in \cref{modelProb}. We find that this yields central values and uncertainties on the final results are essentially the same as before, with at most minor changes to the weights in the third pie-chart from the top in \cref{fig:BMACompareW,fig:BMACompareW2}. This result is not unexpected, because in our case, $N_{\rm cut} \leq 1$, and only a small fraction of the total variations in our averages have $N_{\rm cut} \neq 0$.

In order to cross check our main continuum-limit extrapolations and the subsequent BMA analysis, we use an empirical Bayes approach to perform independent continuum-limit extrapolations (see \cref{eq:GPLfunc}). In the comparison of the two approaches, we use the $\amu$ data sets obtained from integrating the correlation functions with Trapezoidal rule and corrected using the CM scheme, with and without first correcting for taste splittings. When performing continuum extrapolations using all the terms in \cref{eq:GPLfunc}, we observe that most of the posterior coefficients are small and consistent with zero --- only the linear $\sim c_{11} a^2 \alpha_s$ and quadratic $\sim c_{20} a^4$ terms in \cref{eq:GPLfunc} are needed to describe the data.  This observation is consistent with our main continuum-extrapolation analysis, described by \cref{eq:fitfunc,eq:discfunc}. \Cref{table:empBayesCrossCheck} shows the comparison of the BMA analysis (restricted to the same CM-corrected data sets) with the empirical Bayes fits for both $\amuLW$ and $\amuLWTwo$. We find good agreement in central values and error bars, after considering the spread between the empirical Bayes results from data sets with and without first correcting for taste splittings.

\subsection{Results and error budgets}\label{sec:errors}

\begin{table}
\centering
\caption{Approximate error budgets for $\amuLW$ and $\amuLWTwo$. \vspace{1mm}}
\label{table:WUncertainty}
\begin{tabular}{l|c|c}
\hline \hline
Source & \quad $\delta \amuLW$ (\%)  \quad  & \quad $\delta \amuLWTwo$ (\%)  \quad  \\ 
\hline
Monte Carlo statistics  & 0.19 & 2.44 \\
Continuum extrapolation ($a \to 0$, $\Delta_{\textrm{TB}}$) & 0.34  & 1.05\\
Finite-volume correction ($\Delta_{\textrm{FV}}$)  & 0.16  & 0.23 \\
Pion-mass adjustment ($\Delta_{M_\pi}$)  & 0.06  & 0.96  \\ 
Scale setting ($w_0$ (fm), $w_0 / a$) & 0.21  & 1.28 \\
Current renormalization ($Z_V$) $\quad \quad \quad  \quad \quad \quad \quad \quad \quad $ & 0.17  & 0.16 \\ 
\hline
Total & 0.50\%  & 3.18\% \\ 
\hline \hline
\end{tabular}
\end{table}

Our results for the light-quark-connected contributions to $\amuW$ and $\amuWTwo$ are
\begin{equation}
\amuLWRes \label{eq:amuLW}
\end{equation}
and
\begin{equation}
\amuLWTwoRes \;, \label{eq:amuLWTwo}
\end{equation}
where the errors are those obtained from the BMA procedure described in the previous section, and include both statistical and systematic uncertainties.

Although Bayesian model averaging provides a robust estimate of the total uncertainties in our results, the construction of detailed error budgets from the BMA is not straightforward.  We start from the expression for the BMA variance in \cref{eq:BMAVar}. The first term on the right-hand side is linear in the variances, and hence can be trivially separated into individual contributions from Monte Carlo statistics and each of the parametric inputs $w_0$, $\Delta_{M_\pi}$, and $Z_V$. For example, the statistical uncertainty is given by 
\begin{align}
\sigma_{a_{\mu}}^{2}(\textrm{stat.})&= \sum_{i=1}^{N_{M}} \sigma_{a_{\mu}, i}^{2}(\textrm{stat.}) \mathrm{pr}\left(M_{i} \mid D\right) \label{eq:BMAVarStat} 
\end{align}
where we average over all analysis variations using the probability weights of \cref{modelProb}. Repeating this procedure for all the above-mentioned contributions yields the error estimates in \cref{table:WUncertainty} in the rows marked ``Monte Carlo statistics'', ``Scale setting'', ``Pion-mass adjustment'' and ``Current renormalization''. 
The second and third terms in \cref{eq:BMAVar} depend solely and non-linearly on the central value of each variation, with the latter term including pairwise differences between all possible model pairs in the full BMA result. This makes it impossible to strictly disentangle the contribution from only a subset of model variations, ({\it e.g.}, finite-volume corrections or treatment of discretization effects). We can obtain an approximate error budget, however, as follows.

\begin{figure}[t]
\centering
\includegraphics[width=.605\textwidth]{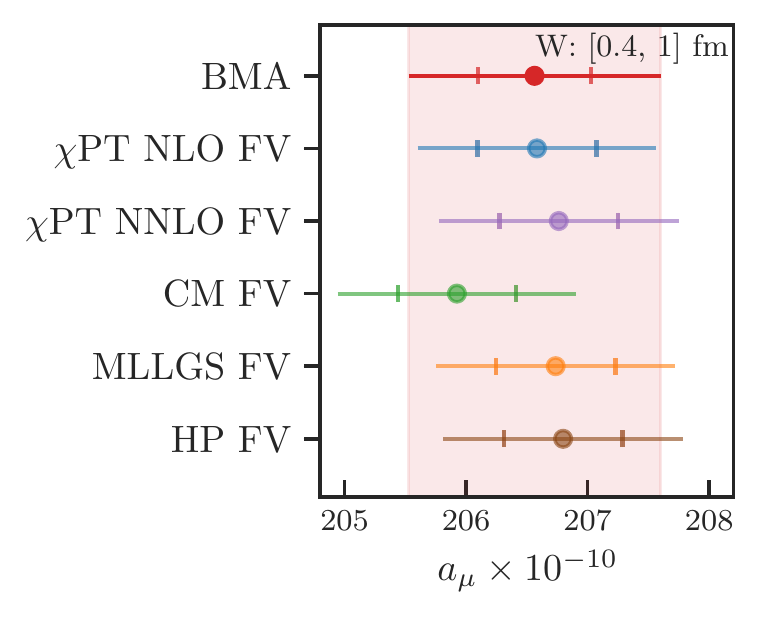}\includegraphics[width=.4\textwidth]{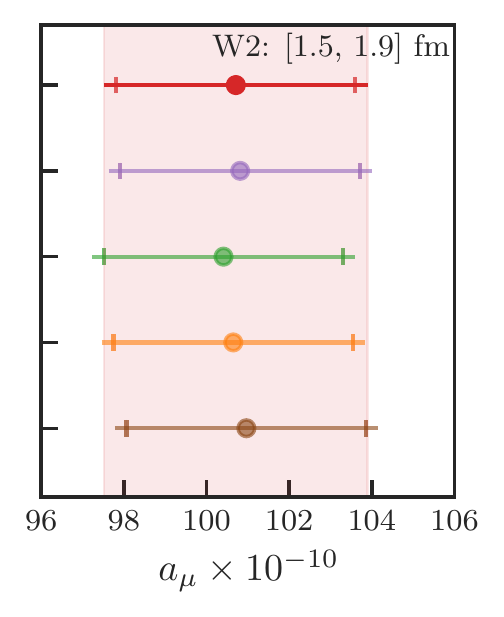}
\vspace{-3mm}
\caption{Breakdown of the BMA result into subsets that contain only one choice of finite-volume correction for $\amuLW$ (left) and $\amuLWTwo$ (right).}
\label{fig:BMA_FVBreakdown}
\end{figure}

First, to estimate the systematic uncertainty associated with the finite-volume correction, we perform subset model averages separately for each finite-volume correction scheme. These results are shown in \cref{fig:BMA_FVBreakdown}. Taking the variance in central values of these results yields the ``finite-volume correction'' error in \cref{table:WUncertainty}. Next, we subtract (in quadrature) the so-estimated finite-volume error from the total model variance. The remaining uncertainty is associated with variations in the treatment of oscillating states in $C(t)$, the taste-breaking corrections, and the continuum-extrapolation fit function. Combining this uncertainty (in quadrature) with those on the fit-function coefficient posteriors yields the ``continuum extrapolation'' error in \cref{table:WUncertainty}.

Table~\ref{table:WUncertainty} presents the approximate error budgets for $\amuLW$ and $\amuLWTwo$ obtained from the above approach.
For $\amuLW$, the largest error is from the continuum extrapolation, and is driven by the spread in results using different taste-breaking correction schemes. Here we note that the consistency between quadratic and cubic continuum extrapolations (as illustrated in \cref{fig:BMACompareW,fig:BMACompareW2}) as well as between our main results and those from the empirical Bayes approach (see \cref{table:empBayesCrossCheck}) indicate that systematic errors due residual higher-order discretization effects are well encompassed by our uncertainties.
Next is the parametric uncertainty from the gradient-flow scale, which is about 30\% smaller. Errors from Monte-Carlo statistics, finite-volume corrections, and current renormalization are also non-neglible, and are roughly commensurate.  For $\amuLWTwo$, Monte-Carlo statistics are by far the largest source of uncertainty. Following that, the contributions from scale setting, the continuum extrapolation, and the pion-mass adjustment, which are $\sim 50$--60\% smaller.  Although finite-volume and current-renormalization errors are negligible compared with these other uncertainties, they will be important for calculations of $\amuHVP$ aiming for $\lesssim 0.5\%$ precision.

\section{Summary and outlook}\label{sec:conclusion}

\begin{figure}
\centering
\hspace{-20mm}
\includegraphics[width=.80\textwidth]{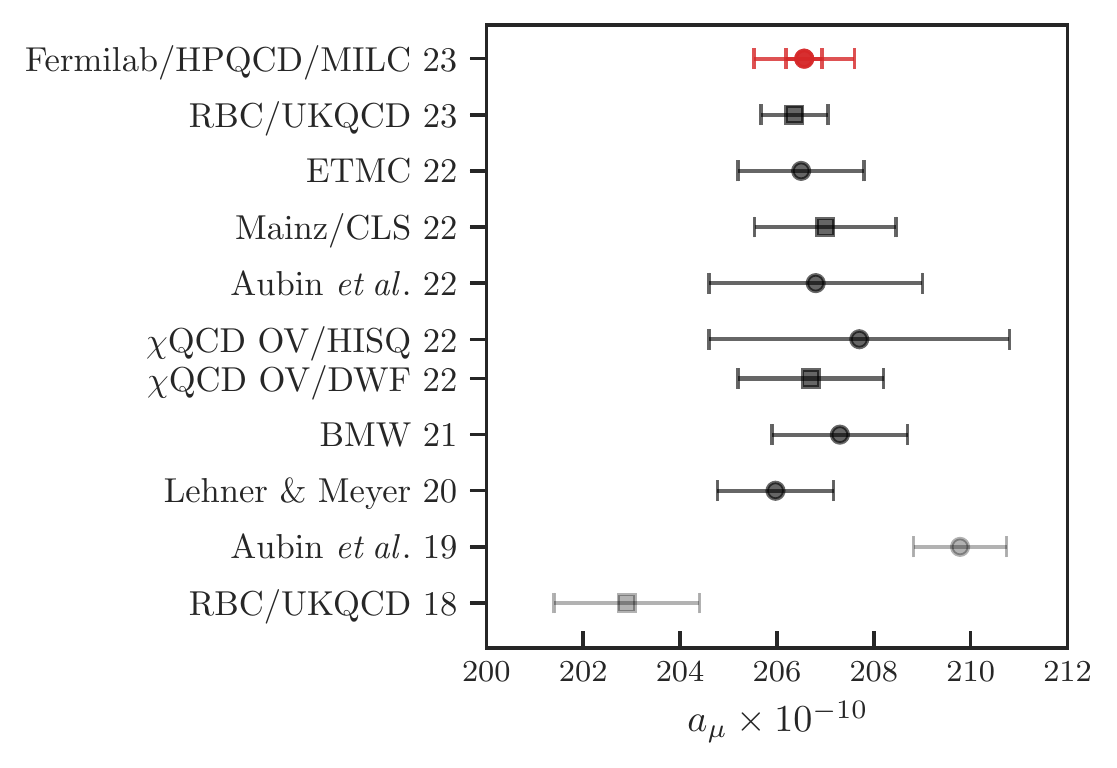}
\caption{Comparison of our lattice determination of $\amuLW$ (red circle) labeled ``Fermilab/HPQCD/MILC~23'' to $n_f=2+1+1$ (black circles) and $n_f=2+1$ (black squares) lattice-QCD calculations by RBC/UKQCD~23 \cite{Blum:2023qou}, ETMC~22 \cite{Alexandrou:2022amy}, Mainz/CLS~22 \cite{Ce:2022kxy}, Aubin {\it et al.}~22 \cite{Aubin:2022hgm}, $\chi$QCD 22 \cite{Wang:2022lkq}, BMW~21 \cite{Borsanyi:2020mff} and Lehner \& Meyer~20 \cite{Lehner:2020crt}. Results by Aubin {\it et al.}~19 \cite{Aubin2019} and RBC/UKQCD~18 \cite{RBC:2018dos}, shown in grey, are superseded by Aubin {\it et al.}~22 and RBC/UKQCD~23, respectively. The inner error bar shown for our result is from Monte Carlo statistics.}
\label{fig:WCompare}
\end{figure}

In \cref{fig:WCompare}, we compare our intermediate-window result, \cref{eq:amuLW}, with other lattice-QCD calculations of this quantity~\cite{Alexandrou:2022amy,Ce:2022kxy,Aubin:2022hgm,Wang:2022lkq,Borsanyi:2020mff,Lehner:2020crt,RBC:2018dos,Blum:2023qou,Aubin2019}, which were obtained using different lattice actions and analysis methods.
Of the results to date, ours has the smallest statistical uncertainty, 0.19\%.
Ours is also the first result for $\amuLW$ obtained from a blind analysis.
While some form of EFT-inspired correction schemes were employed in every calculation, our analysis is the first to include all of them.
Because we incorporate uncertainties due to analysis choices via Bayesian model averaging~\cite{Jay:2020jkz,Neil:2022joj}, our systematic error estimate is robust without being overly conservative.

\begin{figure}[t]
\centering
\hspace{-20mm}
\includegraphics[width=.70\textwidth]{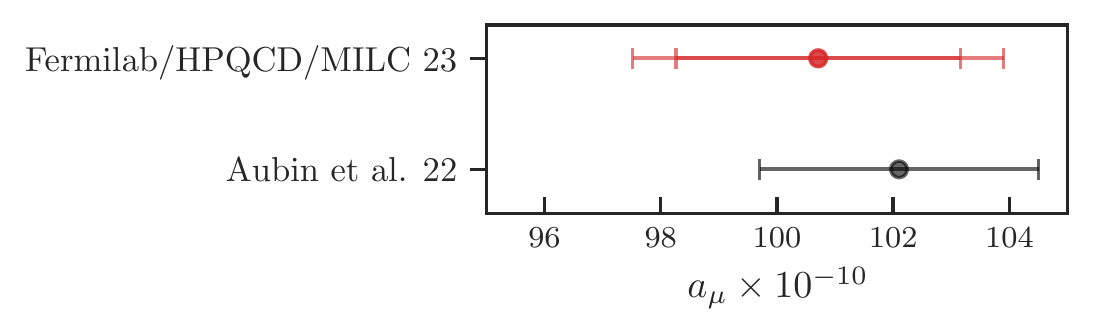}
\caption{Comparison of our lattice determination of $\amuLWTwo$ (red circle) labeled ``Fermilab/HPQCD/MILC 23'' with the result of Ref.~\cite{Aubin:2022hgm} (black circle) labeled Aubin {\it et al.} 22. The inner error bar shown for our result is from Monte Carlo statistics.}
\label{fig:W2Compare}
\end{figure}

In \cref{fig:W2Compare}, we compare our result for the ``W2'' window observable, \cref{eq:amuLWTwo}, with the only other available lattice-QCD result for this quantity~\cite{Aubin:2022hgm}.
Although the results appear consistent, they are not wholly independent because the analysis in Ref.~\cite{Aubin:2022hgm} is based on some of the same ensembles as employed in this work.
Statistical and systematic correlations due to the shared configurations must be taken into account to make a quantitative comparison.
Other independent lattice-QCD calculations of $\amuLWTwo$ would provide welcome consistency checks.

Before our results for $\amuLW$ and $\amuLWTwo$ can be directly compared with data-driven determinations,
the contributions from heavier flavors must be added as well as those from quark-line disconnected contractions and isospin-breaking corrections (QED and $m_u\neq m_d$).
The $s$-, $c$-, and $b$-quark-connected contributions to $\amuHVP$ have already been computed on the HISQ ensembles with high precision~\cite{Chakraborty:2014mwa,Hatton:2020qhk,Hatton:2021dvg}; windowing these results will be straightforward.
The remaining contributions are being computed in ongoing projects; see Refs.~\cite{Yamamoto:2018cqm, FermilabLattice:2019dbx, FermilabLattice:2021hzx, Ray:2022ycg}. 

Looking at the big picture, the observed consistency between so many different, largely independent, results for the light-quark connected contribution to the intermediate-window observable (see \cref{fig:WCompare}) indicates that the systematic errors in lattice-QCD calculations of this quantity are under reasonable control.
It is therefore unlikely that the differences between the lattice-QCD calculations reported in Refs.~\cite{Borsanyi:2020mff, Ce:2022kxy, Alexandrou:2022amy,Blum:2023qou} and the data-driven result of Ref.~\cite{Colangelo:2022vok} will be resolved by further improvements in lattice-QCD calculations of~$\amuLW$.  
Lattice-QCD calculations of the quark-connected contributions from heavier flavors are also unlikely causes of the difference, since their uncertainties are smaller by an order of magnitude~\cite{Borsanyi:2020mff,Hatton:2020qhk,Hatton:2021dvg}.
The quark-disconnected and isospin-breaking contributions to $\amuHVP$, however, have been computed by only a few collaborations~\cite{FermilabLattice:2017wgj,RBC:2018dos,Gerardin:2019rua,Giusti:2019xct,Borsanyi:2020mff,Blum:2023qou}.\footnote{Indeed, only the BMW collaboration~\cite{Borsanyi:2020mff} has presented a complete calculation of all contributions to $\amuHVP$ including the disconnected QED and disconnected strong-isospin-breaking corrections.}
Although these contributions are too small to change $\amuW$ substantially, additional independent lattice-QCD calculations are needed to solidify the central value and uncertainty in order to better quantify the significance of the difference. 

In Ref.~\cite{Davies:2022epg}, we pointed out that other windowed observables can provide more stringent comparisons between lattice-QCD and data-driven results right now.
Because intermediate-window observables cut out low-$t$ contributions to $\amuHVP$ where lattice-QCD statistical errors are smallest, ``one-sided windows" without a lower bound on the Euclidean time can capture a larger fraction of the total $\amuHVP$ while retaining controlled uncertainties.
We are currently repeating the analysis of Ref.~\cite{Davies:2022epg} using the larger data set employed in this work.

The light-quark connected contribution to the intermediate-window observable represents only around a third of the total leading-order HVP contribution to the muon's anomalous magnetic moment.
Thus, the work presented in this paper is only a part of a multi-year project to compute $\amuHVP$ with $\lesssim0.5\%$ precision.
Several of our ongoing efforts aim to reduce the dominant sources of uncertainty in our published result for the total light-quark connected contribution to $\amuHVP$~\cite{Davies:2019efs}; these will also improve our determinations of the intermediate-window observables in this work. 
For example, recently we introduced a ``low-mode-improved'' method into our analysis that substantially reduces statistical errors at large Euclidean times~\cite{LynchLat22Poster}.
The  uncertainty on the scale-setting quantity $w_0$ is an important source of uncertainty not only for all $\amuHVP$ observables, but also for many other analyses based on the MILC HISQ ensembles.
We are therefore working to compute precisely the $\Omega$-baryon mass on these ensembles~\cite{Lin:2019pia}, as well as the relative scale $w_0/a$, and plan to use the results to determine the scale in physical units with reduced uncertainty.

With these ongoing efforts, we expect to obtain $\amuHVP$ with sub-percent-level precision in the near future.
In order to further reduce the precision to match that of the Fermilab~\cite{PhysRevLett.126.141801} and JPARC~\cite{Abe:2019thb,E34webpage} experiments,  however, it seems likely that considerable exascale computing resources will be needed. In particular, the inclusion of MILC's physical-mass HISQ ensemble with $a\approx 0.042$~fm would enable more robust continuum extrapolations of all $\amuHVP$ observables and provide better control over this important source of systematic error. A direct finite-volume study is needed to better quantify the finite-volume corrections and reduce the corresponding uncertainty. This would require the generation and analysis of new ensembles with different spatial volumes and all other parameters held fixed. Finally, further control over long-distance effects and statistical noise could be achieved by computing directly the two-pion contributions to the vector current correlation functions~\cite{Lahert:2021xxu,Erben:2019nmx,Bruno:2019nzm}.

\begin{acknowledgments}

We thank Claude Bernard, Urs Heller, Jack Laiho, Bob Sugar, and Doug Toussaint for their scientific leadership and collaboration. In particular, we are grateful to Bob for his tireless efforts to obtain computational resources, to Claude for guidance on chiral perturbation theory, to Doug for his invaluable expertise in creating so many of our gauge-field ensembles, and to Jack and Urs for essential contributions to previous projects that formed the basis for this work. We thank Anthony Grebe for his contributions to the chiral perturbation theory codes used in this work. We thank Maarten Golterman for useful comments and suggestions.
Computations for this work were carried out in part with resources provided by the USQCD Collaboration, the National Energy Research Scientific Computing Center (Cori), the Argonne Leadership Computing Facility (Mira) under the INCITE program, and the Oak Ridge Leadership Computing Facility (Summit) under the Innovative and Novel Computational Impact on Theory and Experiment (INCITE) and the ASCR Leadership Computing Challenge (ALCC) programs, which are funded by the Office of Science of the U.S.\ Department of Energy. 
This work used the Extreme Science and Engineering Discovery Environment (XSEDE) supercomputer Stampede 2 at the Texas Advanced Computing Center (TACC) through allocation TG-MCA93S002. The XSEDE program is supported by the National Science Foundation under grant number ACI-1548562.
Computations on the Big Red II+ and Big Red 3 supercomputers were supported in part by Lilly Endowment, Inc., through its support for the Indiana University Pervasive Technology Institute.
The parallel file system employed by Big Red II+ is supported by the National Science Foundation under Grant No.~CNS-0521433.
This work utilized the RMACC Summit supercomputer, which is supported by the National Science Foundation (awards ACI-1532235 and ACI-1532236), the University of Colorado Boulder, and Colorado State University. The Summit supercomputer is a joint effort of the University of Colorado Boulder and Colorado State University.
Some of the computations were done using the Blue Waters sustained-petascale computer, which was supported by the National Science Foundation (awards OCI-0725070 and ACI-1238993) and the state of Illinois. Blue Waters was a joint effort of the University of Illinois at Urbana-Champaign and its National Center for Supercomputing Applications. 
Some computations also used the Cambridge Service for Data Driven Discovery (CSD3) operated by University of Cambridge Research Computing on behalf of the STFC DiRAC HPC Facility. The DiRAC component of CSD3 was funded by BEIS and STFC under grants ST/P002307/1, ST/R002452/1 and ST/R00689X/1. 

This work was supported in part by the U.S.~Department of Energy, Office of Science, under Awards
No.~DE-SC0010005 (E.T.N.),
No.~DE-SC0010120 (S.G.), 
Nos.~DE-SC0011090 and~DE-SC0021006 (W.J.),
No.~DE-SC0015655 (A.X.K., S.L., M.L., A.T.L.), and
the Funding Opportunity Announcement Scientific Discovery through Advanced Computing: High Energy Physics, LAB 22-2580; 
by the National Science Foundation under Grants Nos.~PHY17-19626 and PHY20-13064 (C.E.D., A.V.) and from their Graduate Research Fellowship under Grant DGE 2040434 (C.T.P); 
by the Simons Foundation under their Simons Fellows in Theoretical Physics program (A.X.K.); 
by the Universities Research Association Visiting Scholarship awards 20-S-12 and 21-S-05 (S.L.);  
by SRA (Spain) under Grant No.\ PID2019-106087GB-C21 / 10.13039/501100011033 (E.G.);
by the Junta de Andalucía (Spain) under Grants No.\ FQM-101, A-FQM-467-UGR18 (FEDER), and P18-FR-4314 (E.G.);
by AEI (Spain) under Grant No.\ RYC2020-030244-I / AEI / 10.13039/501100011033 (A.V.);
and by UK Science and Technology Facilities Council under Grant ST/T000945/1 (C.T.H.D). 
This document was prepared by the Fermilab Lattice, HPQCD, and MILC Collaborations using the resources of the Fermi National Accelerator Laboratory (Fermilab), a U.S. Department of Energy, Office of Science, HEP User Facility.
Fermilab is managed by Fermi Research Alliance, LLC (FRA), acting under Contract No.~DE- AC02-07CH11359.

\end{acknowledgments}

\appendix

\section{Cross-checks of window determinations from staggered correlation functions}\label{sec:appendix:fit}

\begin{figure}[t]
\centering
\includegraphics[width=.85\textwidth]{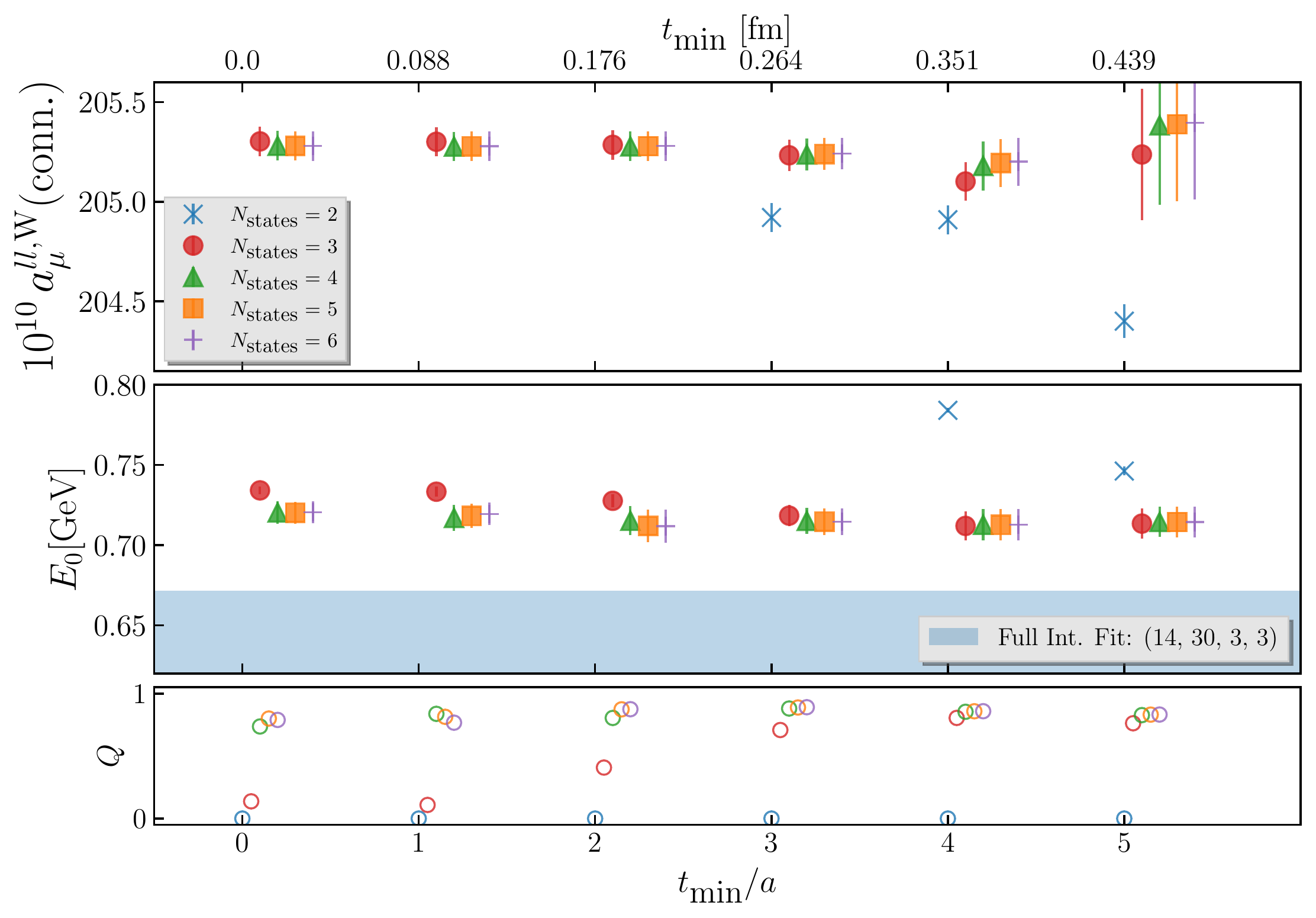}
\caption{Fit and reconstruction results on the $0.09$~fm ensemble. (Top) $\amuLW$ from the reconstruction for a range of $t_\textrm{min}$ and number of exponentials in the fit function values with $t_\textrm{max}=1.3$~fm fixed. (Middle) Ground state energies determined by the fit and fit result for the late-time correlation function obtained from a fit with three states, $t_\textrm{min}/a=14$ and $t_\textrm{max}/a=30$ (blue band). (Bottom) Fit quality, Q, coming from the augmented chi squared fit. }
\label{fig:fitStab}
\end{figure}

\begin{figure}[t]
\centering
\includegraphics[scale=0.55]{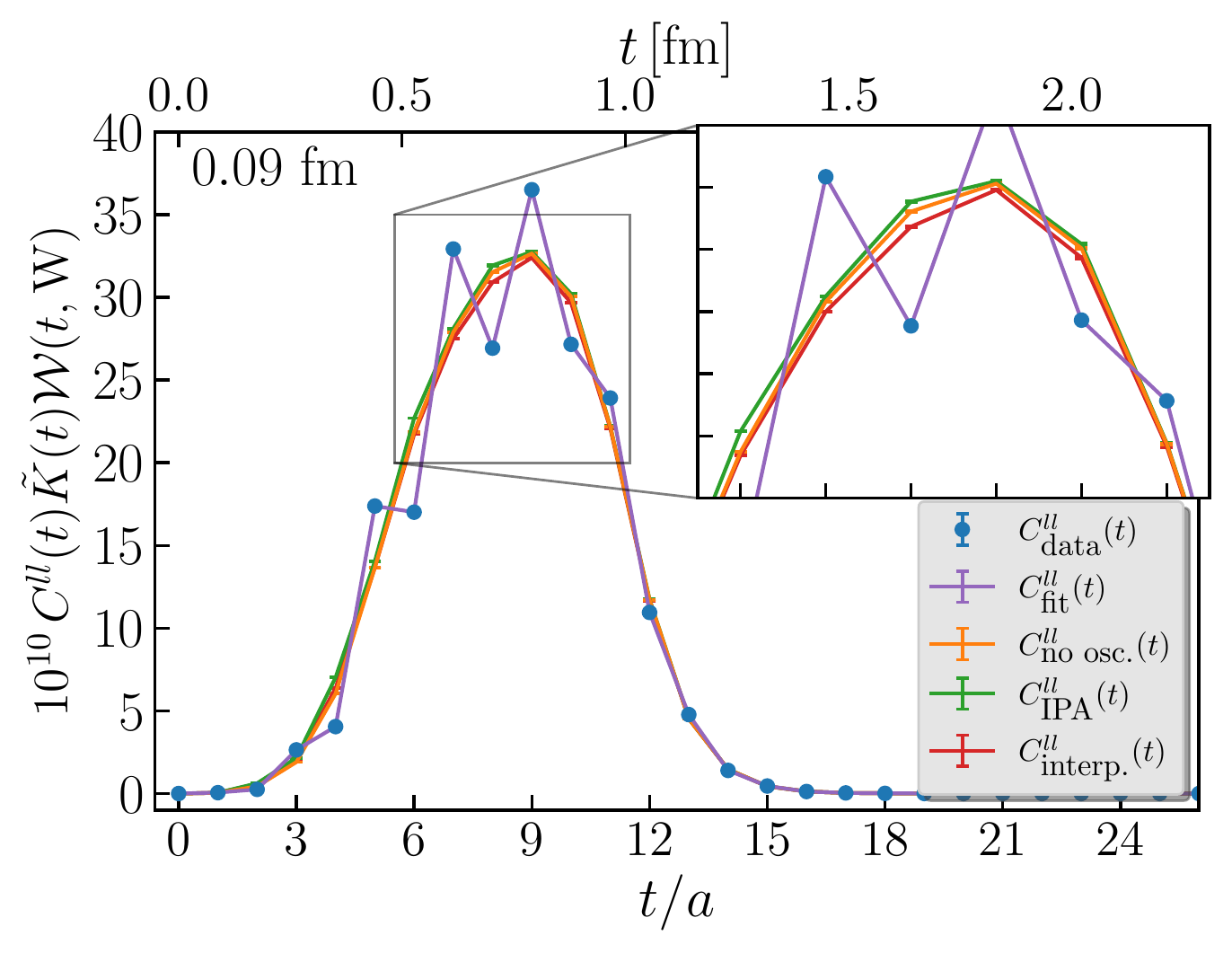} \includegraphics[scale=0.50]{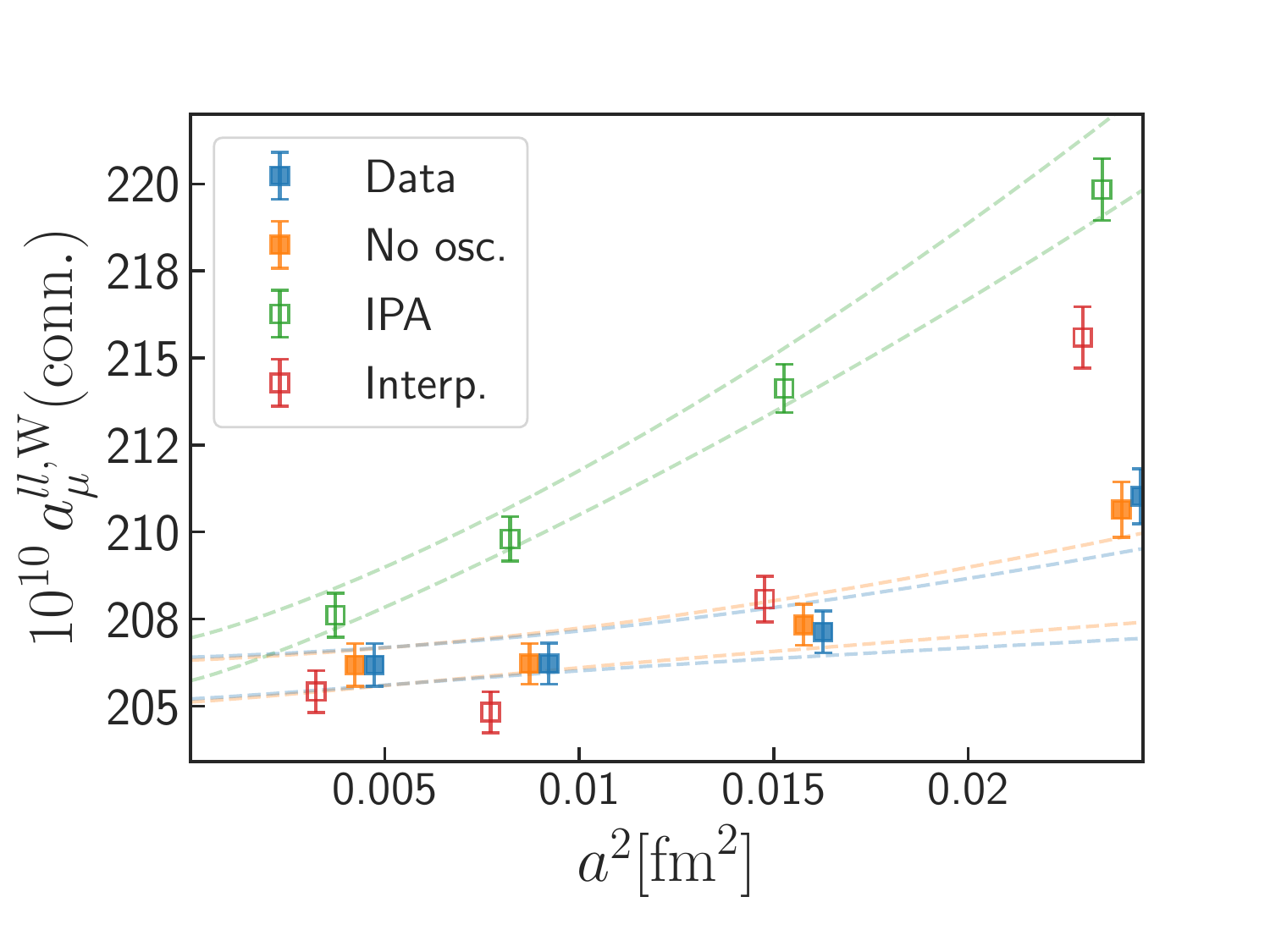}\\
\includegraphics[scale=0.55]{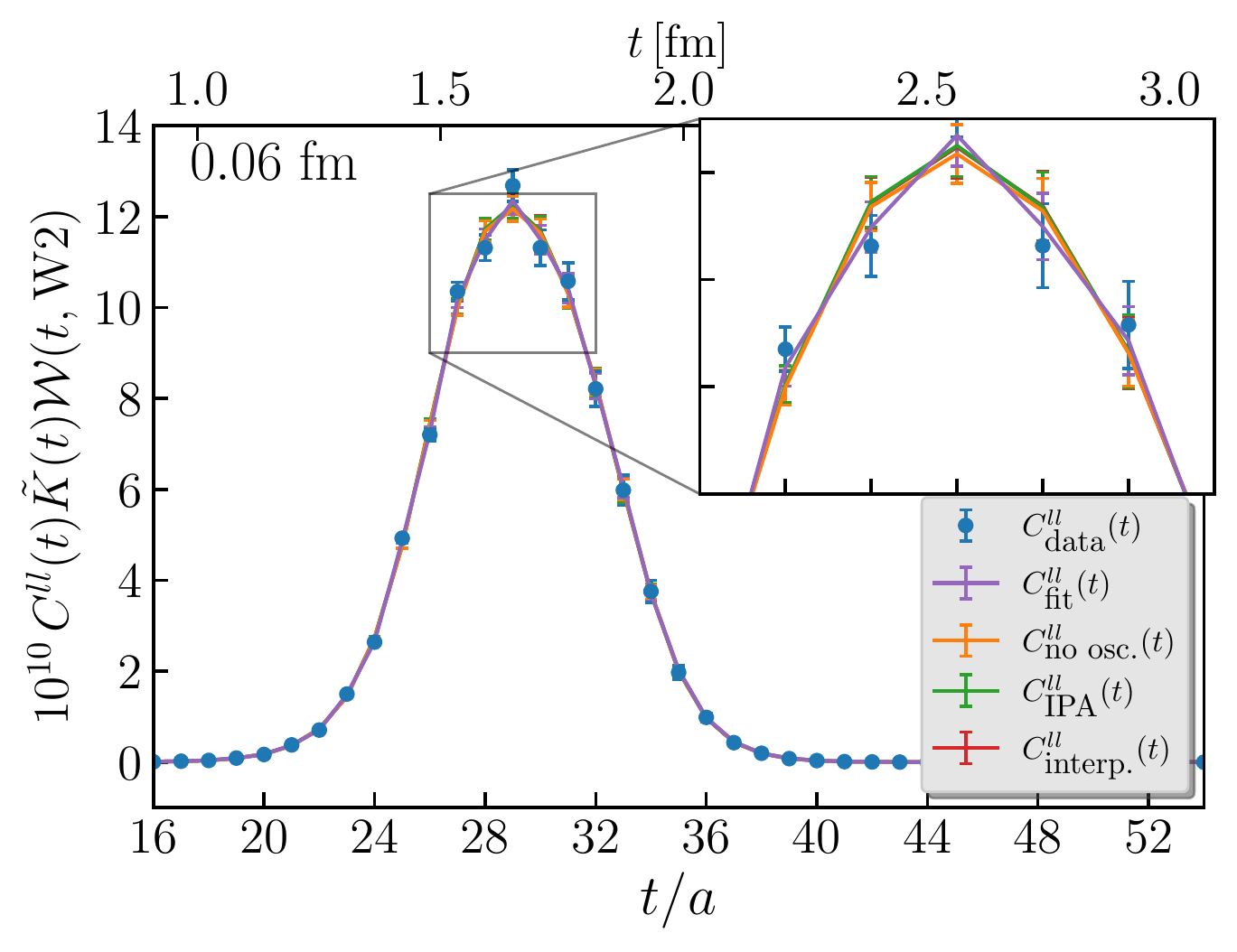} \includegraphics[scale=0.50]{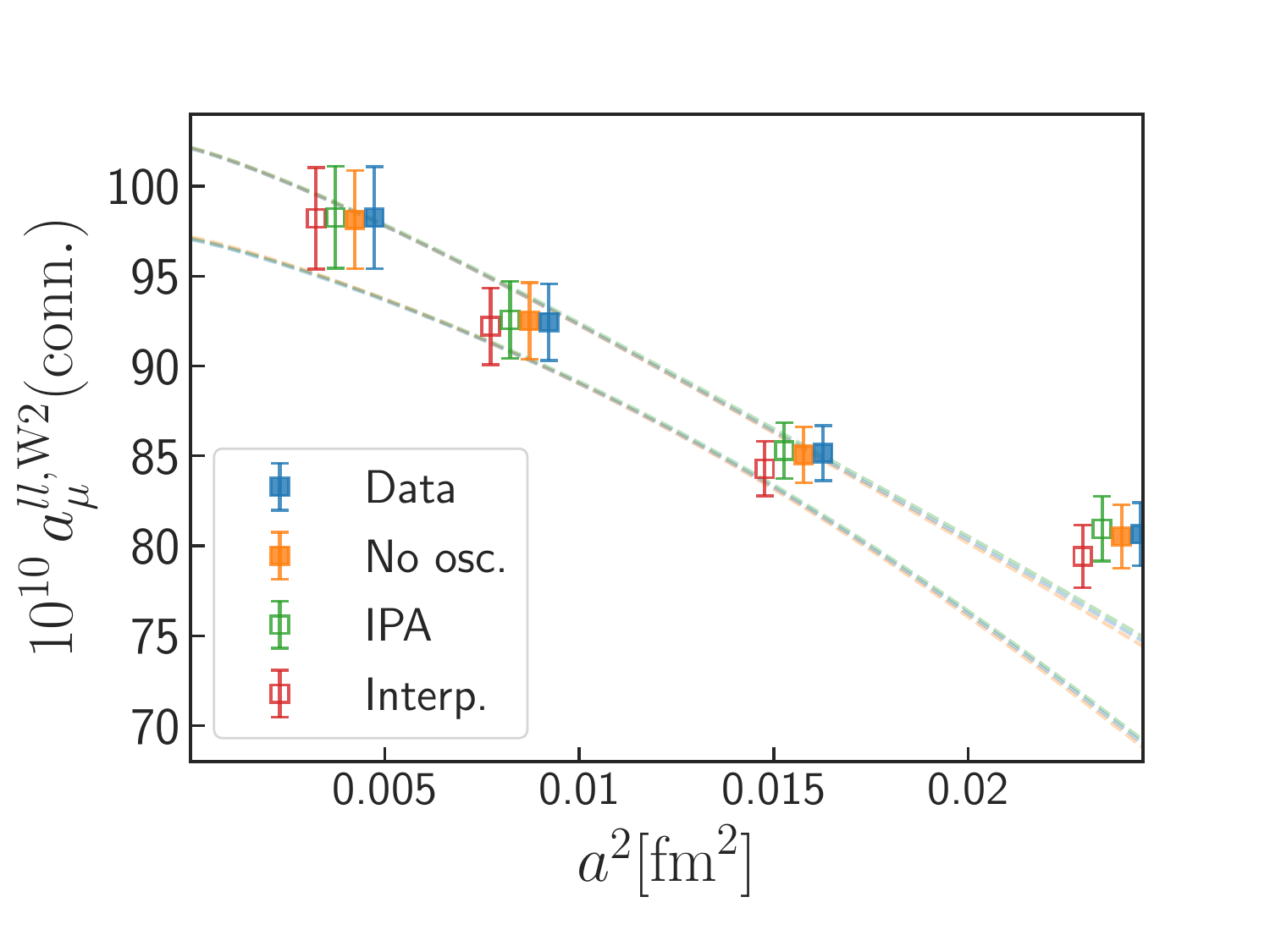}
\caption{(Left) Comparison of the methods used to remove the oscillating contribution to the integrand for $\amuLW$ at 0.09~fm (top left) and $\amuLWTwo$ at 0.06~fm (bottom left). Shown are the integrands obtained with raw correlation-function data $C(t)$ (blue circles), the reconstruction from the fit including oscillating states $C_{\rm fit}(t)$ (purple), without oscillating states $C_{\textrm{no osc.}}(t)$ (orange), improved parity averaged correlator $C_{\mathrm{IPA}}(t)$, (\cref{eq:IPA}) (green), and interpolated correlator  $C_{\mathrm{interp}}(t)$ (\cref{eq:interp}) (red). (Right) Lattice-spacing dependence of $\amuLW$ (top right) and $\amuLWTwo$ (bottom right) data obtained from the correlation functions modified with the oscillation removal techniques discussed. All data sets are corrected for finite-volume effects using the Chiral Model and pion-mass mistuning effects using the data-driven approach, described in \cref{sec:DeltaFV,sec:DeltaMpi}. The data points are slightly displaced horizontally for clarity. A linear fit function (see \cref{sec:cont_extrap}) is used to fit the $\amuLW$ and $\amuLWTwo$ data at the three finest lattice spacings.}
\label{fig:oscRemovalW}
\end{figure}

In this section, we detail the methods used to obtain the windowed $\amuHVP$ from the staggered correlation function, $C(t)$. First, we compare three different approaches for treating the oscillating contribution to the correlator. The first method, which we use in our main analysis, is to fit $C(t)$ over the region of interest and reconstruct it from the fit posteriors excluding the oscillatory contribution. The correlation function has the spectral representation, 
\begin{align}
C(t) &= \sum_n\left[ Z^2_n e^{-E_n t} + (-1)^t Z^2_{n,\textrm{osc}} e^{-E_{n,\textrm{osc}} t} \right],\label{eq:corrfitfunc}
\end{align}
where the sum is over all possible contributing states. We use this expression to craft a model fit function that separates the oscillating and nonoscillating contributions.  For this purpose, we truncate the sum:
\begin{align}
C_{\textrm{fit}}(t) &= \sum^{N_{\textrm{states}}}_n \left[Z^2_n e^{-E_n t} + (-1)^t Z^2_{n,\textrm{osc}} e^{-E_{n,\textrm{osc}} t}\right]. \label{eq:fitRecon} 
\end{align}
For simplicity, we keep the same number $N_{\textrm{states}}$ of regular and oscillating states. We restrict the fit range $[t_\textrm{min}, t_\textrm{max}]$ to cover the window region of interest. The use of $t_\textrm{min}$ justifies the truncation to a finite $N_\textrm{states}$ by suppressing contributions from states with large energy.  On the other hand, the use of a fixed $t_\textrm{max}$ carries a risk that the lowest-lying energies and amplitudes may not be accurately resolved with finite statistical precision.  However, since  we are simply using the expression as a useful model for removing the unwanted oscillations, it is not critical that our estimates of all energy levels are asymptotically correct for $t_{\rm max} \rightarrow \infty$. For the energies and amplitudes of the light-quark-connected correlator, we take the Gaussian priors associated with the local (unsmeared) data in Eqs.~(A3) and~(A4) of Ref.~\cite{Chakraborty:2016mwy}. We then have the corresponding fit reconstruction of the correlation function without the oscillating contribution
\begin{align}
C_{\textrm{no osc.}}(t) &= \sum^{N_{\textrm{states}}}_n Z^2_n e^{-E_n t}. \label{eq:fitReconOsc} 
\end{align}
Results for $\amuLW$ computed from the fit reconstruction on the $0.09$~fm ensemble are shown in \cref{fig:fitStab} (top). Here, for the fit range, we fix $t_\textrm{max}=1.3$~fm ($t_1+ 2\Delta$) and vary $t_\textrm{min}$. We also fit up to six states with good stability obtained at four, which we take to be our value for $N_{\textrm{states}}$ on all ensembles. The second panel of \cref{fig:fitStab} shows the ground state energies obtained from these fits; shown also is the ground state energy obtained from a fit to the full correlation function (blue band). We see a significant difference in these energies, perhaps because the full fit picks up some mixture of the hard-to-determine two-pion states in the large-time region. Nonetheless, we observe in \cref{fig:oscRemovalW} (as described below) that $C_{\textrm{fit}}$ accurately reconstructs the correlation function data in the window region of interest. For $\amuLW$ and $\amuLWTwo$ we take $t_\textrm{min}$ and $t_\textrm{max}$ to be $2\Delta$ beyond the $t_0$ and $t_1$ boundaries in the corresponding window definition. For the coarsest two ensembles, this would correspond to a $t_\textrm{min}/a=0,1$. To avoid possible staggered-operator complications at small $t/a$, we take $t_{\rm min}/a = 2$ for those two ensembles.
As a test of the fidelity of this method, we show results for the correlated differences of $\amuLW$ and $\amuLWTwo$ computed from the fit reconstruction with the oscillating states, $C_{\textrm{fit}}(t)$, and the original correlation function in \cref{table:dataVSfit}. One can see tiny differences on the coarsest ensembles for $\amuLW$, likely due to the restriction of not using the first two time-slices; however, these differences are well within the uncertainties of the results for $\amuLW$.

The second method we examined is improved parity averaging (IPA) as employed in Ref.~\cite{Lehner:2020crt} for computing $\amuHVP$, a modification of the method developed in Ref.~\cite{Bailey:2008wp}. Here, the correlation function is replaced by the following equation:
\begin{align}
C_{\mathrm{IPA}}(t)=\frac{e^{-m_\rho t}}{4}\left[\frac{C(t-1)}{e^{-m_\rho(t-1)}}+2 \frac{C(t)}{e^{-m_\rho(t)}}+\frac{C(t+1)}{e^{-m_\rho(t+1)}}\right] \label{eq:IPA}
\end{align}
The exponent used is the PDG  value of the $\rho$ meson mass \cite{ParticleDataGroup:2022pth}, to give the best cancellation in the $\rho$ resonance peak which dominates in the regions of W and W2. This approach introduces additional discretization effects; however, one expects a consistent continuum limit as the oscillations become small at finer lattice spacing. 

The final approach, originally used in Ref.~\cite{Aarts:2007wj}, is performed by interpolating the even- and odd-site correlation functions separately, then averaging the two interpolations to obtain a new correlation function where the oscillating contribution has been removed.
\begin{align}
C_{\mathrm{interp}}(t)=\frac{1}{2}\left(C^{\text{even.\ interp}}(t)+C^{\text{odd.\ interp}}(t)\right) \label{eq:interp}
\end{align}
We use a cubic-spline interpolation with the Steffen algorithm implemented in the \textsc{gvar} Python package \cite{gvarGitHub} to interpolate the correlation functions.  

\begin{table}[t]
\centering
\caption{$\amuLW$ and $\amuLWTwo$ computed from the raw data (columns two and five), the fit reconstruction with oscillating states (columns three and six) and the correlated difference between them (columns four and seven). \vspace{1mm}}
\label{table:dataVSfit}
\begin{tabular}{l|c|c|c|c|c|c}
\hline \hline
$\approx a$ &$a^{ll,\mathrm {W}}_{\mu}(\mathrm{conn.})$ & $a^{ll,\mathrm{W}}_{\mu,\, \mathrm{fit}}(\mathrm{conn.})$ & $\Delta \amuLW$ &$a^{ll,\mathrm {W2}}_{\mu}(\mathrm{conn.})$ & $a^{ll,\mathrm{W2}}_{\mu,\, \mathrm{fit}}(\mathrm{conn.})$ & $\Delta \amuLWTwo$ \\
\hline
0.15 & 211.01(79) & 211.15(80) & $-$0.14(11) & 80.3(1.7) & 80.1(1.7) & 0.20(19) \\
0.12 & 207.13(60) & 207.16(60) & $-$0.025(29) & 84.7(1.5) & 84.6(1.5) & 0.09(10)   \\
0.09 & 206.56(55) & 206.58(55) & $-$0.016(10) & 92.7(1.8) & 92.7(1.8) & $-$0.07(22) \\
0.06 & 206.22(61) & 206.22(61) & 0.003(61) & 95.6(2.8) & 95.5(2.7) & 0.12(73) \\
\hline\hline
\end{tabular}
\end{table}

In \cref{fig:oscRemovalW} (top left), we compare the $\amuLW$ integrand on the 0.09~fm ensemble obtained from the raw correlator data (blue circles), $C_{\textrm{fit}}(t)$ (purple line), $C_{\textrm{no osc.}}(t)$ (orange line), $C_{\mathrm{IPA}}(t)$ (green line), and $C_{\mathrm{interp}}(t)$ (red line). We find that the $C_{\textrm{fit}}(t)$ integrand is in excellent agreement with the raw data in the region of interest, suggesting that $C_{\textrm{no osc.}}(t)$ is an accurate representation of the correlation function without the oscillating contribution. However, we see some differences between the $C_{\mathrm{IPA}}(t)$ and $C_{\mathrm{interp}}(t)$ integrands and the $C_{\textrm{no osc.}}(t)$ integrand, especially at shorter times where a large number of excited states contribute.

\Cref{fig:oscRemovalW} (top right) examines the lattice spacing dependence of the $\amuLW$ data obtained with all of the different oscillation removal techniques. In the case of  $\amuLW$ data obtained from $C_{\textrm{no osc.}}(t)$ we see only small deviations compared to the $\amuLW$ from the raw data which are more significant at coarser lattice spacing. As a result, the continuum extrapolations (which use a simple linear fit in $a^2 \alpha_s(2/a)$ to the three finest ensembles, leaving out the 0.15~fm data point) of the two data sets are in excellent agreement. While the IPA method does yield a consistent result in the continuum limit, it exibits much larger discretization effects. The interpolation method modifies the lattice spacing dependence so significantly that a linear fit is not enough to describe the observed behavior. This is likely due to the interpolation scheme not capturing the high energy state contributions sufficiently. In \cref{fig:oscRemovalW} (bottom), we compare these methods applied to $\amuLWTwo$; here the different methods give nearly identical results because the oscillations are less pronounced and fewer excited states contribute significantly.

\begin{table}[t]
\centering
\caption{$\amuLW$ and $\amuLWTwo$ computed from the raw data (columns two and five), the fit reconstruction without oscillating states (columns three and six) and the correlated difference between them (columns four and seven). \vspace{1mm}}
\label{table:OscVSNoOsc}
\begin{tabular}{l|c|c|c|c|c|c}
\hline \hline
$\approx a$ &$a^{ll,\mathrm {W}}_{\mu}(\mathrm{conn.})$ & $a^{ll,\mathrm{W}}_{\mu,\, \textrm{No osc.}}(\mathrm{conn.})$ & $\Delta \amuLW$ &$a^{ll,\mathrm {W2}}_{\mu}(\mathrm{conn.})$ & $a^{ll,\mathrm{W2}}_{\mu,\, \textrm{No osc.}}(\mathrm{conn.})$ & $\Delta \amuLWTwo$ \\
\hline
0.15 & 211.01(79) & 210.62(79) & 0.39(20) & 80.3(1.7) & 80.2(1.7) & 0.13(19) \\
0.12 & 207.13(60) & 207.34(59) & $-$0.204(34) & 84.7(1.5) & 84.6(1.5) & 0.10(11)   \\
0.09 & 206.56(55) & 206.56(55) & 0.001(10) & 92.7(1.8) & 92.7(1.8) & $-$0.07(22) \\
0.06 & 206.22(61) & 206.22(61) & 0.003(60) & 95.6(2.8) & 95.5(2.7) & 0.12(73) \\
\hline\hline
\end{tabular}
\end{table}

In order to quantify the effects of the oscillations in $\amuLW$ and $\amuLWTwo$, we use the fit approach, our preferred method of removing them, in \cref{table:OscVSNoOsc},  
where we compare results and correlated differences obtained using the trapezoidal rule (see \cref{sec:integrals}) for the raw correlation function data vs. $C_{\textrm{no osc.}}(t)$. For $\amuLW$, we find the differences to be small but statistically significant on the coarsest two ensembles and statistically zero on the finer ones. For $\amuLWTwo$, we find the differences to be zero on all ensembles, which is expected because the oscillating contributions are from heavier states which contribute significantly less in the long time region.

Finally, we examine the truncation effects associated with the trapezoidal rule by comparing $\amu$ observables computed from it to results obtained with Simpson's rule. Simpson's rule cannot be applied to the raw correlation function data because of the presence of oscillatory contributions. Hence, the comparisons in \cref{table:trapVSsimp} employ the $C_{\textrm{no osc.}}(t)$ correlation functions. Here, the differences are within errors on $\amuLW$ and $\amuLWTwo$ and decrease much faster than $a^2$.

In summary, truncation effects from numerical integration and discretization effects due to the oscillatory contributions are clearly well-controlled and small compared to other systematic effects.  To make certain that any systematic error due to integration and removal of oscillatory states is included, we include variations on both numerical integration and removal of oscillatory contributions in our main analysis, as described in \cref{sec:BMA}.

\begin{table}
\centering
\caption{$\amuLW$ and $\amuLWTwo$ computed from the fit reconstruction without oscillating states with the trapezoidal rule (columns two and five), Simpson's rule (columns three and six) and the correlated difference between them (columns four and seven). \vspace{1mm}}
\label{table:trapVSsimp}
\begin{tabular}{l|c|c|c|c|c|c}
\hline \hline
$\approx a$ &$a^{ll,\mathrm {W}}_{\mu,\, \text{ Simp.}}(\text{conn.})$ & $a^{ll,\text{W}}_{\mu,\, \text{ Trap.}}(\text{conn.})$ & $\Delta \amuLW$ &$a^{ll,\mathrm {W2}}_{\mu,\, \text{ Simp.}}(\mathrm{conn.})$ & $a^{ll,\mathrm{W2}}_{\mu,\, \text{ Trap.}}(\mathrm{conn.})$ & $\Delta \amuLWTwo$ \\
\hline
0.15 & 210.62(79) & 210.07(77) & 0.55(26) & 80.2(1.7) & 79.5(1.7) & 0.70(16) \\
0.12 & 207.34(59) & 206.96(61) & 0.373(49) & 84.6(1.5) & 84.9(1.5) & 0.254(59)  \\
0.09 & 206.56(55) & 206.60(55) & $-$0.039(12) & 92.7(1.8) & 92.7(1.8) & $-$0.01(13) \\
0.06 & 206.22(61) & 206.22(61) & 0.0002(691) & 95.5(2.7) & 95.5(2.7) & $-$0.0003(4285) \\
\hline\hline
\end{tabular}
\end{table}

\section{Chiral-model expressions for the Euclidean-space vacuum polarization function }\label{sec:chiralModel}

In this appendix, we provide expressions for calculating lattice corrections to $\amuHVP$ (and windows thereof) within the chiral model of pions, photons, and $\rho$ mesons denoted ``CM'' in Sec.~\ref{sec:analysis} and employed in our 2019 work~\cite{,Davies:2019efs}.  

We begin with Blum's formulation of the ${\mathcal O}(\alpha^2)$ Standard-Model HVP contribution as an integral over the Euclidean-space momentum transfer $Q^2$~\cite{Blum:2013xva}
\be
\amuHVP = 4\alpha^{2} \int_{0}^{\infty} \mathrm{d}Q^{2} K_{E}\big(Q^{2}\big) \widehat{\Pi}\big(Q^{2}\big), \label{em-rep}
\ee
where $\hat{\Pi}\left(Q^{2}\right)=\Pi\left(Q^{2}\right)-\Pi(0)$ is the renormalized vacuum polarization function and the integration kernel $K_{E}(Q^{2})$ depends on the muon's mass: 
\begin{align}
    K_E\left(Q^2\right) = \frac{m_\mu^2Q^2Z^3 (1-Q^2Z)}{1+m_\mu^2},  \quad\quad
    Z = - \frac{Q^2-(Q^4 + 4m_\mu^2 Q^2)^{1/2}}{2m_\mu^2Q^2}.
    \label{eq:KE}
\end{align}
In the chiral model~\cite{JegerlehnerModel,Chakraborty:2016mwy},
the renormalized light-quark hadronic vacuum polarization function is given by
\be
\hat{\Pi}\left(Q^{2}\right)=-\hat{\Sigma}\left(Q^{2}\right) + \frac{\hat{f}^{2}_{\rho}}{2 \hat{m}_{\rho}^{2}} \frac{q^{2}\left(1+g_{\rho} g_{\rho \pi \pi} \hat{\Sigma}\left(Q^{2}\right)\right)^{2}}{Q^{2}\left(1+g_{\rho \pi \pi}^{2} \hat{\Sigma}\left(Q^{2}\right)\right)+\hat{m}_{\rho}^{2}}, \label{cmVacPol}
\ee
where $\hat{\Sigma}\left(Q^{2}\right) \equiv \operatorname{Re} \Sigma\left(Q^{2}\right)-\Sigma(0)$ is the renormalized photon self energy and $\hat{m}_{\rho}$ ($\hat{f}_{\rho})$ are the renormalized $\rho$-meson mass (decay constant).
In the chiral model, the leading contribution to $\Sigma\left(Q^{2}\right)$ arises from $\pi\pi$ loops, and is given by the integral
\be
\begin{aligned}
&-\hat{\Sigma}\left(Q^{2}, m_{a}, m_{b}\right) \equiv
&\frac{4 Q^{2}}{3} \int \frac{\mathrm{d}^{3} \mathbf{k}}{(2 \pi)^{3} 2 E_{a} E_{b}} \frac{\mathbf{k}^{2}}{\left(E_{a}+E_{b}\right)^{3}\left(Q^{2}+\left(E_{a}+E_{b}\right)^{2}\right)} \;,
\end{aligned}
\label{eq:bubble_int}
\ee
where $m_{a}, m_{b}$ are the masses of the two pions in the loop.
The renormalized $\rho$ parameters can be expressed in terms of the bare mass, $\rho\gamma$ coupling, $\rho\pi\pi$ coupling, and $\Sigma(0)$ as
\bea
\hat{m}_{\rho}^{2} & \equiv & m_{0 \rho}^{2}\left(1-g_{\rho \pi \pi}^{2} \Sigma(0)\right) \\
\frac{\hat{f}_{\rho}}{\hat{m}_{\rho}} & \equiv & \frac{\sqrt{2}}{g_{\rho}}\left(1+g_{\rho} g_{\rho \pi \pi} \Sigma(0)-\frac{1}{2} g_{\rho \pi \pi}^{2} \Sigma(0)\right) \;.
\eea
We take the values of the bare parameters from Ref.~\cite{Chakraborty:2016mwy}:
\bea
m_{0 \rho}=0.766 \mathrm{GeV} \quad g_{\rho}=5.4 \quad g_{\rho \pi \pi}=6.0.
\eea

In the chiral model, lattice effects are incorporated by modifying the pion self energy in two ways. First, to account for the finite volume, the continuous momentum integrals in \cref{eq:bubble_int} are replaced by sums over the discrete lattice momenta, {\it i.e.,}
\be 
\int \frac{\mathrm{d}^{3} \mathbf{k}}{(2 \pi)^{3}} \rightarrow \frac{1}{L^{3}} \sum_{k_{x}=-\infty}^{\infty} \sum_{k_{y}=-\infty}^{\infty} \sum_{k_{z}=-\infty}^{\infty} \label{eq:discrete}
\ee
where $k_i = \frac{2\pi}{L}n_i$, $n_i =1,2,\ldots$. Second, taste-breaking effects are incorporated by replacing the renormalized photon self energy with an average over sea-pion tastes~\cite{ Chakraborty:2016mwy, Davies:2019efs}
\bea 
\hat{\Sigma}\left(Q^{2}, m_{\pi}, m_{\pi}\right) &\rightarrow& \frac{1}{16} \sum_{\xi_{a}, \xi_{b}} \hat{\Sigma}\left(Q^{2}, m_{\pi}\left(\xi_{a}\right), m_{\pi}\left(\xi_{b}\right)\right) . \label{eq:renorm}
\eea
As stated in \cref{sec:lat_corrections}, for the analysis in this work we also include taste-breaking contributions to $\Sigma(0)$ via the replacement
\be 
\Sigma\left(0, m_{\pi}, m_{\pi}\right) \rightarrow \frac{1}{16} \sum_{\xi_{a}, \xi_{b}} \Sigma\left(0, m_{\pi}\left(\xi_{a}\right), m_{\pi}\left(\xi_{b}\right)\right)\;. \label{eq:sigma}
\ee

Finally, the windowed HVP can be computed in the chiral model via~\cite{Borsanyi:2020mff}
\begin{equation} 
    \hat{\Pi}\left(Q^{2}\right) \rightarrow \hat{\Pi}_{\mathrm{win.}}\big(Q^2\big)=\int_{-\infty}^{\infty} \frac{\mathrm{d}P}{2\pi} \frac{1}{Q^2}\bigg[\widetilde{\mathcal{W}}(P-Q)-\widetilde{\mathcal{W}}(P)-\frac{Q^2}{2}\frac{\mathrm{d}^2 \widetilde{\mathcal{W}}(P)}{\mathrm{d}P^2}\bigg]P^2\hat{\Pi}\big(P^2\big),
    \label{eq:pihatq2}
\end{equation}
where $\hat{\Pi}$ is given by \cref{cmVacPol} and $\widetilde{\mathcal{W}}$ is the Fourier transform of the window function defined in \cref{eq:windofunc}.

\bibliographystyle{apsrev4-2}
\bibliography{refs}

\end{document}